\documentclass[12pt]{article}
\usepackage[usenames]{color}
\usepackage{amsmath,amssymb,euscript,cite,amsfonts,graphicx}
\usepackage{pstricks,fancybox}
\fontfamily{pnc}\selectfont

\newcommand{\SP}[1]{\begin{equation}\begin{split} #1 \end{split}\end{equation}}

\begin{document}
\setlength{\parskip}{2ex} \setlength{\parindent}{0em}
\setlength{\baselineskip}{3.6ex}
\newcommand{\onefigure}[2]{\begin{figure}[htbp]
         \caption{\small #2\label{#1}(#1)}
         \end{figure}}
\newcommand{\onefigurenocap}[1]{\begin{figure}[h]
         \begin{center}\leavevmode\epsfbox{#1.eps}\end{center}
         \end{figure}}
\renewcommand{\onefigure}[2]{\begin{figure}[htbp]
         \begin{center}\leavevmode\epsfbox{#1.eps}\end{center}
         \caption{\small #2\label{#1}}
         \end{figure}}
\newcommand{\comment}[1]{}
\newcommand{\myref}[1]{(\ref{#1})}
\newcommand{\secref}[1]{sec.~\protect\ref{#1}}
\newcommand{\figref}[1]{Fig.~\protect\ref{#1}}
\def\sl2z{SL(2,\Z)}
\newcommand{\be}{\begin{equation}}
\def\bp{\,{\hbox{P}\!\!\!\!\!\hbox{I}\,\,\,}}
\def\br{\,{\hbox{R}\!\!\!\!\!\hbox{I}\,\,\,}}
\def\bc{\,{\hbox{C}\!\!\!\hbox{I}\,\,\,}}
\def\bz{\,{\hbox{Z}\!\!\hbox{Z}\,\,\,}}

\newcommand{\ee}{\end{equation}}
\newcommand{\bea}{\begin{eqnarray}}
\newcommand{\eea}{\end{eqnarray}}
\newcommand{\nn}{\nonumber}
\newcommand{\unit}{1\!\!1}
\newcommand{\R}{\bf R}
\newcommand{\X}{{\bf X}}
\newcommand{\T}{{\bf T}}
\newcommand{\PP}{\bf P}
\newcommand{\CC}{\bf C}\def\Z{{\bf Z}}
\newdimen\tableauside\tableauside=1.0ex
\newdimen\tableaurule\tableaurule=0.4pt
\newdimen\tableaustep
\def\phantomhrule#1{\hbox{\vbox to0pt{\hrule height\tableaurule width#1\vss}}}
\def\phantomvrule#1{\vbox{\hbox to0pt{\vrule width\tableaurule height#1\hss}}}
\def\sqr{\vbox{%
\phantomhrule\tableaustep
\hbox{\phantomvrule\tableaustep\kern\tableaustep\phantomvrule\tableaustep}%
\hbox{\vbox{\phantomhrule\tableauside}\kern-\tableaurule}}}
\def\squares#1{\hbox{\count0=#1\noindent\loop\sqr
\advance\count0 by-1 \ifnum\count0>0\repeat}}
\def\tableau#1{\vcenter{\offinterlineskip
\tableaustep=\tableauside\advance\tableaustep by-\tableaurule
\kern\normallineskip\hbox
    {\kern\normallineskip\vbox
      {\gettableau#1 0 }%
     \kern\normallineskip\kern\tableaurule}%
  \kern\normallineskip\kern\tableaurule}}
\def\gettableau#1 {\ifnum#1=0\let\next=\null\else
  \squares{#1}\let\next=\gettableau\fi\next}

\tableauside=1.0ex \tableaurule=0.4pt

\setcounter{page}{1} \pagestyle{plain}

\definecolor{myorange}{rgb}{1,0.5,0}
\definecolor{darkslategray}{rgb}{0.18431,0.3098,0.3098}
\definecolor{darkslategrey}{rgb}{0.18431,0.3098,0.3098}
\definecolor{dimgray}{rgb}{0.41176,0.41176,0.41176}
\definecolor{dimgrey}{rgb}{0.41176,0.41176,0.41176}
\definecolor{slategray}{rgb}{0.43921,0.50195,0.5647}
\definecolor{slategrey}{rgb}{0.43921,0.50195,0.5647}
\definecolor{lightslategray}{rgb}{0.46666,0.53333,0.59999}
\definecolor{lightslategrey}{rgb}{0.46666,0.53333,0.59999}
\definecolor{gray}{rgb}{0.74509,0.74509,0.74509}
\definecolor{grey}{rgb}{0.74509,0.74509,0.74509}
\definecolor{lightgrey}{rgb}{0.82744,0.82744,0.82744}
\definecolor{lightgray}{rgb}{0.82744,0.82744,0.82744}

\begin{titlepage}
\begin{center}
\hfill hep-th/0701156\\ \vskip 1cm

{\Large{\bf  The Refined Topological Vertex }} \vskip 0.5cm
{\large {{Amer\,\,Iqbal$^{1}$, Can Koz\c{c}az$^{2}$, Cumrun Vafa$^{3,4}$}}\\
\vskip 0.5cm {{$^{1}$Department of Mathematics,\\ University of
Washington,\\ Seattle, WA, 98195, U.S.A.\\} \vskip 0.5cm
{$^{2}$Department of Physics,\\
University of Washington,\\
Seattle, WA, 98195, U.S.A.\\} \vskip 0.5cm {$^{3}$Jefferson Physical
Laboratory,\\
Harvard University,\\
Cambridge, MA, 02138, U.S.A.\\} \vskip 0.5cm {$^{4}$Center for
Theoretical Physics,\\
Massachusetts Institute of Technology,\\
 Cambridge, MA, 02139, U.S.A.\\}}}
\end{center}
\vskip 0.7 cm
\begin{abstract} We define a refined topological
vertex which depends in addition on a parameter, which physically
corresponds to extending the self-dual graviphoton field strength to
a more general configuration. Using this refined topological vertex
we compute, using geometric engineering, a two-parameter
(equivariant) instanton expansion of gauge theories which reproduce
the results of Nekrasov. The refined vertex is also expected to be
related to Khovanov knot invariants.
\end{abstract}

\end{titlepage}
 \baselineskip 0.01cm \tableofcontents

 \baselineskip 0.63cm
\section{Introduction}
The study of topological strings on Calabi-Yau manifolds has been
the topic of intense research for many years now. There are a number
of conjectures relating the topological string amplitudes with
various generating functions of interest to both physicists and
mathematicians.

The Calabi-Yau threefold (CY3-fold) $X$ gives rise to the corresponding
compactified theory via M-theory compactification. In this way gauge
theories with certain gauge groups and matter content can be
geometrically engineered using CY3-folds
\cite{KKV1,Katz:1997eq}. The topological string partition function
on such spaces is expected to be related to instanton sums in gauge
theories.  This conjecture has been sharpened, thanks to the work of
Nekrasov \cite{Nekrasov:2002qd}, which provides the tool to directly
compute the partition function of 5D supersymmetric gauge theory on
$\mathbb{C}^{2}\times S^{1}$.

On the other hand, using the topological vertex formalism
\cite{AKMV, iqbal} the partition function of topological string can
be evaluated on such backgrounds.  In particular, for $U(N)$ gauge
theories with and without hypermultiplets, the equivalence of gauge
theory and the corresponding topological string partition function
has been proven using the topological vertex formalism
\cite{KI1,KI2,HIV, KI3}. However, as was noted in \cite{HIV}, the
instanton calculus \cite{Nekrasov:2002qd} which was used to
calculate the gauge theory partition function has more refined
information. Recall that on the gauge theory side the partition
function is calculated using localization in equivariant K-theory
\cite{NY1,NY2} with respect to an $r+2$ dimensional torus
$\mathbb{T}^{2}\times \mathbb{K}$, where $\mathbb{K}$ is the $r$
dimensional maximal torus of the gauge group and $\mathbb{T}^{2}$
acts on the $\mathbb{C}^{2}$, \bea
\mathbb{T}^{2}:\,\,(z_{1},z_{2})\mapsto
\,\,(e^{i\epsilon_{1}}z_{1},e^{i\epsilon_{2}}z_{2})\,. \eea The
$\mathbb{T}^{2}$ action on $\mathbb{C}^{2}$ lifts to an action on
the instanton moduli space such that the fixed points are labeled
by the colored partitions (Young diagrams) of certain instanton charge
\cite{nakajima-book}.

The gauge theory partition function is a function of two equivariant
parameters $\epsilon_{1,2}$. For $\epsilon_{1}=-\epsilon_{2}=g_{s}$,
the gauge theory partition function reduces to the A-model
topological string partition function with genus parameter $g_{s}$
\cite{KI1, KI2}. In this limit ($\epsilon_{1}+\epsilon_{2}=0$), the
topological vertex formalism can be used to calculate
the partition function from the toric geometry of the corresponding
CY3-fold. However, the usual topological vertex formalism, needs
to be extended to deal with the case
$\epsilon_{1}+\epsilon_{2}\neq
0$.

Recall that the topological string partition function is the
generating function of the Gromov-Witten invariants. Therefore a
natural question to ask is whether the partition function with
$\epsilon_{1}+\epsilon_{2}\neq 0$ is the generating function of some
invariants more refined than the Gromov-Witten invariants. The
Gopakumar-Vafa (GV) reformulation \cite{GV}\,of the topological
string amplitudes suggests such a possibility which was explored in
\cite{HIV}. Given a CY3-fold $X$, the M-theory compactification on
$X$ gives (in an appropriate limit) a 5D supersymmetric gauge theory
with eight supercharges. The BPS particles in the 5D theory have a
geometric origin as the M2-branes wrapped on holomorphic curves in
$X$. The mass of such a particle coming from the holomorphic curve
$C\in H_{2}(X,\mathbb{Z})$ is given by $\int_{C}\omega$, where
$\omega$ is the K\"{a}hler form on $X$. The spin of these particles
is classified by the little group of massive particles which in 5D
is $SO(4)\simeq SU(2)_{L}\times SU(2)_{R}$. Compactifying on a
circle to get Type IIA on $X$, the wrapped M2-branes with some
momentum in the compact direction become the bound states of
D2-branes with D0-branes. The number of particles with charge $C\in
H_{2}(X,\mathbb{Z})$ and $SU(2)_{L}\times SU(2)_{R}$ spin
$(j_{L},j_{R})$, $N^{(j_{L},j_{R})}_{C}$, is equal to the number of the cohomology
classes of the moduli space of D2-brane wrapped on $C$. For generic
CY3-folds, $N^{(j_{L},j_{R})}_{C}$ is not an invariant and can change
as we change the complex structure. But
$N^{j_{L}}_{C}=\sum_{j_{R}}(-1)^{2j_{R}}(2j_{R}+1)N^{(j_{L},j_{R})}_{C}$,
which sums over all $j_R$'s with alternating signs, remains
\textit{invariant}. For the case of non-compact toric CY3-folds, there are no
complex structure deformations. Therefore, one would expect no jumps
in the $N^{(j_{L},j_{R})}_{C}$ degeneracies, and so one would hope
to be able to compute these as well.

Because the D-brane has a $U(1)$ gauge field living on its
worldvolume, the moduli space of supersymmetric configurations
includes not only the curve moduli but also the moduli of the flat
connections on the curve coming from the gauge field. Since the
moduli space of flat connections on a smooth curve of genus $g$ is
$T^{2g}$,  the moduli space of the D-brane is a $T^{2g}$ fibration
over the moduli space of the curve. The total space is a K\"ahler
manifold and the Lefschetz action by the K\"ahler class is the
diagonal $SU(2)_D\subset SU(2)_{L}\times SU(2)_{R}$ action on the
moduli space.  The $SU(2)_{L}\times SU(2)_{R}$ action on the moduli
space is such that $SU(2)_{L}$ acts on the fiber direction and the
$SU(2)_{R}$ acts in the base direction.

The topological string partition function is the generating function
of the invariants $N^{j_{L}}_{C}$, \bea
Z(\omega,g_{s})&:=&\mbox{exp} \Big(\sum_{g\geq
0}g_{s}^{2g-2}F_{g}(\omega)\Big)\\\nn &=&\prod_{C\in
H_{2}(X,\mathbb{Z})}\prod_{j_{L}}
\prod_{k_{L}=-j_{L}}^{+j_{L}}\prod_{m=0}^{\infty}
\Big(1-q^{2\,k+m+1}
\,Q^{C}\Big)^{(-1)^{2j_{L}+1}(m+1)N^{j_{L}}_{C}}\,, \eea where
$q=e^{ig_{s}}$ and $Q^{C}=e^{-\int_{C}\omega}$. The parameters $Q$
give the charge under $H_{2}(X,\mathbb{Z})$ whereas the parameter
$q$ couples with the $SU(2)_{L}$ spin.

As mentioned before, for Calabi-Yau manifolds which do not admit any
complex structure deformations, such as non-compact toric
threefolds, the multiplicities $N_{C}^{(j_{L},j_{R})}$ themselves
are invariants. Using these multiplicities we can define a refined
topological string partition function with a product structure
similar to the one given above \cite{HIV}, \bea\nn
Z(\omega,q,t)&:=&\prod_{C\in H_{2}(X,{\mathbb
Z})}\prod_{j_{L},j_{R}}
\prod_{k_{L}=-j_{L}}^{+j_{L}}\prod_{k_{R}=-j_{R}}^{+j_{R}}
\prod_{m_{1},m_{2}=1}^{\infty}
\Big(1-t^{k_{L}+k_{R}+m_{1}-\frac{1}{2}}\,q^{k_{L}-k_{R}+m_{2}-\frac{1}{2}}
Q^{C}\Big)^{M^{(j_{L},j_{R})}_{C}}\,,\\\\\nn
M^{(j_{L},j_{R})}_{C}&=&(-1)^{2(j_{L}+j_{R})+1}N^{(j_{L},j_{R})}_{C}\,,
\eea where the parameters $\sqrt{q\,t}$ and $\sqrt{\frac{t}{q}}$
couple with $SU(2)_{L}$ and $SU(2)_{R}$ spin, respectively.

It was argued in \cite{HIV} that for Calabi-Yau manifolds which give
rise to ${\cal N}=2$ supersymmetric gauge theories via geometric
engineering, the refined topological string partition function is
equal to the partition function of the compactified 5D gauge theory,
\textit{i.e.}, the K-theoretic version of the Nekrasov's instanton partition
functions \cite{NY1,NY2,ON} with
$q=e^{i\epsilon_{1}},t=e^{-i\epsilon_{2}}$.

The topological vertex formalism \cite{AKMV} provides a powerful method
to calculate the topological string partition function for
non-compact toric CY3-folds. A similar formalism to calculate the
refined partition functions will be very interesting providing a
refinement of the Gromov-Witten and Donaldson-Thomas theories of
toric CY3-folds. The purpose of this paper is to develop such a
formalism. We will define a refined topological vertex
$C_{\lambda\,\mu\,\nu}(t,q)$ which now depends on one extra
parameter compared to the ordinary topological vertex, where
together with the usual gluing algorithm for toric CY3-folds, gives
the refined topological string partition. However, the refined
vertex can be used to define the refined invariants only when the
toric Calabi-Yau threefold is made of vertices, all of which contain
a fixed locus $(p,q)$ of vanishing cycle in $T^2 $ (which is a
subset of the $T^3$ fibration of toric geometries).  This implies
that we can compute the refined topological string amplitudes only
for toric threefolds which are somewhat special.  However, one can
also obtain a generic toric case from the refined vertex by using
analytic continuation and doing flops on the vertices. This in
particular means that the refined vertex is not cyclically symmetric
as the usual topological vertex. The toric CY3-folds for which the
refined vertex works are exactly those which give rise to gauge
theories via geometric engineering. This implies that the refined
vertex contains no more information than the K-theoretic version of
the instanton partition functions. However, the refined vertex
provides a combinatorial interpretation of the instanton partitions
functions. Since the refined vertex is not cyclically symmetric a
certain choice of direction in the toric diagram of the CY3-fold has
to be made.

The fact that the topological vertex has a combinatorial
interpretation in terms of counting certain 3D partitions with
fixed asymptotes is a well known fact \cite{ORV}. {\it As a guiding
principal in formulating the refined topological vertex we will
demand a similar combinatorial interpretation in terms of 3D
partitions for the refined vertex}.\footnote{For another attempt at
defining a refined topological vertex see \cite{AK}.}

This paper is organized as follows. In section 2, we will review GV
formulation of the topological string amplitudes and their
computation using the topological vertex formalism. In section 3, we
propose the refined topological vertex.  In section 4, we discuss
the connection between the refined vertex and stacks of branes,
and motivate the gluing rules for the refined vertex.
  In section 5, we will calculate the refined
partition functions for certain geometries using the refined vertex.
In particular, we show how one recovers Nekrasov's results using the
refined vertex.  We also compute the degeneracy of the BPS states in
these geometries and explain the $SU(2)_L\times SU(2)_R$ content of
the states. In Appendix A, we will give the complete derivation of
the refined vertex in terms of 3D partitions. In Appendix B, we will
show that the refined partition function of ${\cal O}(-1)\oplus
{\cal O}(-1)\mapsto \mathbb{P}^{1}$ can be obtained by appropriately
weighting the contribution of the holomorphic maps to the two fixed
points of the geometry. We will also show that for $\mathbb{C}^{3}$
by appropriately weighting the contribution of the maps to the torus
invariant fixed point gives a generalization of the MacMahon
function which also has a combinatorial interpretation.

\section{GV Formulation and Topological Vertex}

In this section, we will briefly review the Gopakumar-Vafa
reformulation of the topological string amplitudes and their
calculation using the topological vertex.

\subsection{Topological string amplitudes and GV reformulation}

The topological string amplitudes $F_{g}$ arise in the A-twisted
topological theory as integrals over the genus $g$ moduli space of
Riemann surfaces and are related to the generating functions of the
genus $g$ Gromov-Witten invariants. The general form of these
amplitudes is given by \bea F_{0}(\omega)&=&
\frac{1}{3!}\int_{X}\omega\wedge \omega\wedge \omega+\sum_{C\in
H_{2}(X,\mathbb{Z})}\,{\cal
N}^{\,0}_{C}\,e^{-\int_{C}\omega}\,,\\\nn
F_{1}(\omega)&=&-\frac{1}{24}\int_{X}\omega\wedge
c_{2}(X)+\sum_{C\in H_{2}(X,\mathbb{Z})}\,{\cal
N}^{\,1}_{C}\,e^{-\int_{C}\omega},\\\nn F_{g\geq
2}(\omega)&=&(-1)^{g}\frac{\chi(X)}{2}\int_{\overline{{\cal M}}_{g}}\lambda_{g-1}^{3}+\sum_{C\in
H_{2}(X,\mathbb{Z})}\,{\cal N}^{\,g}_{C}\,e^{-\int_{C}\omega}\,,
\eea where $\omega$ is the K\"ahler form, ${\cal N}_{C}^{g}$ is the
genus g Gromov-Witten invariant of $C$, $\overline{{\cal M}}_{g}$ is the
moduli space of genus $g$ Riemann surfaces and $\lambda_{g-1}$ is
the $g^{th}$ Chern class of the Hodge bundle over $\overline{{\cal M}}_{g}$
(see Appendix B). The topological string amplitudes can be compactly
organized into the generating function, the topological string
partition function \bea Z(\omega,g_{s})=\mbox{exp}
\Big(\sum_{g=0}^{\infty}g_{s}^{2g-2}F_{g}(\omega)\Big)\,.\eea

From the worldsheet perspective, the genus $g$ amplitude, $F_{g}$,
is the generating function of the ``number'' of maps from a genus
$g$ Riemann surface to CY3-fold $X$. However, the target space
viewpoint provides a more physical interpretation of the generating
function $F(\omega,g_{s})$ \cite{GV}. We will briefly review this
interpretation since it is crucial in understanding the refined
partition functions. Recall that in M-theory compactification on
CY3-fold $X$ we get a 5D field theory with eight
supercharges. The particles in this theory come from quantization of
the moduli space of wrapped M2-branes on various 2-cycles of $X$.
These particles carry $SU(2)_{L}\times SU(2)_{R}$ quantum numbers
where $SU(2)_{L}\times SU(2)_{R}=SO(4)$ is the little group of
massive particles in 5D. If we compactify one direction, then the
particles can be interpreted as wrapped D2-branes and the Kaluza-Klein modes
as bound D0-branes. These charged particles when integrated out give
rise to the F-terms in the effective action. The contribution of a
particle of mass $m$ and in representation ${\cal R}$ of the
$SU(2)_{L}\times SU(2)_{R}$ to ${F}$ is given by \bea
S=\mbox{log}\,\mbox{det}(\Delta+m^2+2e\,\sigma_{L}{\cal F})\,=
\int_{\epsilon}^{\infty}\frac{ds}{s} \frac{\mbox{Tr}_{{\cal
R}}(-1)^{\sigma_{L}+\sigma_{R}} e^{-sm^2}e^{-2s e\sigma_{L}{\cal
F}}}{(2\sinh (se{\cal F}/2))^{2}}\,, \eea where $\sigma^{L}$ is the
Cartan of $SU(2)_{L}$ and arises because the graviphoton field
strength is self-dual. $e$ is the charge of the particle, and is
equal to its mass and we identify the graviphoton field strength
${\cal F}=g_s$. The mass of the particle is given by the area of the
curve on which the D2-brane is wrapped. An extra subtlety arises due
to D0-branes. In the lift to M-theory, we see that a wrapped
M2-brane comes with momentum in the circle direction, and therefore,
if we denote the mass of the M2-brane wrapping a curve class $C\in
H_{2}(X,{\mathbb Z})$ by $T_{C}$ then the mass of the M2-brane with momentum
$n$ is given by taking $T_{C}$ to $T_{C}+2\pi in/g_{s}$. Let us
denote by $N^{(j_{L},j_{R})}_{C}$ the number of BPS states coming
from an M2-brane wrapped on the holomorphic curve $C$, and the
left-right spin content under $SU(2)_{L}\times SU(2)_{R}$ given by
$(j_{L},j_{R})$. Then the total contribution coming from all
particles is obtained by summing over the momentum, the holomorphic
curves and the left-right spin content, \bea F&=&\sum_{C\in
H_{2}(X,{\mathbb Z})}\sum_{n\in {\mathbb Z}}\sum_{j_{L},j_{R}}N^{(j_{L},j_{R})}_{C}
\int_{\epsilon}^{\infty}\frac{ds}{s}\frac{\mbox{Tr}_{(j_{L},j_{R})}
(-1)^{\sigma_{L}+\sigma_{R}}e^{-sT_{C}-2\pi
in} e^{-2s\sigma_{L}\lambda_s}}{(2\sinh (s\lambda_s/2))^{2}}\,\\
\nn &=&\sum_{C\in
H_{2}(X,{\mathbb Z})}\sum_{k=1}^{\infty}\sum_{j_{L},j_{R}}N^{(j_{L},j_{R})}_{C}
e^{-kT_{C}}\frac{\mbox{Tr}_{(j_{L},j_{R})}(-1)^{\sigma_{L}+\sigma_{R}}
e^{-2k\lambda_s\sigma_{L}}}{k(2\sinh (k\lambda_s/2))^{2}}\,\\
\nn &=&\sum_{C\in
H_{2}(X,{\mathbb Z})}\sum_{k=1}^{\infty}\sum_{j_{L}}N^{j_{L}}_{C}
e^{-kT_{C}}\frac{\mbox{Tr}_{j_{L}}(-1)^{\sigma_{L}}
e^{-2k\lambda_s\sigma_{L}}}{k(2\sinh
(k\lambda_s/2))^{2}}\,,\,\,\,\,\mbox{where}\,\,\,\,
N^{j_{L}}_{C}=\sum_{j_{R}}N^{(j_{L},j_{R})}_{C}(-1)^{2j_{R}}(2j_{R}+1)\,.
\eea In terms of these integers $N_{C}^{j_{L}}$ one can write $F$ as
\bea F=\sum_{C\in
H_{2}(X,{\mathbb Z})}\sum_{k=1}^{\infty}\sum_{j_{L}}(-1)^{2j_{L}}N^{j_{L}}_{C}
e^{-kT_{C}}\Big(\frac{q^{-2j_{L}k}+\cdots
+q^{+2j_{L}k}}{k(q^{k/2}-q^{-k/2})^2}\Big)\,,\,\,\,q=e^{ig_{s}}\,.\\
\nn \eea If we turn on a constant graviphoton field strength which
is not self-dual $F=F_{1}\,dx^{1}\wedge dx^{2}+F_{2}\,dx^{3}\wedge
dx^{4}$, then we can write the contribution that comes from
integrating out the particle in representation ${\cal R}$ of
$SU(2)_{L}\times SU(2)_{R}$ as \bea
S:=\int_{\epsilon}^{\infty}\frac{ds}{s} \frac{\mbox{Tr}_{{\cal
R}}(-1)^{\sigma_{L}+\sigma_{R}}e^{-sm^2}e^{-2s e(\sigma_{L}F_{+}
+\sigma_{R}F_{-})}}{(2\sinh (seF_{1}/2))(-2\sinh (seF_{2}/2))}\,.
\eea Summing over the contributions from all particles as before we
get \SP{ &F(\omega, t,q)=
\\&\sum_{C\in H_{2}(X,{\mathbb Z})}\sum_{n=1}^{\infty}\sum_{j_{L},j_{R}}
\frac{(-1)^{2j_{L}+2j_{R}}\,N^{(j_{L},j_{R})}_{C}\Big((t\,q)^{-nj_{L}}+\cdots+
(t\,q)^{nj_{L}}\Big)\Big((\frac{t}{q})^{-nj_{R}}+\cdots +
(\frac{t}{q})^{nj_{R}}\Big)}{n(t^{n/2}-t^{-n/2})
(q^{n/2}-q^{-n/2})}\,e^{-nT_{C}}\,, } where
$q=e^{F_{1}},t=e^{F_{2}}$. The integers $N^{(j_{L},j_{R})}_{C}$ give
the degeneracy of particles with spin content $(j_{L},j_{R})$, and
charge $C$ and  are the number of cohomology classes with spin
$(j_{L},j_{R})$ of the moduli space of a D-brane wrapped on a
holomorphic curve in the class $C$
 \cite{GV}.

As an example, consider the local ${\mathbb P}^1\times {\mathbb
P}^{1}$ which we will denote by $X$. M-theory compactification on
$S^{1}\times X$ gives $SU(2)$ gauge theory with eight supercharges.
In this case, the gauge theory partition function was calculated in
\cite{Nekrasov:2002qd}. As we will show in the last section this
partition function can be obtained from the refined topological
vertex as well and is given by\footnote{$\nu_{1,2}$ are 2D
partitions, $\nu^{t}$ is the transpose partition and
$||\nu||^2=\sum_{i}\nu_{i}^2$.} \bea
Z(Q_{b},Q_{f},t,q)&=&\sum_{\nu_{1},\nu_{2}}Q_{b}^{|\nu_{1}|+|\nu_{2}|}Z(\nu_{1},\nu_{2};Q_{f},
t,q)\\\nn
Z(\nu_{1},\nu_{2};Q,t,q)&:=&\left(\frac{t}{q}\right)^{|\nu_{1}|+|\nu_{2}|}q^{||\nu_{1}^{t}||^{2}}\,
t^{||\nu_{2}||^{2}}\widetilde
{Z}_{\nu_{1}^{t}}(t,q)\widetilde{Z}_{\nu_{1}}(q,t)\widetilde{Z}_{\nu_{2}^{t}}(t,q)\widetilde{Z}_{\nu_{2}}(q,t)\,G(\nu_{1},\nu_{2},Q,t,q)\\\nn
G(\nu_{1},\nu_{2},Q,t,q)&=&\prod_{i,j=1}^{\infty}\frac{(1-Q\,q^{j-1}t^{i})(1-Q\,(q/t)\,q^{j-1}t^{i})}
{(1-Q\,q^{-\nu_{2,i}^{t}+j-1}\,t^{-\nu_{1,j}+i})(1-Q\,(q/t)\,q^{-\nu_{2,i}^{t}+j-1}\,t^{-\nu_{1,j}+i})}\eea
\bea
\widetilde{Z}_{\nu}(t,a)=\prod_{(i,j)\in\,\nu}(1-t^{a(i,j)+1}q^{\ell(i,j)})^{-1}\,,\,\,\,
a(i,j)=\nu^{t}_{j}-i\,,\,\,\ell(i,j)=\nu_{i}-j\,, \eea where
$-\mbox{log}(Q_{b,f})=T_{b,f}$ are the K\"ahler parameters
associated with the base and the fiber ${\mathbb P}^{1}$'s.

We can use the above partition function to calculate the BPS
degeneracies of various states corresponding to charge $C \in
H_{2}(X,{\mathbb Z})$. For example, consider the curve $2B+2F$, the canonical
class of the $\mathbb{P}^{1}\times \mathbb{P}^{1}$. This is a genus
one curve and therefore the corresponding moduli space will admit
non-trivial $SU(2)_{L}$ action. The spin content can be extracted
from the refined partition function and is given by \bea
\sum_{j_{L},j_{R}}N^{(j_{L},j_{R})}_{2B+2F}(j_{L},j_{R})=(\tfrac{1}{2},4)\oplus(0,\tfrac{7}{2})
\oplus(0,\tfrac{5}{2})\,. \eea To see that this is the correct
result note that the moduli space of $2B+2F$ together with its
Jacobian is given by a ${\mathbb P}^{7}$ bundle over ${\mathbb
P}^{1}\times {\mathbb P}^{1}$: pick a point in ${\mathbb
P}^{1}\times {\mathbb P}^{1}$, the moduli space of curves passing
through that point in the class $2B+2F$ is given by ${\mathbb
P}^{7}$. Thus the diagonal $SU(2)_{L}\times SU(2)_{R}$ action which
is just the Lefshetz action is given by \bea (\tfrac{1}{2})\otimes
(\tfrac{1}{2})\otimes (\tfrac{7}{2})=
(\tfrac{5}{2})\oplus2(\tfrac{7}{2})\oplus (\tfrac{9}{2})\,.\eea Note
that since $2B+2F$ is a genus one curve, the corresponding Jacobian
is also genus one, and therefore $j_{L}$ can only be $0$ or
$\frac{1}{2}$. From this restriction on $j_{L}$ and the above
diagonal action, we see that the unique left-right spin content is
given by \bea
(\tfrac{1}{2},4)\oplus(0,\tfrac{7}{2})\oplus(0,\tfrac{5}{2})\,, \eea
exactly as predicted by the partition function calculation

\subsection{Partition function from the topological vertex}
The topological vertex formalism \cite{AKMV} completely solves the
problem of calculating the topological string partitions for toric
CY3-folds. Consider the topological A-model with a toric non-compact
Calabi-Yau manifold \textit{X} as its target space. The amplitude of
this model is the sum over the holomorphic maps from a Riemann
surface $\Sigma_{g}$ of genus \textit{g} to the target Calabi-Yau
manifold \textit{X} where each term is weighted by the area of the
surface in \textit{X}. One can use the so-called toric diagrams (or
web diagrams) to encode the geometry of the target space as a
tri-valent graph on the plane. These diagrams show the degeneration
loci of the toric action on \textit{X}, \textit{i.e.}, along each edge of the
web one of the 1-cycles of the fiber $\mathbb{T}^{2}$ shrinks
leaving the dual cycle $S^{1}$. The basic idea behind the
topological vertex is to divide the corresponding toric diagram of
\textit{X} into tri-valent vertices, which, from physics point of
view, should be considered as placing Lagrangian
D-brane/anti-D-brane pairs to ``cut'' \textit{X}. Each tri-valent
vertex corresponds to a $\mathbb{C}^{3}$ patch.

\begin{figure}\begin{center}
$\begin{array}{c@{\hspace{1in}}c} \multicolumn{1}{l}{\mbox{}} &
    \multicolumn{1}{l}{\mbox{}} \\ [-0.53cm]
 \includegraphics[width=2in]{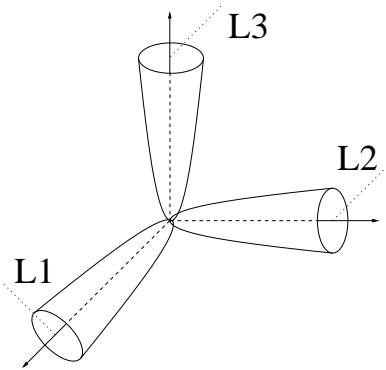}
\end{array}$
\caption{The holomorphic maps wrapping the disks along the
degeneration loci with boundaries on the Lagrangian branes.}
\label{f7}\end{center}
\end{figure}

The separation of the target space into ${\mathbb C}^{3}$ patches
results in cuts in the holomorphic maps from the worldsheet to the
target space as well. In other words, one ends up with Riemann
surfaces (not to be confused with $\Sigma_{g}$) with boundaries over
the point on the edge where the cut is made. The boundaries live on
stack of D-branes (or anti-D-branes) along the three edges of the
web. Closed string amplitudes on a given toric Calabi-Yau manifold are
obtained by an appropriate gluing procedure. The rules
to calculate the topological string amplitude on general toric,
non-compact Calabi-Yau manifolds, given the toric diagram are the
following:

\begin{itemize}
\item After dividing the toric diagram into vertices, associate each edge
described by an integer vector $v_{i}$ with a representation $\mu_{i}$.
\item The orientation of the vectors $(v_{i},v_{j},v_{k})$ describing the
degeneration loci is important in order to write down correctly the
associated vertex to each patch: if all vectors $v_{i}$ are incoming
then $C_{\mu_{i}\mu_{j}\mu_{k}}$ is the correct factor, otherwise
we replace the partition with its transpose $\mu^{t}$ for any
outgoing edge.
\item Once we have set up all vertices and the associated factors
$C_{\mu_{i}\mu_{j}\mu_{k}}$, we can glue them along their common
edges. Assume that two vertices have the same $v_{i}$, we can take one of the
$v_{i}$'s to be incoming on one vertex and outgoing on the other
one. Then ``gluing'' turns out to be the following summation
\bea
\sum_{\mu_{i}}(-1)^{(n_{i}+1)\ell(\mu_{i})}q^{-n_{i}\frac{\kappa(\mu_{i})}{2}}e^{-\ell(\mu_{i})t_{i}}C_{\mu_{j}\mu_{k}\mu_{i}}C_{\mu_{i}^{t}\mu'_{j}\mu'_{k}}
\eea with the integer $n_{i}=|v'_{k}\wedge v_{k}|$. The appearance
of this factor signals the equality of the framing along an edge on
both vertices.
\item The K\"{a}hler parameter $T_{i}$ associated to an edge
described by $v_{i}=(p_{i},q_{i})$ is given by
$T_{i}=x_{i}/\sqrt{p_{i}^{2}+q_{i}^{2}}$ where $x_{i}$
is the length in the plane.
\item The partition $\mu$ along any non-compact direction
is a trivial one and denoted by ``$\emptyset$''.
\end{itemize}

A useful representation of the vertex is given using the skew-Schur
functions \cite{ORV}, \bea C_{\lambda\, \mu \,\nu}(q)=
q^{\frac{\kappa(\mu)}{2}} s_{\nu^{t}}(q^{-\rho})\sum_{\eta}
s_{\lambda^{t}/\eta}(q^{-\nu-\rho})
s_{\mu/\eta}(q^{-\nu^{t}-\rho})\,, \label{tv1} \eea where
$q^{-\nu-\rho}=\{q^{-\nu_{1}+1/2},q^{-\nu_{2}+3/2},q^{-\nu_{3}+5/2},\cdots\}$,
and $s_{\mu/\eta}(x)$ is the skew-Schur function\footnote{For a
brief overview see Appendix D} defined, using the
Littlewood-Richardson coefficients $c^{\mu}_{\eta\,\lambda}$, in
terms of the Schur functions, \bea
s_{\mu/\eta}(x)=\sum_{\lambda}c^{\mu}_{\eta
\lambda}s_{\lambda}(x)\,. \eea

\section{The Refined Topological Vertex}

In this section, we will explain the combinatorial interpretation of
the refined vertex in terms of 3D partitions leaving the complete
derivation to Appendix A where the relevant notation is also
reviewed.

Recall that the generating function of the 3D partitions is given by
the MacMahon function, \bea M(q)=\sum_{n\geq
0}C_{n}q^{n}=\prod_{k=1}^{\infty}(1-q^{n})^{-n}\,. \eea The
topological vertex $C_{\lambda\,\mu\,\nu}(q)$ has the following
combinatorial interpretation \cite{ORV} \bea
M(q)C_{\lambda\,\mu\,\nu}(q)=\sum_{\pi(\lambda,\mu,\nu)}q^{|\pi(\lambda,\mu,\nu)|-|\pi_{\bullet}(\lambda,\mu,\nu)|}\,,
\eea where $\pi(\lambda,\mu,\nu)$ is a 3D partition such that along
the three axes it asymptotically approaches the three 2D partitions
$\lambda,\mu$ and $\nu$. $|\pi|$ is number of boxes (volume) of the
3D partition $\pi$ and $\pi_{\bullet}$ is the 3D partition with the
least number of boxes satisfying the same boundary
condition\footnote{Since even the partition with the least number of
boxes has infinite number of boxes we need to regularize this by
putting it in an $N\times N\times N$ box as discussed in
\cite{ORV}.}.

The refined vertex also has a similar combinatorial interpretation
in terms of 3D partitions which we will explain now. Recall that the
diagonal slices of a 3D partition, $\pi$, are 2D partitions which
interlace each other. These are the 2D partitions living on
the planes $x-y=a$, where $a\in \mathbb{Z}$. We will denote these 2D
partitions by $\pi_{a}$. For the usual vertex the $a^{th}$ slice is
weighted with $q^{|\pi_{a}|}$, where $|\pi_{a}|$ is the number of
boxes cut by the slice (the number of boxes in the 2D partition
$\pi_{a}$). The 3D partition is then weighted by\bea \prod_{a\in
\mathbb{Z}}q^{|\pi_{a}|}=q^{\sum_{a\in
\mathbb{Z}}|\pi_{a}|}=q^{\mbox{$\#$ of boxes in the $\pi$}} \eea In
the case of the refined vertex, the 3D partition is weighted in a
different manner. Given a 3D partition $\pi$ and its diagonal slices
$\pi_{a}$ we weigh the slices for $a<0$ with parameter $q$ and the
slices with $a\geq 0$ with parameter $t$ so that the measure
associated with $\pi$ is given by \bea
\Big(\prod_{a<0}q^{|\pi_{a}|}\Big)\,\Big(\prod_{a\geq
0}t^{|\pi_{a}|}\Big)=q^{\sum_{i=1}^{\infty}|\pi(-i)|}\,t^{\sum_{j=1}^{\infty}|\pi(j-1)|}\,.
\eea The generating function for this counting is a generalization
of the MacMahon function and is given by \bea M(t,q):=
\sum_{\pi}q^{\sum_{i=1}^{\infty}|\pi(-i)|}\,t^{\sum_{j=1}^{\infty}|\pi(j-1)|}=\prod_{i,j=1}^{\infty}(1-q^{i-1}t^{j})^{-1}\,.
\eea We can think of this assignment of $q$ and $t$ to the slices in
the following way. If we start from large positive $a$ and move
towards the slice passing through the origin then everytime we move
the slice towards the left we count it with $t$ and everytime we
move the slice up (which happens when we go from $a=i$ to $a=i-1$,
$i=0,1,2\cdots$) we count it with $q$.

Since we are slicing the skew 3D partitions with planes $x-y=a$ we
naturally have a preferred direction given by the $z$-axis. Let us take
the 2D partition along the $z$-axis to be $\nu$. The case we discussed
above, obtaining the refined MacMahon function, corresponds to $\nu=\emptyset$.
For a non-trivial, $\nu$ the assignment of $q$ and $t$ to various
slices is different and depends on the shape of $\nu$. As we go from
$+\infty$ to $-\infty$ the slices are counted with $t$ if we go
towards the left and is counted with $q$ if we move up. An example
is shown in \figref{ex}.

\begin{figure}\begin{center}
$\begin{array}{c@{\hspace{1in}}c} \multicolumn{1}{l}{\mbox{}} &
    \multicolumn{1}{l}{\mbox{}} \\ [-0.53cm]
{
\begin{pspicture}(7,0)(4,5)
\psframe[unit=0.75cm, linestyle=none, fillstyle=solid,
fillcolor=lightgray](0,0)(1,1) \psframe[unit=0.75cm, linestyle=none,
fillstyle=solid, fillcolor=lightgray](1,0)(2,1)
\psframe[unit=0.75cm, linestyle=none, fillstyle=solid,
fillcolor=lightgray](0,1)(1,2) \psframe[unit=0.75cm, linestyle=none,
fillstyle=solid, fillcolor=lightgray](1,1)(2,2)
\psframe[unit=0.75cm, linestyle=none, fillstyle=solid,
fillcolor=lightgray](2,0)(3,1)\psframe[unit=0.75cm, linestyle=none,
fillstyle=solid, fillcolor=lightgray](0,2)(1,3)
\psframe[unit=0.75cm, linestyle=none, fillstyle=solid,
fillcolor=lightgray](0,3)(1,4) \psframe[unit=0.75cm, linestyle=none,
fillstyle=solid, fillcolor=lightgray](1,2)(2,3)

 \psgrid[unit=0.75cm, subgriddiv=1,
gridcolor=myorange, %
gridlabelcolor=white]%
(0,0)(6,6) \psline[unit=0.75cm,
linecolor=red,linestyle=dashed](5,0)(6.2,1.2) \psline[unit=0.75cm,
linecolor=red, linestyle=dashed](4,0)(6.2,2.2) \psline[unit=0.75cm,
linecolor=red,linestyle=dashed](3,0)(6.2,3.2) \psline[unit=0.75cm,
linecolor=blue](3,1)(6.2,4.2) \psline[unit=0.75cm,
linecolor=red,linestyle=dashed](2,1)(6.2,5.2) \psline[unit=0.75cm,
linecolor=blue](2,2)(6.2,6.2) \psline[unit=0.75cm,
linecolor=blue](2,3)(5.2,6.2) \psline[unit=0.75cm,
linecolor=red,linestyle=dashed](1,3)(4.2,6.2) \psline[unit=0.75cm,
linecolor=blue](1,4)(3.2,6.2) \psline[unit=0.75cm,
linecolor=red,linestyle=dashed](0,4)(2.2,6.2) \psline[unit=0.75cm,
linecolor=blue](0,5)(1.2,6.2) \psline[unit=0.75cm,
linecolor=red,linestyle=dashed](6,0)(6.2,0.2) \psline[unit=0.75cm,
linecolor=blue](0,6)(.2,6.2) \psline[unit=0.75cm,
linecolor=myorange](0,0)(0,7) \psline[unit=0.75cm,
linecolor=myorange](0,0)(7,0)

\put(5,3){$\pi\rightarrow
\,\prod_{a\in\mathbb{Z}}q_{a}^{|\pi(a)|}=q^{\sum_{i=1}^{\infty}|\pi(\nu^{t}_{i}-i)|}\,t^{\sum_{j=1}^{\infty}|\pi(-\nu_{j}+j-1)|}$}
\put(5,2){$q=\mbox{blue}\,(\mbox{solid
line}),\,t=\mbox{red}\,(\mbox{dashed line})$}

\put(5,1){$\nu=(4,3,1)$}
\end{pspicture}
}
\\ [-0.5cm] \mbox{} & \mbox{}
\end{array}$
\caption{Slices of the 3$D$ partitions are counted with parameters
$t$ and $q$ depending on the shape of $\nu$.} \label{ex}\end{center}
\end{figure}

After taking into account the framing and the fact that the slices
relevant for the topological vertex are not the perpendicular slices
\cite{ORV} the generating function is given by \bea\nn
G_{\lambda\,\mu\,\nu}(t,q)=\Big(\frac{q}{t}\Big)^{\frac{||\mu||^2+||\nu||^2}{2}}\,t^{\frac{\kappa(\mu)}{2}}\,M(t,q)
P_{\nu^{t}}(t^{-\rho};q,t)\,
\sum_{\eta}\Big(\frac{q}{t}\Big)^{\frac{|\eta|+|\lambda|-|\mu|}{2}}s_{\lambda^{t}/\eta}(t^{-\rho}q^{-\nu})
s_{\mu/\eta}(t^{-\nu^{t}}q^{-\rho}), \eea and the refined vertex is
given by \bea
C_{\lambda\,\mu\,\nu}(t,q)&=&\frac{G_{\lambda\,\mu\,\nu}(t,q)}{M(t,q)}\\\nn
&=&\Big(\frac{q}{t}\Big)^{\frac{||\mu||^2+||\nu||^2}{2}}\,t^{\frac{\kappa(\mu)}{2}}\,
P_{\nu^{t}}(t^{-\rho};q,t)\,
\sum_{\eta}\Big(\frac{q}{t}\Big)^{\frac{|\eta|+|\lambda|-|\mu|}{2}}s_{\lambda^{t}/\eta}(t^{-\rho}q^{-\nu})
s_{\mu/\eta}(t^{-\nu^{t}}q^{-\rho}). \eea

In the above expression, $P_{\nu}({\bf x};q,t)$ is the Macdonald
function such that \bea
P_{\nu^{t}}(t^{-\rho};q,t)&=&t^{\frac{||\nu||^{2}}{2}}\,\widetilde{Z}_{\nu}(t,q)\,,\\\nn
\widetilde{Z}_{\nu}(t,q)&=&\prod_{(i,j)\in
\,\nu}(1-t^{a(i,j)+1}q^{\ell(i,j)})^{-1}\,,\,\,\,\,
a(i,j)=\nu^{t}_{j}-i\,,\,\,\,\ell(i,j)=\nu_{i}-j\,. \eea

\section{Open String Partition Function and $\mbox{Sym}^{\bullet}(\mathbb{C})$}
In order to gain some insight into the proposed refined vertex and
its gluing rules (to be discussed below) it is useful to recall the
connection between topological vertex and open string amplitudes in
the presence of stack of A-branes.

Let us consider the connection between  open string partition
function and the topological vertex. When a stack of Lagrangian
D-branes is ending on one of the legs of the $\mathbb{C}^{3}$ the
partition function is given by \bea
Z(q;V)=\sum_{\nu}C_{\emptyset\,\emptyset\,\nu}(q^{-1})\,\mbox{Tr}_{\nu}V\,.
\eea Since $\mbox{Tr}_{\nu}V=s_{\nu}({\bf x})$ where ${\bf
x}=\{x_{1},x_{2},\cdots\}$ are the eigenvalues of the holonomy
matrix $V$. \bea
Z(q;V)&=&\sum_{\nu}C_{\emptyset\,\emptyset\,\nu}(q^{-1})\,s_{\nu}({\bf
x})\\\nn &=& \sum_{\nu}s_{\nu^{t}}(q^{\rho})s_{\nu}({\bf
x})=\prod_{i,j=1}^{\infty}(1+q^{-i+\frac{1}{2}}x_{j})\,.\eea

In the case of a single Lagrangian brane ${\bf x}=(-Q,0,0,0,\cdots)$
we get the well known partition function \bea
Z(q;Q)=\prod_{i=1}^{\infty}(1-Q\,q^{-i+\frac{1}{2}})\,. \eea We will now show
that the above partition function of a single Lagrangian brane can
be interpreted in terms of the Hilbert series of the the symmetric
product of $\mathbb{C}$.

Recall that the Schur functions have the property that\bea
s_{\nu/\lambda}(Q)=\left\{
                     \begin{array}{ll}
                       Q^{|\nu|-|\lambda|}, & \nu \succ \lambda\\
                       0, & \mbox{otherwise}.
                     \end{array}
                   \right.
\eea This implies that $s_{\nu}(Q)$ is non-zero only for those
partitions for which $\ell(\nu)=1$, \textit{i.e.}, $\nu=(\nu_{1},0,0,\cdots)$.
These are exactly the partitions which label the fixed points of the
symmetric product of $\mathbb{C}$, \textit{i.e.}, $\mbox{Sym}^{k}(\mathbb{C})$
has a single fixed point labelled by the partition
$\nu=(k,0,0,\cdots)$. We can construct a generating function of the
Hilbert series of the symmetric products \cite{NY1}, \bea
G(\phi,q):=\sum_{k=0}^{\infty}\phi^{k}\,H[\,\mbox{Sym}^{k}(\mathbb{C})](q)
\eea Since the symmetric product $\mbox{Sym}^{k}(\mathbb{C})$ can be
identified with the ring
$R_{k}:=\mathbb{C}[z_{1},z_{2},\cdots,z_{k}]/S_{k}$ therefore the
Hilbert series is given by \cite{NY1}\bea
H[R](q)&=&\sum_{n=0}^{\infty}q^{n}\,r_{n}(R)\\\nn
r_{n}(R)&=&\mbox{\# of monomials in $R$ of charge $n$}\,\eea where
on $\mathbb{C}$ $q$ acts as a $\mathbb{C}^{\times}$ action $z\mapsto
qz$. To determine $H[\,\mbox{Sym}^{k}(\mathbb{C})](q)$ note that the
$R_{k}$ is just the ring of symmetric functions in the variables
$(z_{1},z_{2},\cdots,z_{k})$ and therefore the Schur functions
provide a basis of $R_{k}$, \bea
R_{k}=<\,s_{\nu}(z_{1},\cdots,z_{k})|\ell(\nu)\leq k\,>\,, \eea
where the condition $\ell(\nu)\leq k$ is necessary since
$s_{\nu}(z_{1},\cdots,z_{k})=0$ for $\ell(\nu)>k$. $R_{k}$ is
isomorphic to the Hilbert space ${\cal H}_{k}$ generated by bosonic
oscillator up to charge $k$. Recall that the bosonic oscillators
satisfying the commutation relation \bea
[\alpha_{n},\alpha_{m}]=n\delta_{n+m,0} \eea generate the Hilbert
space, ${\cal H}$, isomorphic to the ring of symmetric functions in
infinite variables $R$ . This essentially follows from the
identification \bea p_{\nu}({\bf
x})\leftarrow\,\,\,\,\nu=1^{m_{1}}2^{m_{2}}\cdots\,\,\,\,\rightarrow
(\alpha_{-1})^{m_{1}}(\alpha_{-2})^{m_{2}}\cdots |0\rangle. \eea
Under the above identification \bea R_{k}\simeq {\cal
H}_{k}=<(\alpha_{-1})^{m_{1}}\cdots
(\alpha_{-k})^{m_{k}}|0\rangle\,|\,\{m_{1},\cdots m_{k}\geq 0\}> \eea
and the Hilbert spaces ${\cal H}_{k}$ form a nested sequence \bea
{\cal H}_{0}\subset {\cal H}_{1}\subset {\cal H}_{2}\subset {\cal
H}_{3}\subset \cdots \eea which corresponds to the nested sequence
of Young diagrams of increasing number of rows.

The $\mathbb{C}^{\times}$ action, which lifts to an action on the $\mbox{Sym}^{\bullet}(\mathbb{C})$ such that the Schur functions
$s_{\nu}(z_{1},\cdots,z_{k})$ are  eigenfunctions with eigenvalue
$q^{|\nu|}$,  becomes the action of $q^{L_{0}}$ on the states in ${\cal H}$ $(L_{0}=\sum_{n>0}\alpha_{-n}\alpha_{n}$), \bea
H[R_{k}](q)=\mbox{Tr}_{{\cal H}_{k}}q^{L_{0}}=\sum_{\nu|\ell(\nu)\leq
k}\,q^{|\nu|}=\prod_{n=1}^{k}(1-q^{n})^{-1}\,. \eea The Hilbert
series of $R_{k}$ in this case turns out to be the generating
function of the number of partitions with at most $k$ parts. We can
express $H[R_{k}](q)$ in terms of the Schur functions, \bea
H[R_{k}](q)=
\prod_{n=1}^{k}(1-q^{n})^{-1}=s_{(k)}(1,q,q^{2},\cdots)\,.\eea The
generating functions $G(\phi,q)$ is then given by \bea
G(\phi,q)&=&\sum_{k=0}^{\infty}\phi^{k}H[R_{k}](q)=\sum_{k=0}^{\infty}\phi^{k}\mbox{Tr}_{{\cal H}_{k}}q^{L_{0}}\\\nn
&=& \sum_{k=0}^{\infty}\phi^{k}s_{(k)}(1,q,q^{2},\cdots)\\\nn
&=&
\sum_{k=0}^{\infty}s_{(k)}(\phi)s_{(k)}(1,q,q^{2},\cdots)=\sum_{\nu}s_{\nu}(\phi)s_{\nu}(1,q,q^{2},\cdots)\,\\\nn
&=&\sum_{\nu}s_{\nu}(q^{-\rho})s_{\nu}(\phi\,q^{-\frac{1}{2}})=\sum_{\nu}s_{\nu^{t}}(q^{\rho})\,s_{\nu}(-q^{-\frac{1}{2}}\phi)\\\nn
&=&\sum_{\nu}C_{\emptyset\,\emptyset\,\nu}(q^{-1})\,\mbox{Tr}_{\nu}V=Z(q;V),\eea
where $\mbox{Tr}_{\nu}V=s_{\nu}(Q)$ and $Q=q^{-\frac{1}{2}}\phi$.

Thus we see that as we move the brane to infinity ($Q=e^{-t}\mapsto
0$) the contribution of the higher modes is suppressed. On the other
hand as the brane moves towards the origin ($Q\mapsto 1$) higher
oscillator modes starts contributing with equal weight to the
partition function.

From the above discussion it also follows that the topological
vertex $C_{\emptyset\,\emptyset\,(k)}(q)$ has an interpretation as
counting the number of states of a given energy in the Hilbert space
${\cal H}_{k}$. It is tempting to conjecture that the topological
vertex with all three partitions non-trivial has a similar
interpretation. This is supported by the fact that topological
vertex when expanded as a powers series in $q$ has non-negative
integer coefficients.

It is easy to see that the recursion relation \bea \mbox{Tr}_{{\cal
H}_{k}}q^{L_{0}}=\frac{1}{1-q^{k}}\mbox{Tr}_{{\cal
H}_{k-1}}q^{L_{0}} \eea implies that the partition function $Z(q;Q)$
satisfies the equation \bea
(q^{-\partial_{u}}-1+\,q^{\frac{1}{2}}\,e^{-u})Z(q;e^{-u})=0\,. \eea
It is easy to determine the disk contribution using this
differential equation. Since
$Z(q;Q)=\mbox{exp}(\frac{F_{0}}{g_{s}}+F_{1}+g_{s}F_{2}+\cdots)$
therefore \bea
(g_{s}\partial_{u})^{n}e^{\frac{F}{g_{s}}}=e^{F/g_{s}}\{(\partial_{u}F)^{n}+O(g_{s})\}
\eea Therefore \bea
Z^{-1}(q;e^{-u})q^{-\partial_{u}}Z(q;e^{-u})=e^{-(\partial_{u}F)}+O(g_{s})
\eea which implies in the limit $g_{s}\mapsto 0$ \bea
\partial_{u}F_{0}=-\mbox{log}(1-q^{1/2}e^{-u})\,.
\eea
This relation was noted in \cite{ADKMV} where it was
related to the non-commutative geometry of the coordinates
on the local mirror geometry.  Below we will obtain a similar
equation in the context of the refined vertex, whose geometric
understanding is an important open question.

\subsection{Stack of Branes} In the previous subsection, we
considered of a single Lagrangian brane ending on one of the legs.
Now we will consider the case of multiple Lagrangian branes on the
one of legs of $\mathbb{C}^{3}$.

The partition function is given by \bea Z({\bf
x},q)&=&\sum_{\nu}C_{\emptyset\,\emptyset\,\nu}(q^{-1})\,s_{\nu}({\bf
x})\,,\,\,\,\,\,\,{\bf x}=\{x_{1},x_{2},\cdots,x_{N}\}\,,\\\nn
&=&\prod_{i=1}^{N}\prod_{j=1}^{\infty}(1+q^{-j+\frac{1}{2}}\,x_{i})\,.\eea

The above partition function is the generating function of the
Hilbert series of product of symmetric products of $\mathbb{C}$. To
see consider the following generating function \bea G(
\varphi_{1},\cdots,\varphi_{N},q)=\sum_{k_{1},\cdots,k_{N}}\varphi_{1}^{k_{1}}\varphi_{2}^{k_{2}}\cdots\varphi_{N}^{k_{N}}\,H[M_{k_{1}\cdots
k_{N}}](q)\,,\,\,\\\nn M_{k_{1}k_{2}\cdots
k_{N}}=\mbox{Sym}^{k_{1}}(\mathbb{C})\times \cdots\times
\mbox{Sym}^{k_{N}}(\mathbb{C})\,. \eea

The ring of functions on $M_{k_{1}\cdots k_{N}}$ is spanned by
\bea\nn s_{\nu_{1}}(x_{1,1},\cdots,
x_{1,k_{1}})s_{\nu_{2}}(x_{2,1},\cdots,x_{2,k_{2}})\cdots
s_{\nu_{N}}(x_{N,1},\cdots,x_{N,k_{N}})\,,\,\,\,\,\,\ell(\nu_{a})\leq k_{a}\,,a=1,2,\mathellipsis ,N.\eea

In terms of the bosonic oscillators $\alpha_{n}^{(a)}$ satisfying
the commutation relations \bea
[\alpha^{(a)}_{n},\alpha^{(b)}_{m}]=n\,\delta_{a,b}\delta_{m+n,0}\,.
\eea the above ring is isomorphic to the Hilbert space ${\cal
H}_{k_{1}\cdots k_{N}}$ spanned by \bea
\prod_{a=1}^{N}(\alpha^{(a)}_{-1})^{m_{a,1}}\cdots
(\alpha^{(a)}_{-k_{a}})^{m_{a,k_{a}}}|0\rangle\,. \eea The Hilbert
series of $M_{k_{1}\cdots k_{N}}$ is then given by the trace of
${\cal H}_{k_{1}\cdots k_{N}}$,\ \bea H[M_{k_{1}\cdots
k_{N}}](q)=\mbox{Tr}_{{\cal H}_{k_{1}\cdots k_{N}}}q^{L_{0}} \eea
where
$L_{0}=\sum_{a=1}^{N}\sum_{n>0}\alpha^{(a)}_{-n}\alpha^{(a)}_{n}$.
This implies that \bea\nn H[M_{k_{1}\cdots
k_{N}}](q)=\sum_{m_{1,1},m_{1,2}\cdots
m_{N,k_{N}}}q^{\sum_{a=1}^{N}\sum_{i=1}^{k_{a}}im_{a,i}}=\prod_{a=1}^{N}\sum_{m_{a,1},\cdots,m_{a,k_{a}}}q^{\sum_{i=1}^{k_{a}}im_{a,i}}=\prod_{a=1}^{N}\prod_{i=1}^{k_{a}}(1-q^{i})^{-1}\,.
\eea Since the Hilbert series is given by the product of the Hilbert
series therefore \bea G( \varphi_{1},\cdots,
\varphi_{N},q)=\prod_{a=1}^{N}G(\varphi_{a},q)=\prod_{a=1}^{N}\prod_{i=1}^{\infty}(1-q^{-i}\varphi_{a})=Z(q;{\bf
x})\,,\,\,\,\,\,\,x_{a}=-q^{-\frac{1}{2}}\,\varphi_{a}\,.\eea

\subsection{Refined vertex and open string partition function}
It is natural to expect that the refined vertex also has an
interpretation in terms of generalized open topological string
amplitudes in the presence of stacks of A-brane.  In fact the
results of \cite{GSV} suggests that the Khovanov knot invariants are
related to this refinement of the open string amplitude.  It thus
suggests that we should view the refined vertex as building blocks
for computation of Khovanov knot invariants that can be obtained from
local toric Calabi-Yau manifolds. The first step in motivating this interpretation
is to show that the stack of D-branes in the context of refined
vertex can also be related to the symmetric product of $\mathbb{C}$,
as it was possible in the context of ordinary vertex. This we will
show here.

When using the refined vertex the open string partition function
depends on the leg on which the stack of branes is put essentially
because the refined vertex is not cyclically symmetric. Thus we have
three choices corresponding to the three legs of $\mathbb{C}^{3}$.
We will consider all three cases, \bea
C_{\lambda\,\emptyset\,\emptyset}(t,q)&=&\Big(\tfrac{q}{t}\Big)^{\tfrac{|\lambda|}{2}}\,s_{\lambda^{t}}(t^{-\rho})\,,\\\nn
C_{\emptyset\,\mu\,\emptyset}(t,q)&=&\Big(\tfrac{q}{t}\Big)^{\tfrac{||\mu^{t}||^{2}-|\mu|}{2}}\,s_{\mu^{t}}(q^{-\rho})\,,\\\nn
C_{\emptyset\,\emptyset\,\nu}(t,q)&=&
\frac{q^{||\nu||^{2}/2}}{\prod_{s\in\nu}(1-t^{1+a(s)}\,q^{\ell(s)})}\,.\eea

{\bf $\mathbb{I}$:}\,The open string partition function is given by
\bea
Z(t,q,x)&=&\sum_{\lambda}C_{\lambda\,\emptyset\,\emptyset}(t^{-1},q^{-1})s_{\lambda}(x)=\sum_{\lambda}\Big(\tfrac{t}{q}\Big)^{\tfrac{|\lambda|}{2}}
s_{\lambda^{t}}(t^{\rho})s_{\lambda}(x)\\\nn
&=&\prod_{i=1}^{\infty}(1+x\,\sqrt{\tfrac{t}{q}}\,t^{-i+\frac{1}{2}})=\prod_{i=1}^{\infty}(1-Q\,t^{-i+\frac{1}{2}})\,,\,\,\,\,Q=-x\sqrt{\tfrac{t}{q}}\,.\eea

{\bf $\mathbb{II}$:}\,The open string partition function in this
case is given by \bea
Z(t,q,x)&=&\sum_{\mu}C_{\emptyset\,\mu\,\emptyset}(t^{-1},q^{-1})s_{\mu}(x)=\sum_{\mu}\Big(\tfrac{t}{q}\Big)^{\tfrac{||\mu^{t}||-|\mu|}{2}}
s_{\mu^{t}}(q^{\rho})s_{\mu}(x)\\\nn
&=&\prod_{i=1}^{\infty}(1+x\,q^{-i+\frac{1}{2}})=\prod_{i=1}^{\infty}(1-Q\,q^{-i+\frac{1}{2}})\,,\,\,\,\,Q=-x\,.\eea

Thus we see that in both these case the partition function is the
same as the partition function obtained from the ordinary vertex
except that the partition function depends on either $t$ or $q$
depending on the leg on which the brane ends.

{\bf $\mathbb{III}$:} The more interesting case is the third one in
which the brane ends on the prefered leg.  In this case the open
string amplitude using the refined vertex is given by \bea
Z(V,t,q)&=&\sum_{\nu}C_{\emptyset\,\emptyset\,\nu}(t^{-1},q^{-1})\mbox{Tr}_{\nu}V\\\nn
&=&\sum_{\nu}C_{\emptyset\,\emptyset\,\nu}(t^{-1},q^{-1})s_{\nu}({\bf
x}) \eea where ${\bf x}=\{x_{1},x_{2},\cdots\}$. Since \bea
C_{\emptyset\,\emptyset\,\nu}(t,q)=\frac{q^{||\nu||^{2}/2}}{\prod_{s\in\nu}(1-t^{1+a(s)}\,q^{\ell(s)})}\,,\,\,C_{\emptyset\,\emptyset\,\nu}(t^{-1},q^{-1})=\frac{(-1)^{|\nu|}\Big(\frac{t}{q}\Big)^{|\nu|/2}\,t^{||\nu^{t}||^{2}/2}}{\prod_{s\in
\nu}(1-t^{1+a(s)}\,q^{\ell(s)})}\eea therefore for ${\bf
x}=\{-Q,0,0,\cdots\}$ we get \bea
Z(Q,t,q)&=&\sum_{\nu}C_{\emptyset\,\emptyset\,\nu}(t^{-1},q^{-1})\,s_{\nu}(-Q)
=\sum_{k=0}^{\infty}C_{\emptyset\,\emptyset\,(k)}(t^{-1},q^{-1})(-Q)^{k}\\\nn
&=&\sum_{k=0}^{\infty}\Big(Q\frac{t}{\sqrt{k}}\Big)^{k}\,\prod_{n=1}^{k}(1-t\,q^{n-1})^{-1}\,.
\eea

The above partition function can also be written using a more
refined Hilbert series of the symmetric product of $\mathbb{C}$. The
Schur functions provide a basis of $R_{k}$. A Schur function
$s_{\nu}(z_{1},\cdots,z_{k})$ has charge $q^{|\nu|}$ under the
$\mathbb{C}^{\times}$ action for $\ell(\nu)\leq k$. We define a
second $\mathbb{C}^{\times}$ action such that \bea
\mathbb{C}^{\times}:\,s_{\nu}(z_{1},z_{2},\cdots,z_{k})\mapsto
\,\vartheta^{\ell(\nu^{t})}\,s_{\nu}(z_{1},z_{2},\cdots,z_{k})\,,\,\,\,\ell(\nu)\leq
k\,. \eea Note that this second $\mathbb{C}^{\times}$ action has a
simple interpretation in terms of the bosonic oscillators. On ${\cal
H}_{k}$ this second $\mathbb{C}^{\times}$ acts as $\vartheta^{N}$
where $N=\sum_{n>0}\frac{\alpha_{-n}\alpha_{n}}{n}$ is the operator
that counts the number of total number of particles in a given state
$|\nu\rangle$ . On a state
$|\nu\rangle=|1^{m_{1}}2^{m_{2}}\cdots\rangle$ the operator $N$ acts
as \bea N|1^{m_{1}}2^{m_{2}}\cdots
\rangle=(m_{1}+m_{2}+\cdots)|1^{m_{1}}2^{m_{2}}\cdots\rangle \eea
Using this second $\mathbb{C}^{\times}$ action  the refined Hilbert
series of $R_{k}$ \bea H[R_{k}](q,\vartheta)=\mbox{Tr}_{{\cal
H}_{k}}q^{L_{0}}\vartheta^{N}=\sum_{\nu|\ell(\nu)\leq
k}q^{|\nu|}\,\vartheta^{\ell(\nu^{t})}\,. \eea For $\ell(\nu)\leq k$
the partition $\nu$ can be written as
$\nu=1^{m_{1}}2^{m_{2}}3^{m_{3}}\cdots k^{m_{k}}$ such that
$|\nu|=\sum_{i=1}^{k}im_{i}$ and $\ell(\nu^{t})=\sum_{i=1}^{k}m_{i}$
therefore \bea H[R_{k}](q,\vartheta)&=&\sum_{\nu|\ell(\nu)\leq
k}q^{|\nu|}\vartheta^{\ell(\nu)}\\\nn &=&
\sum_{m_{1},m_{2},\cdots,m_{k}}q^{m_{1}+2m_{2}+3m_{3}+\cdots+km_{k}}\,\vartheta^{m_{1}+m_{2}+\cdots
m_{k}}=\prod_{i=1}^{k}(1-\vartheta\,q^{i})^{-1}\,.\eea Then the
generating function of the refined Hilbert series is given by \bea
G(\phi,q,\vartheta)&=&\sum_{k=0}^{\infty}\phi^{k}\,H[R_{k}](q,\vartheta)=\sum_{k=0}^{\infty}\phi^{k}\,\prod_{i=1}^{k}(1-\vartheta\,q^{i})^{-1}\,\\\nn
&=&Z(\phi\frac{\sqrt{q}}{t},\vartheta\,q,q)\,. \eea

The refined partition function also satisfies an equation similar to the one satisfied by the quantum dilogarithm,
\bea
(\vartheta\,q^{-\partial_{u}}-1+\vartheta\,q^{\frac{1}{2}}\,e^{-u}\,)Z(e^{-u},\vartheta\,q,q)=\vartheta-1\,.
\eea
where $Q=e^{-u}$.

Understanding the geometric meaning of this relation is an open
problem which is important for a deeper understanding of the refined
vertex.

\subsection{Brane orientation and the gluing rule}
\label{gluerule} We have seen previously that both the topological
vertex and its refinement can be understood in terms of symmetric
products of $\mathbb{C}$. The appearance of the
$\mbox{Sym}^{\bullet}(\mathbb{C})$ can be understood if we embed the
topological string in the physical Type IIA string theory
\cite{vafa}. In this case, the Lagrangian branes become the D4-branes
wrapping the Lagrangian 3-cycle in the CY3-fold and filling up two
dimensions of the transverse four dimensions. The appearance of the
symmetric product can then be interpreted as counting particles in
two dimensions.

The refined topological vertex depends on two parameters $t,q$ which
in the instanton calculus corresponds to the $U(1)$ rotation
parameters of the two orthogonal planes in $\mathbb{C}^{2}$. For the
branes on the two unpreffered directions the open topological string
partition function depends only on either $t$ or $q$. This suggests
that branes on two unpreferred directions actually fill the two
different orthogonal planes in $\mathbb{C}^{2}$. To obtain the
closed string partition function we have to glue two edges of the
$\mathbb{C}^{3}$ vertex. From the instanton calculus we know that
the closed string partition function can be obtained by counting
points in $\mathbb{C}^{2}$. Therefore the gluing to obtain the
closed string expression must be such that the two stack of branes,
on the two legs which are to be glued, fill up orthogonal two
dimensional planes in the $\mathbb{C}^{2}$ transverse to the
CY3-fold.

Even though this gluing rule is very natural and we will see that it
works, a deeper explanation of this is needed.  In particular the
asymmetry of the refined vertex is a feature that has to be
explained in terms of the orientation of the Lagrangian branes:  the
unpreferred directions have branes that fill two orthogonal
subspaces of $\mathbb{C}^2$.  But we also need to have an
explanation of the Lagrangian brane on the preferred direction. This
we leave for future work.  This is also related to the Khovanov knot
invariant interpretation of the refined vertex:  It should be
possible to compute the Khovanov invariants (for toric knots at
least) using the refined vertex, as we noted above.  This is
currently under investigation \cite{future}.

\section{Refined Partition Functions from the Refined Vertex}

In this section, we will use the refined topological vertex to
determine the generalized partition function for various local toric
CY3-folds.

\subsection{${\cal O}(-1)\oplus {\cal O}(-1)\mapsto \mathbb{P}^{1}$}
The compactification of Type IIA string theory on the Calabi-Yau
threefold ${\cal X}={\cal O}(-1)\oplus {\cal O}(-1)\mapsto
\mathbb{P}^{1}$ gives rise to $U(1)$ ${\cal N}=2$ gauge theory on
the transverse $\mathbb{C}^{2}$ in a particular limit \cite{KKV1}.
Using the topological vertex formalism, the topological string
partition function is given by \bea
Z(q,Q)&=&\sum_{\nu}Q^{|\nu|}(-1)^{|\nu|}\,C_{\emptyset\,\emptyset\,\nu}(q)\,C_{\emptyset\,\emptyset\,\nu^{t}}(q)\\\nn
&=&
\sum_{\nu}Q^{|\nu|}(-1)^{|\nu|}s_{\nu^{t}}(q^{-\rho})s_{\nu}(q^{-\rho})\\\nn
&=&\prod_{i,j=1}^{\infty}\Big(1-Q\,q^{i+j-1}\Big)=\prod_{k=1}^{\infty}\Big(1-q^{k}\,Q\Big)^{k}\,,\eea
where $T=-\mbox{ln}(Q)$ is the K\"{a}hler parameter, the size
of the $\mathbb{P}^{1}$.

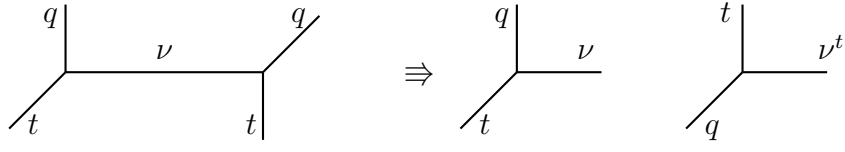
\begin{figure}\begin{center}
$\begin{array}{c@{\hspace{1in}}c} \multicolumn{1}{l}{\mbox{}} &
    \multicolumn{1}{l}{\mbox{}} \\ [-0.53cm]
{
\begin{pspicture}(6,-1)(4,3)
\psline[unit=0.75cm](2,1.5)(5.5,1.5) \psline[unit=0.75 cm](2,
1.5)(2,2.7)\psline[unit=0.75cm](2,1.5)(1,0.5)
\psline[unit=0.75cm](5.5,1.5)(5.5,0.3)
\psline[unit=0.75cm](5.5,1.5)(6.5,2.5) \put(2.7, 1.3){$\nu$}

\put(6,1){$\Rrightarrow$}

\put(1.2,1.8){$q$} \put(1.0,0.3){$t$} \put(4.5,1.8){$q$}
\put(3.9,0.3){$t$} \psline[unit=0.75cm](10,1.5)(11.5,1.5)
\psline[unit=0.75cm](10,1.5)(10,2.7)
\psline[unit=0.75cm](10,1.5)(9,0.5) \put(8.3,1.3){$\nu$}
\put(7.2,1.8){$q$} \put(7,0.3){$t$}

\psline[unit=0.75cm](14,1.5)(15.5,1.5)
\psline[unit=0.75cm](14,1.5)(14,2.7)
\psline[unit=0.75cm](14,1.5)(13,0.5)
\put(11.5,1.3){$\nu^{t}$}\put(10.2,1.8){$t$} \put(10,0.3){$q$}
\label{conifold}
\end{pspicture}
}
\\ [-1.0cm] \mbox{} & \mbox{}
\end{array}$
\caption{Toric diagram of ${\cal O}(-1)\oplus {\cal O}(-1)\mapsto
\mathbb{P}^{1}$. The vertices are glued along the preferred direction $\nu$.} \label{f15}\end{center}
\end{figure}

We can use the refined topological vertex to determine the refined
partition function. The toric diagram of ${\cal X}$ and the gluing
of the refined vertex are shown in \figref{f15}.
From the gluing of the vertices in \figref{f15}, we see that the
refined topological string partition function is given by \bea
Z(t,q,Q):=\sum_{\nu}Q^{|\nu|}(-1)^{|\nu|}\,C_{\emptyset\,\emptyset\,\nu}(t,q)\,\,C_{\emptyset\,\emptyset\,\nu^{t}}(q,t)\,.\eea
Since \bea
C_{\emptyset\,\emptyset\,\nu}(t,q)=q^{\frac{||\nu||^{2}}{2}}\,\widetilde{Z}_{\nu}(t,q)
= q^{\frac{||\nu||^{2}}{2}}\,\prod_{s\in
\nu}(1-t^{a(s)+1}q^{\ell(s)})^{-1}\,, \eea the refined partition
function becomes\footnote{An equivalent expression obtained by
$\nu\mapsto \nu^{t}$ is given by
$$\sum_{\nu}\frac{Q^{|\nu|}(-1)^{|\nu|}\,q^{\frac{||\nu^{t}||^{2}}{2}}\,t^{\frac{||\nu||^{2}}{2}}}
{\prod_{s\in
\nu}(1-t^{\ell(s)+1}q^{a(s)})(1-t^{\ell(s)}q^{a(s)+1})}$$.} \bea
Z(t,q,Q)&=&\sum_{\nu}\frac{Q^{|\nu|}(-1)^{|\nu|}\,q^{\frac{||\nu||^{2}}{2}}\,t^{\frac{||\nu^{t}||^{2}}{2}}}
{\prod_{s\in
\nu}(1-t^{a(s)+1}q^{\ell(s)})(1-t^{a(s)}q^{\ell(s)+1})}\,.\eea

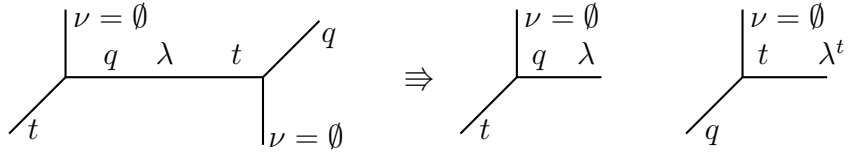
\begin{figure}\begin{center}
$\begin{array}{c@{\hspace{1in}}c} \multicolumn{1}{l}{\mbox{}} &
    \multicolumn{1}{l}{\mbox{}} \\ [-0.53cm]
{
\begin{pspicture}(6,-1)(4,3)
\psline[unit=0.75cm](2,1.5)(5.5,1.5) \psline[unit=0.75 cm](2,
1.5)(2,2.7)\psline[unit=0.75cm](2,1.5)(1,0.5)
\psline[unit=0.75cm](5.5,1.5)(5.5,0.3)
\psline[unit=0.75cm](5.5,1.5)(6.5,2.5) \put(2.7, 1.3){$\lambda$}
\put(2,1.3){$q$} \put(1.0,0.3){$t$}

\put(1.6,1.8){$\nu=\emptyset$} \put(4.2,.2){$\nu=\emptyset$}
\put(4.9,1.6){$q$}

 \put(6,1){$\Rrightarrow$}

\put(3.7,1.3){$t$} \psline[unit=0.75cm](10,1.5)(11.5,1.5)
\psline[unit=0.75cm](10,1.5)(10,2.7)
\psline[unit=0.75cm](10,1.5)(9,0.5) \put(8.3,1.3){$\lambda$}
\put(7.7,1.3){$q$} \put(7,0.3){$t$} \put(7.6,1.8){$\nu=\emptyset$}

\psline[unit=0.75cm](14,1.5)(15.5,1.5)
\psline[unit=0.75cm](14,1.5)(14,2.7)
\psline[unit=0.75cm](14,1.5)(13,0.5)
\put(11.5,1.3){$\lambda^{t}$}\put(10.7,1.3){$t$} \put(10,0.3){$q$}
\put(10.6,1.8){$\nu=\emptyset$} \label{diff}
\end{pspicture}
}
\\ [-1.0cm] \mbox{} & \mbox{}
\end{array}$
\caption{Toric diagram of ${\cal O}(-1)\oplus {\cal O}(-1)\mapsto
\mathbb{P}^{1}$. The vertices are glued along the unpreferred
direction $\lambda$ } \label{f16}\end{center}
\end{figure}

This is exactly the partition function given in Eq. (4.5) of \cite{NY1}
if we identify $(t_{1},t_{2}, \mathfrak{q})
=(t,q^{-1},Q\sqrt{\frac{t}{q}})$. A different representation of the
partition function can be obtained by choosing different preferred
directions as shown in \figref{f16}.\\

The refined partition function with this choice is given by \bea
Z(t,q,Q)&=&\sum_{\lambda}Q^{|\lambda|}(-1)^{|\lambda|}\,C_{\lambda\,\emptyset\,\emptyset}(t,q)\,
C_{\lambda^{t}\,\emptyset\,\emptyset}(q,t)\\\nn &=&
\sum_{\lambda}(-Q)^{|\lambda|}\Big(\tfrac{q}{t}\Big)^{\frac{|\lambda|}{2}}\,
s_{\lambda^{t}}(t^{-\rho})\,\Big(\tfrac{t}{q}\Big)^{\frac{|\lambda|}{2}}\,s_{\lambda}(q^{-\rho})\,\\\nn
&=&\sum_{\lambda}(-Q)^{|\lambda|}s_{\lambda^{t}}(t^{-\rho})\,s_{\lambda}(q^{-\rho})\,=\prod_{i,j=1}^{\infty}(1-Q\,q^{i-\frac{1}{2}}\,t^{j-\frac{1}{2}})\,\\\nn
&=&\,\mbox{Exp}\left \{
-\sum_{n=1}^{\infty}\frac{Q^{n}}{n(q^{\frac{n}{2}}-q^{-\frac{n}{2}})(t^{\frac{n}{2}}-t^{-\frac{n}{2}})}\right
\}\,. \eea Identifying the above two representations of the
partition function we get the following identity \bea
\sum_{\nu}\frac{Q^{|\nu|}(-1)^{|\nu|}\,q^{\frac{||\nu||^{2}}{2}}\,t^{\frac{||\nu^{t}||^{2}}{2}}}
{\prod_{s\in
\nu}(1-t^{a(s)+1}q^{\ell(s)})(1-t^{a(s)}q^{\ell(s)+1})}=\mbox{Exp}\left\{-\sum_{n=1}^{\infty}\frac{Q^{n}}{n(q^{\frac{n}{2}}-q^{-\frac{n}{2}})(t^{\frac{n}{2}}-t^{-\frac{n}{2}})}\right
\} \eea which is a specialization of the identity Eq. (5.4) of
\cite{macdonald} and was also derived in \cite{NY1}.

Note that in gluing the two vertices, we have have taken the
parameters $q$ or $t$ be the different on the gluing edges as
discussed in section \ref{gluerule}. The parameter does not have to
be the same as the vertices are actually an infinite distance apart.
Actually, one can also check that a combinatorial description of the
partition function requires that the parameters be different on the
two gluing edges.

\subsection{$\chi_{y}$-genus, $\mbox{Sym}^{\bullet}(\mathbb{C}^2$) and the
refined topological vertex} In \cite{HIV}, it was shown that the the generating
function of the equivariant $\chi_{y}$-genus of the Hilbert scheme
of $\mathbb{C}^{2}$, denoted by $\mbox{Hilb}^{k}[\mathbb{C}^{2}]$, is given by the topological string amplitude of
a certain CY3-fold $X_{0}$, which is the partial compactification of ${\cal X}$. The equivariant action of
$\mathbb{C}^{2}$ was given by $(z_{1},z_{2})\mapsto
(q\,z_{1},q^{-1}\,z_{2})$. Here we will show that the refined
partition function of $X_{0}$ is given by similar generating
function for which the equivariant action is given by
$(z_{1},z_{2})\mapsto (q\,z_{1},t^{-1}z_{2})$. The generating
function is given by \bea
G(\varphi,y,t,q)=\sum_{k=0}^{\infty}\varphi^{k}\chi_{y}(\mbox{Hilb}^{k}[\mathbb{C}^{2}])\,.
\eea and will be calculated using the localization. The fixed points
of $\mbox{Hilb}^{k}[\mathbb{C}^{2}]$ under the above two parameter
action are labeled by the 2D partitions of $n$ \cite{nakajima-book}.
The weight at the fixed point labeled by the partition $\nu$ is
given by \cite{haiman,NY1}\bea \sum_{i,j}e^{w_{i,j}}=\sum_{(i,j)\in
\,\nu}(t^{1+a(i,j)}q^{\ell(i,j)}+t^{-a(i,j)}q^{-1-\ell(i,j)}) \eea
Using these weights the $\chi_{y}$-genus of
$\mbox{Hilb}^{k}[\mathbb{C}^{2}]$ is given by \bea
\chi_{y}(\mbox{Hilb}^{k}[\mathbb{C}^{2}])=\sum_{i}\prod_{j=1}^{2k}
\frac{1-ye^{-w_{i,j}}}{1-e^{-w_{i,j}}}\,,\eea where $i$ label the
fixed points (the partitions of $k$ in this case) and $j$ label the
weights at a given fixed point, \bea
\chi_{y}(\mbox{Hilb}^{k}[\mathbb{C}^{2}])=\sum_{\nu,|\nu|=k}\prod_{(i,j)\in\,\nu}
\frac{(1-yt^{1+a(i,j)}q^{\ell(i,j)})(1-yt^{-a(i,j)}q^{-1-\ell(i,j)})}
{(1-t^{1+a(i,j)}q^{\ell(i,j)})(1-t^{-a(i,j)}q^{-1-\ell(i,j)})}\eea
And the generating function is given by \bea
G(\varphi,y,t,q)=\sum_{\nu}\varphi^{|\nu|} \prod_{(i,j)\in\,\nu}
\frac{(1-yt^{1+a(i,j)}q^{\ell(i,j)})(1-yt^{-a(i,j)}q^{-1-\ell(i,j)})}
{(1-t^{1+a(i,j)}q^{\ell(i,j)})(1-t^{-a(i,j)}q^{-1-\ell(i,j)})}\eea
The above generating function can be simplified to an expression
which can be compared with the refined vertex calculation more
easily, \bea G(\varphi,y,t,q)&=&\sum_{\nu}\varphi^{|\nu|}
q^{\frac{||\nu||^{2}}{2}}t^{\frac{||\nu^{t}||^{2}}{2}}\widetilde{Z}_{\nu}(t,q)\widetilde{Z}_{\nu^{t}}(q,t)\\\nn
&&\times
\prod_{(i,j)\in\,\nu}(1-y\,t^{1+a(i,j)}q^{\ell(i,j)})(1-y\,t^{-a(i,j)}q^{-1-\ell(i,j)})
\label{gf2} \eea

Now we will calculate the refined partition function of the CY3-fold
$X_{0}$. The toric diagram of the CY3-fold $X_{0}$ is shown in Fig.
5.
\begin{figure}
\begin{center}
\begin{pspicture}(3,-1)(4,3)
\psline[unit=0.75cm](2,1.5)(5.5,1.5) \psline[unit=0.75 cm](2,
1.5)(2,2.7)\psline[unit=0.75cm](2,1.5)(1,0.5)
\psline[unit=0.75cm](5.5,1.5)(5.5,0.3)
\psline[unit=0.75cm](5.5,1.5)(6.5,2.5) \put(2.7, 1.3){$\nu$}
\psline[unit=0.75cm,linecolor=red](5.7,0.5)(5.3,0.5)
\psline[unit=0.75cm,linecolor=red](5.7,0.4)(5.3,0.4)
\psline[unit=0.75cm,linecolor=red](1.8,2.6)(2.2,2.6)
\psline[unit=0.75cm,linecolor=red](1.8,2.5)(2.2,2.5)
\put(1,1.6){$\lambda$} \put(4.4,0.3){$\lambda$} \label{chiytoric}
\end{pspicture}
\caption{Toric diagram of partially compactified ${\cal O}(-1)\oplus
{\cal O}(-1)\mapsto \mathbb{P}^{1}$.}
\end{center}
\end{figure}
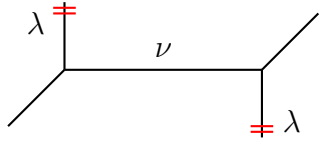
In this case the topological string partition function is given by
\bea \nn
Z(Q_{1},Q_{2})&:=&\sum_{\lambda,\nu}(-Q_{1})^{|\nu|}(-Q_{2})^{|\lambda|}\,C_{\lambda\,\emptyset\,\nu}(t,q)\,C_{\lambda^{t}\,\emptyset\,\nu^{t}}(q,t)\\\nn
&=&\sum_{\nu,\lambda}(-Q_{1})^{|\nu|}(-Q_{2})^{|\lambda|}\,q^{\frac{||\nu||^{2}}{2}}\,\widetilde{Z}_{\nu}(t,q)\,
s_{\lambda^{t}}(t^{-\rho}q^{-\nu})\,t^{\frac{||\nu^{t}||^{2}}{2}}\,\widetilde{Z}_{\nu^{t}}(q,t)\,s_{\lambda}(t^{-\nu^{t}}q^{-\rho})\\\nn
&=&\sum_{\nu}(-Q_{1})^{|\nu|}\,q^{\frac{||\nu||^{2}}{2}}\,t^{\frac{||\nu^{t}||^2}{2}}\widetilde{Z}_{\nu}(t,q)\,\widetilde{Z}_{\nu^{t}}(q,t)\sum_{\lambda}
(-Q)^{|\lambda|}s_{\lambda^{t}}(t^{-\rho}q^{-\nu})\,s_{\lambda}(t^{-\nu^{t}}q^{-\rho})\\\nn
&=&\sum_{\nu}(-Q_{1})^{|\nu|}\,q^{\frac{||\nu||^{2}}{2}}\,t^{\frac{||\nu^{t}||^2}{2}}\widetilde{Z}_{\nu}(t,q)\,\widetilde{Z}_{\nu^{t}}(q,t)
\prod_{i,j=1}^{\infty}(1-Q_{2}\,t^{i-\frac{1}{2}-\nu^{t}_{j}}\,q^{j-\frac{1}{2}-\nu_{i}})\,.
\eea

For $Q_{1}=0$ we get the perturbative (in the gauge theory sense)
contribution. The instanton part is then given by \bea\nn
\frac{Z(Q_{1},Q_{2})}{Z(0,Q_{2})}&=&\sum_{\nu}(-Q_{1})^{|\nu|}\,q^{\frac{||\nu||^{2}}{2}}\,t^{\frac{||\nu^{t}||^2}{2}}\widetilde{Z}_{\nu}(t,q)\,\widetilde{Z}_{\nu^{t}}(q,t)
\prod_{i,j=1}^{\infty}\Big(\frac{1-Q_{2}\,t^{i-\frac{1}{2}-\nu^{t}_{j}}\,q^{j-\frac{1}{2}-\nu_{i}}}{1-Q_{2}t^{i-\frac{1}{2}}q^{j-\frac{1}{2}}}\Big)\,\\\nn
&=&\sum_{\nu}(-Q_{1})^{|\nu|}\,q^{\frac{||\nu||^{2}}{2}}\,t^{\frac{||\nu^{t}||^2}{2}}\widetilde{Z}_{\nu}(t,q)\,\widetilde{Z}_{\nu^{t}}(q,t)
\\\nn
&&\times\prod_{(i,j)\in
\nu}\Big(1-Q_{2}\,t^{i-\frac{1}{2}-\nu^{t}_{j}}\,q^{j-\frac{1}{2}-\nu_{i}}\Big)\Big(
1-Q_{2}t^{\nu^{t}_{j}-i+\frac{1}{2}}q^{\nu_{i}-j+\frac{1}{2}}\Big
)\,.
 \eea
The above partition function is exactly the generating function of
the $\chi_{y}$-genus after the identification \bea
\varphi=-Q_{1}\,,\,\,\,y=Q_{2}\sqrt{\tfrac{q}{t}}\,.\eea

\subsection{${\cal O}(0)\oplus {\cal O}(-2)\mapsto \mathbb{P}^{1}$}
This geometry can be obtained from local $\mathbb{P}^{1}\times
\mathbb{P}^{1}$ by taking the size of one of the $\mathbb{P}^{1}$
very large. This limit gives two copies of ${\cal O}(0)\oplus {\cal
O}(-2)\mapsto \mathbb{P}^{1}$.

In the usual topological vertex formalism the partition function is
given by \bea
Z(q,Q)&=&\sum_{\nu}Q^{|\nu|}(-1)^{|\nu|}C_{\emptyset\,\emptyset\,\nu}(q)\,(-1)^{|\nu|}\,q^{\frac{\kappa(\nu)}{2}}\,C_{\emptyset\,\emptyset\,\nu^{t}}(q)\,\\\nn
&=&\sum_{\nu}Q^{|\nu|}s_{\nu^{t}}(q^{-\rho})\,q^{\frac{\kappa(\nu)}{2}}\,s_{\nu}(q^{-\rho})
=\sum_{\nu}Q^{|\nu|}s_{\nu^{t}}(q^{-\rho})\,s_{\nu^t}(q^{-\rho})\,\\\nn
&=&\,\prod_{i,j=1}^{\infty}\Big(1-Q\,q^{i+j-1}\Big)^{-1}\,=\prod_{k=1}^{\infty}\Big(1-Q\,q^{k}\Big)^{-k}\\\nn
&=&
\mbox{Exp}\left\{\sum_{n=1}^{\infty}\frac{Q^{n}}{n(q^{\frac{n}{2}}-q^{-\frac{n}{2}})^2}\right
\}. \eea

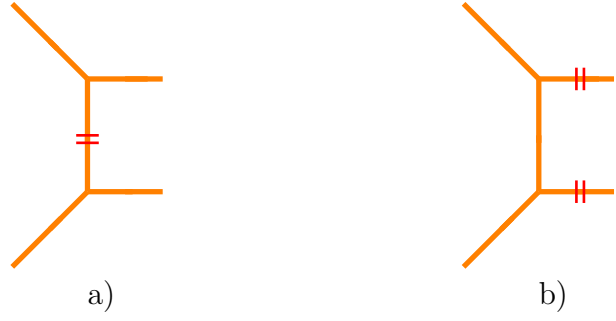
\begin{figure}\begin{center}
$\begin{array}{c@{\hspace{1in}}c} \multicolumn{1}{l}{\mbox{}} &
    \multicolumn{1}{l}{\mbox{}} \\ [-0.53cm]
{
\begin{pspicture}(0,0)(4,4)

\psline[unit=0.5cm,linewidth=2pt,linecolor=myorange](0,2)(0,3.5)
\psline[unit=0.5cm,linewidth=2pt,linecolor=myorange](0,3.3)(0,5)
\psline[unit=0.5cm,linewidth=2pt,linecolor=myorange](0,2)(1.2,2)
\psline[unit=0.5cm,linewidth=2pt,linecolor=myorange](1,2)(2,2)
\psline[unit=0.5cm,linewidth=2pt,linecolor=myorange](0,5)(1.6,5)
\psline[unit=0.5cm,linewidth=2pt,linecolor=myorange](1,5)(2,5)
\psline[unit=0.5cm,linewidth=2pt,linecolor=myorange](0,2)(-1,1)
\psline[unit=0.5cm,linewidth=2pt,linecolor=myorange](-2,0)(-0.7,1.3)
\psline[unit=0.5cm,linewidth=2pt,linecolor=myorange](0,5)(-1,6)
\psline[unit=0.5cm,linewidth=2pt,linecolor=myorange](-0.8,5.8)(-2,7)

\psline[unit=0.5cm,linewidth=2pt,linecolor=myorange](12,2)(12,3.5)
\psline[unit=0.5cm,linewidth=2pt,linecolor=myorange](12,3.3)(12,5)
\psline[unit=0.5cm,linewidth=2pt,linecolor=myorange](12,2)(13.2,2)
\psline[unit=0.5cm,linewidth=2pt,linecolor=myorange](13,2)(14,2)
\psline[unit=0.5cm,linewidth=2pt,linecolor=myorange](12,5)(13.6,5)
\psline[unit=0.5cm,linewidth=2pt,linecolor=myorange](13,5)(14,5)
\psline[unit=0.5cm,linewidth=2pt,linecolor=myorange](12,2)(11,1)
\psline[unit=0.5cm,linewidth=2pt,linecolor=myorange](10,0)(11.3,1.3)
\psline[unit=0.5cm,linewidth=2pt,linecolor=myorange](12,5)(11,6)
\psline[unit=0.5cm,linewidth=2pt,linecolor=myorange](11.2,5.8)(10,7)

\psline[unit=0.5cm,linewidth=1pt,linecolor=red](-0.3,3.5)(0.3,3.5)
\psline[unit=0.5cm,linewidth=1pt,linecolor=red](-0.3,3.3)(0.3,3.3)

\psline[unit=0.5cm,linewidth=1pt,linecolor=red](13,4.7)(13,5.3)
\psline[unit=0.5cm,linewidth=1pt,linecolor=red](13.2,4.7)(13.2,5.3)
\psline[unit=0.5cm,linewidth=1pt,linecolor=red](13,1.7)(13,2.3)
\psline[unit=0.5cm,linewidth=1pt,linecolor=red](13.2,1.7)(13.2,2.3)
\put(0,-0.5){a)}

\put(6,-0.5){b)}

\end{pspicture}}
\\ [-0.2cm] \mbox{} & \mbox{}
\end{array}$
\caption{Two possible choices for the preferred direction, the
internal line (a) and the parallel external lines (b).}
\label{twochoices}\end{center}
\end{figure}

In this case to define the refined partition function we have two
choices for the preferred direction as shown in \figref{twochoices}.
The refined partition function for the case (a) of
\figref{twochoices} is given by \bea
Z(t,q,Q)&=&\sum_{\nu}Q^{|\nu|}(-1)^{|\nu|}\,C_{\emptyset\,\emptyset\,\nu}(t,q)\,f_{\nu}(t,q)\,C_{\emptyset\,\emptyset\,\nu^{t}}(q,t)\\\nn
&=&\sum_{\nu}(-Q)^{|\nu|}\widetilde{Z}_{\nu}(t,q)\widetilde{Z}_{\nu^{t}}(q,t)\,q^{\frac{||\nu||^{2}}{2}}\,t^{\frac{||\nu^{t}||^{2}}{2}}\,f_{\nu}(t,q)\\\nn
&=&\sum_{\nu}\frac{(Q\sqrt{\tfrac{q}{t}})^{|\nu|}\,t^{||\nu^{t}||^{2}}}{\prod_{s\in
\nu}(1-t^{a(s)+1}\,q^{\ell(s)})(1-t^{a(s)}\,q^{\ell(s)+1})}\,.\eea

The partition function for case (b) of \figref{twochoices} is given
by, \bea
Z(t,q,Q)&=&\sum_{\lambda}Q^{|\lambda|}(-1)^{|\lambda|}\,C_{\emptyset\,\lambda\,\emptyset}(t,q)\,f_{\lambda}(t,q)\,\,
C_{\lambda^{t}\,\emptyset\,\emptyset}(t,q)\,\\\nn &=&
\sum_{\lambda}(-Q)^{|\lambda|}\Big(\frac{q}{t}\Big)^{\frac{||\lambda||^2}{2}}
t^{\frac{\kappa(\lambda)}{2}}\,s_{\lambda^{t}}(q^{-\rho})\,f_{\lambda}(t,q)\,s_{\lambda}(t^{-\rho})\,\,\,\\\nn
&=&\sum_{\lambda}(Q\sqrt{\tfrac{q}{t}})^{|\lambda|}\,s_{\lambda^{t}}(t^{-\rho})\,s_{\lambda^{t}}(q^{-\rho})
=\prod_{i,j=1}^{\infty}\Big(1-Q\,q^{i}\,t^{j-1}\Big)^{-1}\\\nn
&=&\,\mbox{Exp}\left \{
\sum_{n=1}^{\infty}\frac{Q^{n}\Big(\frac{q}{t}\Big)^{\frac{n}{2}}}{n(q^{\frac{n}{2}}-q^{-\frac{n}{2}})(t^{\frac{n}{2}}-t^{-\frac{n}{2}})}\right
\}\,. \eea

The partition functions corresponding to the two choices are
actually equal to each other (after scaling $Q$ by $\sqrt{q/t}$) as can be seen by using the summation
formulas for the Macdonald functions.

\subsection{Another toric geometry: $\widetilde{{\mathbb C}^{3}/\mathbb{Z}_{2}\times\mathbb{Z}_{2}}$}
\par
\begin{figure}\begin{center}
\includegraphics[width=2.5in]{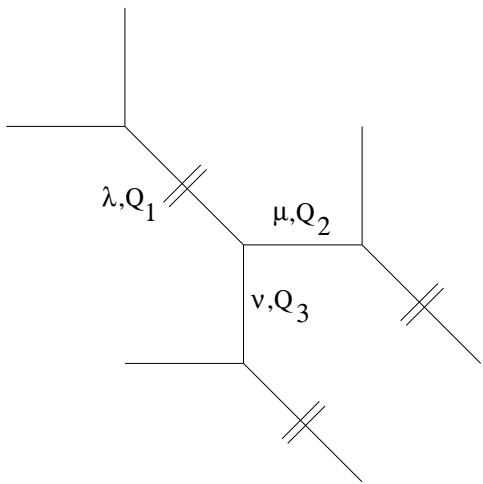}
\caption{A toric geometry with three $\mathbb{P}^{1}$'s.}
\label{geo}
\end{center}\end{figure}

{The geometry in \figref{geo} is an interesting
geometry consisting of a vertex with all non-trivial representations
in the middle. In the limit of vanishing $Q_{1}$ and $Q_{3}$ (with
$\lambda=\emptyset$), \textit{i.e}, sending the lower and upper-most
vertices to infinity, we recover our previous result for the
conifold. }

The refined partition function is given by \bea
Z&=&\sum_{\lambda,\mu,\nu}
(-Q_{1})^{|\lambda|}(-Q_{2})^{|\mu|}(-Q_3{})^{|\nu|}
C_{\emptyset\,\emptyset\,\lambda}(t,q)\,C_{\mu\,\nu\,\lambda^{t}}(q,t)\,C_{\mu^{t}\,\emptyset\,\emptyset}(t,q)\,C_{\emptyset\,\nu^{t}\,\emptyset}(t,q)\\
\nn
&=&\sum_{\lambda,\mu,\nu,\eta}(-Q_{1})^{|\lambda|}q^{\frac{\|\lambda\|^{2}}{2}}t^{\frac{\|\lambda^{t}\|^{2}}{2}}\widetilde{Z}_{\lambda}(t,q)\widetilde{Z}_{\lambda^{t}}(q,t)\left
( \frac{t}{q}\right
)^{\frac{|\eta|}{2}}s_{\mu^{t}/\eta}(q^{-\rho}t^{-\lambda^{t}})s_{\nu/\eta}(q^{-\lambda}t^{-\rho})\\
\nn &\times& s_{\mu}(-Q_{2} t^{-\rho})s_{\nu^{t}}(-Q_{3}q^{-\rho})\\
\nn &=&
\sum_{\lambda}(-Q_{1})^{|\lambda|}q^{\frac{\|\lambda\|^{2}}{2}}t^{\frac{\|\lambda^{t}\|^{2}}{2}}\widetilde{Z}_{\lambda}(t,q)\widetilde{Z}_{\lambda^{t}}(q,t)\prod_{i,j=1}^{\infty}\frac{(1-Q_{2}\,q^{-\rho_{j}}t^{-\lambda^{t}_{j}-\rho_{i}})(1-Q_{3}\,q^{-\lambda_{j}-\rho_{i}}t^{-\rho_{j}})}{(1-Q_{2}Q_{3}\,q^{-\rho_{i}-1/2}t^{-\rho_{i}+1/2})}
\eea

Note that for $Q_{1}=0$ the product representation of the refined
partition function is consistent with having two $\mathbb{P}^{1}$'s
with normal bundle ${\cal O}(-1)\oplus {\cal O}(-1)$ such that the
sum of the two $\mathbb{P}^{1}$'s can be deformed into a
$\mathbb{P}^{1}$ with normal bundle ${\cal O}(0)\oplus {\cal
O}(-2)$.

\subsection{Local $\mathbb{P}^{1}\times \mathbb{P}^{1}$}
\begin{figure}\begin{center}
$\begin{array}{c@{\hspace{1in}}c} \multicolumn{1}{l}{\mbox{}} &
    \multicolumn{1}{l}{\mbox{}} \\ [-0.53cm]
{

\begin{pspicture}(0,0)(4,4)
\pspolygon[unit=0.5cm,
linewidth=2pt,linecolor=myorange](-3,2)(-3,5)(0,5)(0,2)(-3,2)
\psline[unit=0.5cm,
linewidth=1pt,linestyle=dashed,linecolor=black](-1.5,1)(-1.5,6)
\psline[unit=0.5cm, linewidth=2pt,linecolor=myorange](-3,2)(-5,0)
\psline[unit=0.5cm, linewidth=2pt,linecolor=myorange](-3,5)(-5,7)
\psline[unit=0.5cm, linewidth=2pt,linecolor=myorange](0,5)(2,7)
\psline[unit=0.5cm, linewidth=2pt,linecolor=myorange](0,2)(2,0)

\psline[unit=0.5cm,linewidth=2pt,linecolor=myorange](10,2)(10,3.5)
\psline[unit=0.5cm,arrows=<-,linewidth=2pt,linecolor=myorange](10,3.3)(10,5)

\psline[unit=0.5cm,linewidth=2pt,linecolor=myorange](10,2)(11.2,2)
\psline[unit=0.5cm,arrows=<-,linewidth=2pt,linecolor=myorange](11,2)(12,2)

\psline[unit=0.5cm,arrows=->,linewidth=2pt,linecolor=myorange](10,5)(11.6,5)
\psline[unit=0.5cm,linewidth=2pt,linecolor=myorange](11,5)(12,5)

\psline[unit=0.5cm,linewidth=2pt,linecolor=myorange](10,2)(9,1)
\psline[unit=0.5cm,arrows=->,linewidth=2pt,linecolor=myorange](8,0)(9.3,1.3)

\psline[unit=0.5cm,arrows=->,linewidth=2pt,linecolor=myorange](10,5)(9,6)
\psline[unit=0.5cm,linewidth=2pt,linecolor=myorange](9.2,5.8)(8,7)

\put(6.1,2.4){$\nu_{1}$}

\put(6.1,0.9){$\nu_{2}$}

\put(6.5,1.65){\large{$=\,Z_{\nu_{1},\nu_{2}}(t,q,Q_{f})$}}

\end{pspicture}}
\\ [-0.2cm] \mbox{(a)} & \mbox{(b)}
\end{array}$
\caption{(a) Toric diagram of local $\mathbb{P}^{1}\times
\mathbb{P}^{1}$, (b) a slice of the toric diagram used to compute
the partition function.} \label{f0}\end{center}
\end{figure}
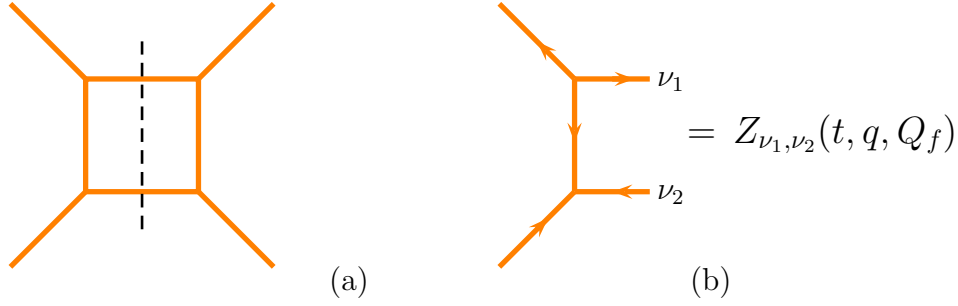

In this section, we will use the refined vertex to calculate the
refined A-model partition function for the local
$\mathbb{P}^{1}\times \mathbb{P}^{1}$. The toric geometry for this
case is shown in \figref{f0}. The parallel horizontal edges in the
rectangle correspond to the base $\mathbb{P}^{1}$. We cut the toric
diagram perpendicular to these parallel lines following \cite{KI1}.
The 2D partitions on the parallel edges will be denoted by
$\nu_{1}$ and $\nu_{2}$. The half of the toric diagram corresponds
to a geometry ${\cal O}(0)\oplus {\cal O}(-2)\mapsto \mathbb{P}^{1}$
with a stack of D-branes on the two parallel edges in the
representation $\nu_{1},\nu_{2}$. We denote the open topological
string partition function by $Z_{\nu_{1},\nu_{2}}(t,q, Q_{f})$ where
$T_{f}=-\mbox{log}\,Q_{f}$ is the K\"{a}hler parameter of the fiber
$\mathbb{P}^{1}$. The two parts of the toric diagram are identical.
This implies that the open topological string partition function
associated with both sides is the same,
$Z_{\nu_{1},\nu_{2}}(t,q,Q_{f})$. The only subtlety arises in how these
two open string partition functions are ``glued'' together to form the
closed string partition function. This gluing information is
contained in the normal geometry of the base curve and determines
the framing factors.

From the toric geometry, \figref{f0}, it is clear that locally
the two $\mathbb{P}^{1}$'s corresponding to the base curve (the
upper and lower parallel edges) are ${\cal O}(0)\oplus {\cal
O}(-2)\mapsto \mathbb{P}^{1}$ and ${\cal O}(-2)\oplus {\cal
O}(0)\mapsto \mathbb{P}^{1}$. Therefore the framing factor with the
top edge is $f_{\nu_{1}}(t,q)$ and with the lower edge is
$f_{\nu_{2}}(q,t)$.

With this choice of framing factors the generalized partition
function is given by \bea
Z(Q_{b},Q_{f},t,q):=\sum_{\nu_{1},\nu_{2}}(-Q_{b})^{|\nu_{1}|+|\nu_{2}|}Z_{\nu_{1},\nu_{2}}(t,q,Q_{f})\,f_{\nu_{1}}(t,q)\,f_{\nu_{2}}(q,t)
\,Z_{\nu_{2},\nu_{1}}(q,t,Q_{f}), \eea where \bea\nn
f_{\nu_{1}}(t,q)\,f_{\nu_{2}}(q,t)&=&(-1)^{|\nu_{1}|}\Big(\frac{t}{q}\Big)^{\frac{||\nu_{1}^{t}||^{2}-|\nu_{1}|}{2}}\,q^{-\frac{\kappa(\nu_{1})}{2}}\,\,(-1)^{|\nu_{2}|}
\Big(\frac{q}{t}\Big)^{\frac{||\nu_{2}^{t}||^{2}-|\nu_{2}|}{2}}\,t^{-\frac{\kappa(\nu_{2})}{2}}\eea
and \bea\nn
Z_{\nu_{1},\nu_{2}}(t,q,Q_{f})&=&\sum_{\lambda}(-Q_{f})^{|\lambda|}C_{\lambda^{t}\,\emptyset\,\nu_{1}}(t,q)\,f_{\lambda}(t,q)\,
 C_{\emptyset\,\lambda\,\nu^{t}_{2}}(t,q)\\\nn
&=&q^{\frac{||\nu_{1}||^{2}}{2}+\frac{||\nu_{2}^{t}||^{2}}{2}}\widetilde{Z}_{\nu_{1}}(t,q)\widetilde{Z}_{\nu_{2}^{t}}(t,q)
\\\nn
&\times& \sum_{\lambda}(-Q_{f})^{|\lambda|}\,
s_{\lambda}(t^{-\rho}q^{-\nu_{1}})\,f_{\lambda}(t,q)\,\Big(\frac{q}{t}\Big)^{\frac{||\lambda||^2}{2}}t^{\frac{\kappa(\lambda)}{2}}
s_{\lambda}(t^{-\nu_{2}}q^{-\rho})\\\nn
&=&q^{\frac{||\nu_{1}||^{2}}{2}+\frac{||\nu_{2}^{t}||^{2}}{2}}\widetilde{Z}_{\nu_{1}}(t,q)\widetilde{Z}_{\nu_{2}^{t}}(t,q)
\sum_{\lambda}(Q_{f}\sqrt{\tfrac{q}{t}})^{|\lambda|}s_{\lambda}(t^{-\rho}q^{-\nu_{1}})\,s_{\lambda}(t^{-\nu_{2}}q^{-\rho})\\
&=&q^{\frac{||\nu_{1}||^{2}}{2}+\frac{||\nu_{2}^{t}||^{2}}{2}}\widetilde{Z}_{\nu_{1}}(t,q)\widetilde{Z}_{\nu_{2}^{t}}(t,q)
\prod_{i,j}\Big(1-Q_{f}\,t^{i-1-\nu_{2,j}}\,q^{j-\nu_{1,i}}\Big)^{-1}\,.\label{zzz}\eea

$Z_{\emptyset,\emptyset}(t,q,Q_{f})$ is the partition function of
${\cal O}(0)\oplus {\cal O}(-2)\mapsto \mathbb{P}^{1}$. We can
separate out the contribution which is independent of $Q_{b}$ and
write the above partition function as \bea\nn
Z(Q_{b},Q_{f},t,q)&=&Z_{pert}(Q_{f},t,q)\,Z_{inst}(Q_{b},Q_{f},t,q)\,\\\nn
Z_{pert}(Q_{f},t,q)&=&Z_{\emptyset,\emptyset}(t,q,Q_{f})\,Z_{\emptyset,\emptyset}(q,t,Q_{f})\\\nn
Z_{inst}(Q_{b},Q_{f},t,q)&=&\sum_{\nu_{1},\nu_{2}}Q_{b}^{|\nu_{1}|+|\nu_{2}|}\,\frac{Z_{\nu_{1},\nu_{2}}(t,q,Q_{f})}{Z_{\emptyset,\emptyset}(t,q,Q_{f})}
\,\Big(\Big(\frac{t}{q}\Big)^{\frac{||\nu_{1}^{t}||^{2}}{2}}\,q^{-\frac{\kappa(\nu_{1})}{2}}\Big)\,\,
\Big(\Big(\frac{q}{t}\Big)^{\frac{||\nu_{2}^{t}||^{2}}{2}}\,t^{-\frac{\kappa(\nu_{2})}{2}}\Big)\,
\frac{Z_{\nu_{2},\nu_{1}}(q,t,Q_{f})}{Z_{\emptyset,\emptyset}(q,t,Q_{f})}\,\\\nn
&=&\sum_{\nu_{1},\nu_{2}}Q_{b}^{|\nu_{1}|+|\nu_{2}|}
\,q^{||\nu_{2}^{t}||^{2}}\,t^{||\nu_{1}^{t}||^{2}}
\widetilde{Z}_{\nu_{1}}(t,q)\widetilde{Z}_{\nu_{2}^{t}}(t,q)\,\widetilde{Z}_{\nu_{2}}(q,t)\widetilde{Z}_{\nu_{1}^{t}}(q,t)\\\nn
&\times&\prod_{i,j=1}^{\infty}\frac{(1-Q_{f}\,t^{i-1}q^{j})(1-Q_{f}\,q^{i-1}t^{j})}
{(1-Q_{f}\,t^{i-1-\nu_{2,j}}q^{j-\nu_{1,i}})(1-Q_{f}\,q^{i-1-\nu_{1,j}}t^{j-\nu_{2,i}})}\label{gpff0}
\eea

\subsubsection{Partition function from instanton calculation}

The partition function of the 4D gauge theory was calculated by
Nekrasov \cite{Nekrasov:2002qd}. The 5D partition function which is
the one A-model topological strings compute is a q-deformation of
the 4D partition function \cite{ON, NY1,NY2}.

The partition function can be calculated once the weights of the
$U(1)\times U(1)$ action at the fixed points are determined. The fixed points are labelled by a set of $N$ 2D partitions (in the
$U(N)$ case), therefore, for a fixed point labelled by
$\{\nu_{1},\cdots,\nu_{N}\}$ the weight $W(\nu_{1},\cdots,\nu_{N})$
is given by \bea
W(\nu_{1},\cdots,\nu_{N})=\prod_{a,b=1}^{N}N(\nu_{a},\nu_{b})\eea
where \bea \nn N(\nu,\nu)&=&\Big[\prod_{s\in
\nu}(1-t^{a(s)}q^{\ell(s)+1})(1-t^{-a(s)-1}q^{-\ell(s)})\Big]^{-1}=
t^{\frac{||\nu^{t}||^{2}+|\nu|}{2}}\,q^{\frac{||\nu||^{2}-|\nu|}{2}}\widetilde{Z}_{\nu}(t,q)\,\widetilde{Z}_{\nu^{t}}(q,t)\\\nn
N(\nu_{a},\nu_{b})&=&\Big[\prod_{(i,j)\in
\nu_{a}}\Big(1-Q_{ba}t^{\nu_{b,j}^{t}-i}\,q^{\nu_{a,i}-j+1}\Big)\prod_{(i,j)\in
\nu_{b}}\Big(1-Q_{ba}t^{-\nu^{t}_{a,j}+i-1}\,q^{-\nu_{b,i}+j}\Big)\Big]^{-1}\\\nn
&=&\prod_{i,j=1}^{\infty}\frac{1-Q_{ab}\,q^{j}\,t^{i-1}}{1-Q_{ab}\,q^{-\nu_{b,i}+j}\,t^{-\nu_{a,j}^{t}+i-1}}\,.
\eea and $-\mbox{log}\,Q_{ab}$ are the K\"ahler parameters. The
partition function is then given by \bea Z^{U(N)}_{\mbox{\tiny gauge
theory}}&:=&\sum_{\nu_{1},\nu_{2},\cdots,\nu_{N}}\widehat{Q}^{|\nu_{1}|+|\nu_{2}|+\cdots+|\nu_{N}|}
\prod_{a,b=1}^{N}N(\nu_{a},\nu_{b}) \label{gt}\eea

The above partition function computed from the gauge theory side is
the A-model partition function of the Calabi-Yau threefold which
gives rise to 5D supersymmetric gauge theory, via M-theory
compactification, with zero Chern-Simons coefficient.

For $N=2$ the partition function in Eq. (\ref{gt}) can be simplified
to the following expression $(Q=Q_{f},
\widehat{Q}Q=\mathfrak{q}\tfrac{q}{t}$) \bea Z^{U(2)}_{\mbox{\tiny
gauge
theory}}&=&\sum_{\nu_{1},\nu_{2}}\mathfrak{q}^{|\nu_{1}|+|\nu_{2}|}Z(\nu_{1}^{t},\nu_{2}^{t};Q,
t,q)\\\nn
Z(\nu_{1},\nu_{2};Q,t,q)&:=&\left(\frac{q}{t}\right)^{|\nu_{1}|+|\nu_{2}|}q^{||\nu_{1}^{t}||^{2}}\,
t^{||\nu_{2}||^{2}}\widetilde
{Z}_{\nu_{1}^{t}}(t,q)\widetilde{Z}_{\nu_{1}}(q,t)\widetilde{Z}_{\nu_{2}^{t}}(t,q)\widetilde{Z}_{\nu_{2}}(q,t)\,G(\nu_{1},\nu_{2},Q,t,q)\\\nn
G(\nu_{1},\nu_{2},Q,t,q)&=&\prod_{i,j=1}^{\infty}\frac{(1-Q\,q^{j-1}t^{i})(1-Q\,q^{j}t^{i-1})}
{(1-Q\,q^{-\nu_{2,i}^{t}+j-1}\,t^{-\nu_{1,j}+i})(1-Q\,q^{-\nu_{2,i}^{t}+j}\,t^{-\nu_{1,j}+i-1})}\eea
\bea
\widetilde{Z}_{\nu}(t,a)=\prod_{s\in\,\nu}(1-t^{a(s)+1}q^{\ell(s)})^{-1}\,,\,\,\,
\ell(i,j)=\nu_{i}-j\,,\,\,\,\,\,a(i,j)=\nu^{t}_{j}-i \eea It is easy
to see that the above gauge theory partition function is the same as
the refined partition function of the last section.

\subsubsection{Spin content of BPS states}

In this section, we list the spin content of various curves obtained
from the refined partition function. A basis of
$H_{2}(\mathbb{P}^{1}\times \mathbb{P}^{1})$ is given by $\{B,F\}$
such that \bea B\cdot B=F\cdot F=0\,,\,\,\,B\cdot F=1\,. \eea The
class $nB+mF$ has a holomorphic representative if $n,m\geq 0$. The
genus of such a curve is given by \bea g(nB+mF)=(n-1)(m-1)\,.\eea
From the refined partition function we can extract the spin content
of the various states coming from $C\in H_{2}(X,{\mathbb Z})$. In
the table below we list the spin content for certain
$C_{n,m}=nB+mF$.

\begin{pspicture}(0,-5)(5,5)

\psline[unit=0.5cm,linewidth=1pt,linecolor=myorange](1.8,9)(30,9)
\psline[unit=0.5cm,linewidth=1pt,linecolor=myorange](1.8,7.4)(30,7.4)
\put(1.5,4){$C_{n,m}$}
\put(8,4){$\sum_{j_{L},j_{R}}N^{(j_{L},j_{R})}_{C}(j_{L},j_{R})$}
\psline[unit=0.5cm,linewidth=1pt,linecolor=myorange](2,9)(2,-9)
\psline[unit=0.5cm,linewidth=1pt,linecolor=myorange](1.8,9)(1.8,-9)
\psline[unit=0.5cm,linewidth=1pt,linecolor=myorange](30,9)(30,-9)
\psline[unit=0.5cm,linewidth=1pt,linecolor=myorange](29.8,9)(29.8,-9)
\psline[unit=0.5cm,linewidth=1pt,linecolor=myorange](8.1,9)(8.1,-9)
\put(1.1,3.2){$B+mF,\,m\geq 0$}\put(9,3.2){$(0,m+\tfrac{1}{2})$}
\psline[unit=0.5cm,linewidth=1pt,linecolor=myorange](2,5.8)(30,5.8)
\put(1.1,2.45){$2B+2F$}\put(8,2.45){$(\tfrac{1}{2},4)\oplus
(0,\tfrac{7}{2})\oplus (0,\tfrac{5}{2})$}

\psline[unit=0.5cm,linewidth=1pt,linecolor=myorange](2,4.3)(30,4.3)
\put(1.1,1.7){$2B+3F$}\put(5,1.7){$(1,\tfrac{11}{2})\oplus
(\tfrac{1}{2},5)\oplus(\tfrac{1}{2},4)\oplus\,2(0,\tfrac{9}{2})\oplus(0,\tfrac{7}{2})\oplus(0,\tfrac{5}{2})$}
\psline[unit=0.5cm,linewidth=1pt,linecolor=myorange](2,2.8)(30,2.8)

\put(1.1,0.9){$3B+3F$}\put(4.5,0.9){$\left(2,\tfrac{15}{2}\right)\oplus
\left(\tfrac{3}{2},7\right)\oplus\left(\tfrac{3}{2},6\right)\oplus\,3\left(1,\tfrac{13}{2}\right)\oplus
\,2\left(1,\tfrac{11}{2}\right)\oplus(1,\tfrac{9}{2})$}
\put(4.5,0.3){$\oplus\left(\tfrac{1}{2},7\right)
\oplus\,3(\tfrac{1}{2},6)\oplus\,3(\tfrac{1}{2},5)
\oplus\,2(\tfrac{1}{2},4) \oplus\,(\tfrac{1}{2},3)\oplus
4(0,\tfrac{11}{2})$}\put(4.5,-0.3){$\oplus\,3(0,\tfrac{9}{2})\oplus\,
3(0,\tfrac{7}{2})\oplus(0,\tfrac{5}{2})\oplus(0,\tfrac{3}{2})$}
\psline[unit=0.5cm,linewidth=1pt,linecolor=myorange](2,-1)(30,-1)

\put(1.1,-1){$3B+4F$}\put(4.5,-1){$\left (3, \tfrac{19}{2} \right
)\oplus\left ( \tfrac{5}{2},9 \right )\oplus\left ( \tfrac{5}{2},8
\right )\oplus3\left (2, \tfrac{17}{2} \right )\oplus\left (
\tfrac{3}{2},9 \right )\oplus2\left (2, \tfrac{15}{2} \right )$}
\put(4.3,-1.65){$\oplus4\left ( \tfrac{3}{2},8 \right )\oplus\left
(1, \tfrac{17}{2} \right ) \oplus\left (2, \tfrac{13}{2} \right
)\oplus\,4\left ( \tfrac{3}{2},7 \right )\oplus7\left (1,
\tfrac{15}{2}\right )\oplus2\left ( \tfrac{1}{2}, 8\right )$}
\put(4.3,-2.25){$\oplus\left (0, \tfrac{17}{2} \right )\oplus2\left
( \tfrac{3}{2},6 \right )\oplus \,6\left (1, \tfrac{13}{2} \right
)\oplus7\left ( \tfrac{1}{2},7 \right ) \oplus\,\left (0,
\tfrac{15}{2} \right )\oplus\,\left ( \tfrac{3}{2},5 \right )$}
\put(4.3,-2.8){$\oplus5\left ( 1,\tfrac{11}{2} \right )\oplus8\left
( \tfrac{1}{2},6 \right )\oplus7\left (0, \tfrac{13}{2} \right
)\oplus2\left (1, \tfrac{9}{2} \right )\oplus6\left ( \tfrac{1}{2},5
\right )$} \put(4.3,-3.4){$\oplus6\left ( 0,\tfrac{11}{2} \right
)\oplus\,\left (1, \tfrac{7}{2} \right )\oplus\,4\left (
\tfrac{1}{2},4 \right )\oplus7\left (0, \tfrac{9}{2} \right
)\oplus2\left ( \tfrac{1}{2},3 \right )$}\put(4.3,-4){$\oplus4\left
( 0,\tfrac{7}{2} \right )\oplus\left ( \tfrac{1}{2},2 \right )
\oplus3\left ( 0,\tfrac{5}{2} \right )\oplus\left ( 0,\tfrac{3}{2}
\right )\oplus\,\left (0, \tfrac{1}{2} \right )$}

\psline[unit=0.5cm,linewidth=1pt,linecolor=myorange](2,-9)(30,-9)

\end{pspicture}

\subsection{Local $\mathbb{F}_{m}$}

In this section, we will use the refined vertex to calculate the
refined partition function for the local $\mathbb{F}_{m}$,
$m=0,1,2$. The case $m=0$ (local $\mathbb{P}^{1}\times
\mathbb{P}^{1}$) has already been discussed in the last section. As
we saw in \cite{KI1}, the partition function for local
$\mathbb{F}_{m}$ differ just by framing factors along the edges
which label the instanton charge (the edges corresponding to the
base $\mathbb{P}^{1}$). This continues to be the case for the
refined partition function for local $\mathbb{F}_{m}$.

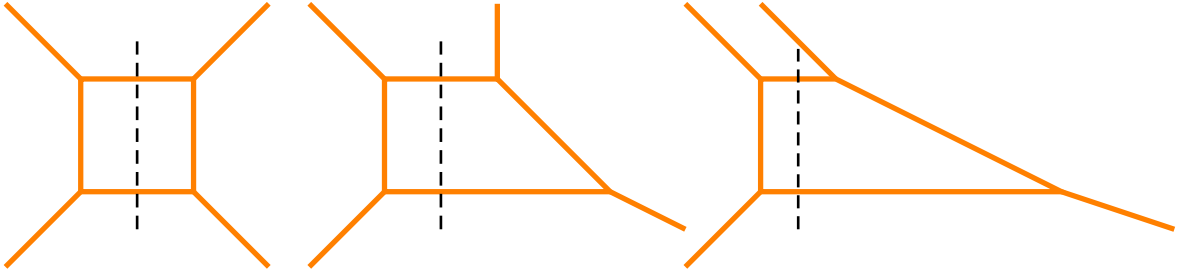
\begin{figure}\begin{center}
$\begin{array}{c@{\hspace{1in}}c} \multicolumn{1}{l}{\mbox{}} &
    \multicolumn{1}{l}{\mbox{}} \\ [-0.53cm]
{

\begin{pspicture}(2.5,0)(4,4)
\pspolygon[unit=0.5cm,
linewidth=2pt,linecolor=myorange](-3,2)(-3,5)(0,5)(0,2)(-3,2)
\psline[unit=0.5cm,
linewidth=1pt,linestyle=dashed,linecolor=black](-1.5,1)(-1.5,6)
\psline[unit=0.5cm, linewidth=2pt,linecolor=myorange](-3,2)(-5,0)
\psline[unit=0.5cm, linewidth=2pt,linecolor=myorange](-3,5)(-5,7)
\psline[unit=0.5cm, linewidth=2pt,linecolor=myorange](0,5)(2,7)
\psline[unit=0.5cm, linewidth=2pt,linecolor=myorange](0,2)(2,0)

\end{pspicture}}&
{\begin{pspicture}(2.5,0)(4,4)
 \pspolygon[unit=0.5cm,
linewidth=2pt,linecolor=myorange](-3,2)(-3,5)(0,5)(3,2)(-3,2)
\psline[unit=0.5cm,linewidth=1pt,linestyle=dashed,linecolor=black](-1.5,1)(-1.5,6)
\psline[unit=0.5cm, linewidth=2pt,linecolor=myorange](-3,2)(-5,0)
\psline[unit=0.5cm, linewidth=2pt,linecolor=myorange](-3,5)(-5,7)
\psline[unit=0.5cm, linewidth=2pt,linecolor=myorange](0,5)(0,7)
\psline[unit=0.5cm, linewidth=2pt,linecolor=myorange](3,2)(5,1)

\pspolygon[unit=0.5cm,
linewidth=2pt,linecolor=myorange](7,2)(7,5)(9,5)(15,2)(7,2)
\psline[unit=0.5cm,
linewidth=1pt,linestyle=dashed,linecolor=black](8,1)(8,5.8)

\psline[unit=0.5cm, linewidth=2pt,linecolor=myorange](7,2)(5,0)
\psline[unit=0.5cm, linewidth=2pt,linecolor=myorange](7,5)(5,7)
\psline[unit=0.5cm, linewidth=2pt,linecolor=myorange](9,5)(7,7)
\psline[unit=0.5cm, linewidth=2pt,linecolor=myorange](15,2)(18,1)

\end{pspicture}}
\\ [-0.2cm] \mbox{} & \mbox{}
\end{array}$
\caption{Local $F_{m}:\,m=0\,,\,\,m=1\,,\,\,m=2\,.$}
\label{fm}\end{center}
\end{figure}

The toric geometry for these cases is shown in \figref{fm}. The
parallel edges in the polygon correspond to the base
$\mathbb{P}^{1}$. We cut the toric diagram perpendicular to these
parallel lines as we did for the local $\mathbb{P}^{1}\times
\mathbb{P}^{1}$. The 2D partitions on the parallel edges will be
denoted by $\nu_{1}$ and $\nu_{2}$. The half of the toric diagram
corresponds to a geometry ${\cal O}(0)\oplus {\cal O}(-2)\mapsto
\mathbb{P}^{1}$ with a stack of D-branes on the two parallel edges in
the representation $\nu_{1},\nu_{2}$. We denote the open topological
string partition function by $Z_{\nu_{1},\nu_{2}}(t,q, Q_{f})$ where
$T_{f}=-\mbox{log}\,Q_{f}$ is the K\"{a}hler parameter of the fiber
$\mathbb{P}^{1}$. $Z_{\nu_{1},\nu_{2}}(t,q,Q_{f})$ was calculated in the
last section and is given by Eq. (\ref{zzz}). Although the two parts
of the toric diagram look different (except for $m=0$ in which case
they are identical) they are related to each other by an $SL(2,{\mathbb Z})$
transformation. This implies that the open topological string
partition function associated with each side is the same,
$Z_{\nu_{1},\nu_{2}}(t,q,Q_{f})$. Thus the difference arises only in how
these two open string partition functions are ``glued'' together to
form the closed string partition function. This gluing information
is contained in the normal geometry of the base curve and is what
determines the framing factors.

From the toric geometry, \figref{fm}, it is clear that locally
the two $\mathbb{P}^{1}$'s corresponding to the base curve (the
upper and lower parallel edges) are ${\cal O}(-m)\oplus {\cal
O}(-2+m)\mapsto \mathbb{P}^{1}$ and ${\cal O}(-2-m)\oplus {\cal
O}(m)\mapsto \mathbb{P}^{1}$. Therefore the framing factor with the
top edge is $f^{-m+1}_{\nu_{1}}(t,q)$ and with the lower edge is
$f^{m+1}_{\nu_{2}}(q,t)$.

Using the above framing factors the generalized partition function
is given by \bea\nn
Z^{(m)}(Q_{b},Q_{f},t,q)=\sum_{\nu_{1},\nu_{2}}(-Q_{b})^{|\nu_{1}|+|\nu_{2}|}Q_{f}^{m\,|\nu_{2}|}\,Z_{\nu_{1},\nu_{2}}(t,q,Q_{f})\,
f_{\nu_{1}}^{-m+1}(t,q)\,f_{\nu_{2}}^{m+1}(q,t)\,
Z_{\nu_{2},\nu_{1}}(q,t,Q_{f}). \eea
 We can write the above partition
function as \bea Z^{(m)}(Q_{b},Q_{f},t,q)=
Z^{(m)}_{pert}(Q_{f},t,q)\,Z^{(m)}_{inst}(Q_{b},Q_{f},t,q) \eea
where $Z^{(m)}_{pert}(Q_{f},t,q)$ is a function only of $Q_{f}$ and, in the
field theory limit, it gives the perturbative prepotential of the
theory. $Z^{(m)}_{inst}(Q_{b},Q_{f},t,q) $ depends on $Q_{b}$ and gives the instanton
correction to the prepotential in the field theory limit. Although
$Q_{f}$ and $Q_{b}$ are on the same footing in the topological
string theory we write the partition function this way to simplify
the expressions and to be able to compare with the $m=0$ case, which
corresponds to
Nekrasov's partition function.\\
 \bea \nn
Z^{(m)}_{pert}(Q_{f},t,q)=Z_{\emptyset,\emptyset}(t,q,Q_{f})\,Z_{\emptyset,\emptyset}(q,t,Q_{f})
=\prod_{i,j=1}^{\infty}(1-Q_{f}\,t^{i}\,q^{j-1})^{-1}\,(1-Q_{f}\,t^{i-1}\,q^{j})^{-1}\,.\eea
\bea\nn
Z^{(m)}_{inst}(Q_{b},Q_{f},t,q) &=&\sum_{\nu_{1},\nu_{2}}Q_{b}^{|\nu_{1}|+|\nu_{2}|}Q_{f}^{m\,|\nu_{2}|}\,\frac{Z_{\nu_{1},\nu_{2}}(t,q,Q_{f})}{Z_{\emptyset,\emptyset}(t,q,Q_{f})}
\Big(\Big(\tfrac{t}{q}\Big)^{\frac{||\nu_{1}^{t}||^{2}}{2}}\,q^{-\frac{\kappa(\nu_{1})}{2}}\Big)^{-m+1}\,\,
\Big(\Big(\tfrac{q}{t}\Big)^{\frac{||\nu_{2}^{t}||^{2}}{2}}\,t^{-\frac{\kappa(\nu_{2})}{2}}\Big)^{m+1}\\\nn
&\times& (-1)^{m(|\nu_{1}|+|\nu_{2}|)}\,
\frac{Z_{\nu_{2},\nu_{1}}(q,t,Q_{f})}{Z_{\emptyset,\emptyset}(q,t,Q_{f})}\,,
\eea where \bea
\frac{Z_{\nu_{1},\nu_{2}}(t,q,Q_{f})}{Z_{\emptyset,\emptyset}(t,q,Q_{f})}=q^{\frac{||\nu_{1}||^{2}}{2}+\frac{||\nu_{2}^{t}||^{2}}{2}}\widetilde{Z}_{\nu_{1}}(t,q)\widetilde{Z}_{\nu_{2}^{t}}(t,q)
\prod_{i,j=1}^{\infty}\frac{1-Q_{f}\,t^{i-1}q^{j}}{1-Q_{f}\,t^{i-1-\nu_{2,j}}\,q^{j-\nu_{1,i}}}\,.
\eea The expression for the $Z_{inst}^{(m)}(Q_{b},Q_{f},t,q)$ can be simplified to
become \bea\nn
Z_{inst}(Q_{b},Q_{f},t,q)&=&\sum_{\nu_{1},\nu_{2}}Q_{b}^{|\nu_{1}|+|\nu_{2}|}Q_{f}^{m|\nu_{2}|}\,(-1)^{m(|\nu_{1}|+|\nu_{2}|)}\,\Big(\frac{q}{t}\Big)^{\frac{m}{2}(||\nu_{1}||^{2}+||\nu_{2}^{t}||^{2})}\,t^{\frac{m}{2}(\kappa(\nu_{1})-\kappa(\nu_{2}))}\\\nn
&\times&\,q^{||\nu_{2}^{t}||^{2}}\,t^{||\nu_{1}^{t}||^{2}}
\widetilde{Z}_{\nu_{1}}(t,q)\widetilde{Z}_{\nu_{2}^{t}}(t,q)\,\widetilde{Z}_{\nu_{2}}(q,t)\widetilde{Z}_{\nu_{1}^{t}}(q,t)\\
&\times&\prod_{i,j=1}^{\infty}\frac{(1-Q_{f}\,t^{i}q^{j-1})(1-Q_{f}\,q^{i}t^{j-1})}{(1-Q_{f}\,t^{i-\nu_{2,j}}q^{j-1-\nu_{1,i}})(1-Q_{f}\,q^{i-\nu_{1,j}}t^{j-1-\nu_{2,i}})}\label{gpffm}.\eea

\subsubsection{Spin content of BPS states: local $\mathbb{F}_{1}$}

In this section, we list the spin content of some curves obtained
from the refined partition function given in Eq. (\ref{gpffm}) for
the case $m=1$. $\mathbb{F}_{1}$ has a two dimensional
$H_{2}(\mathbb{F}_{1})$ with a
 basis given by $B$ and $F$ such that
\bea
 B\cdot B=-1\,,\,\,F\cdot F=0\,,\,\,\,B\cdot F=1\,.
\eea A class $nB+mF$ has a holomorphic curve in it if $m-n,n\geq 0$.
The arithmetic genus of such a curve, $C=nB+mF$, is given by the
adjunction formula,\bea g(nB+(n+k)F)=\frac{(n-1)(n-2)}{2}+k(n-1)\,.
\eea Since $\mathbb{F}_{1}$ is the one point blowup of
$\mathbb{P}^{2}$ there is a different basis $\{H,E\}$ of
$H_{2}(\mathbb{F}_{1})$ which will be useful for us later, \bea
H&=&B+F,\,\,E=B\,\\\nn H\cdot H&=&1\,,\,\,E\cdot E=-1\,,\,\,H\cdot
E=0\,, \eea where $E$ is the exceptional curve obtained by the
blowup and $H$ is the basic class of $\mathbb{P}^{1}$ in
$\mathbb{P}^{2}$ given by linear polynomials. It is clear that the
invariants of the curves $B+F, 2(B+F),3(B+F),\cdots$ will be the
same as the invariants of the curves $H,2H,3H,\cdots$  in local
$\mathbb{P}^{2}$.

\begin{itemize}
\item
\underline{\bf $B+k\,F$:} These curves are of genus zero for all
$k\geq 0$. The moduli space of such curves is given by
$\mathbb{P}^{d-1}$ where $d=-C\cdot K_{F_{1}}=2k+1$. Therefore, the
left spin $j_{L}=0$ and right spin $j_{R}=k$, \bea
N^{(j_{L},j_{R})}_{B+k\,F}=\delta_{j_{L},0}\,\delta_{j_{R},k}\,.\eea
This is exactly what we get from the refined partition function.
\item

\underline{$2B+2F$:}\,This curve is also of genus zero, and since
$2(B+F)=2H$ the moduli space is given by $\mathbb{P}^{5}$, the space
of quadratic polynomials in $\mathbb{P}^{2}$ (up to overall
scaling). Thus \bea
N^{(j_{L},j_{R})}_{2B+2F}=\delta_{j_{L},0}\,\delta_{j_{R},\frac{5}{2}}\,.\eea
This is exactly what we get from the refined partition function once
multicovering has been taken into account.

\item
 \underline{\bf $2B+3F$:} This curve is of genus one. The spin
 content from the refined partition function is
 \bea \nn(\tfrac{1}{2},4)\oplus
(0,\tfrac{7}{2})\oplus (0,\tfrac{5}{2})\,.\eea

\item
\underline{\bf $2B+4F$:} This curve is of genus $2$. The spin
content from the refined partition function is \bea\nn \left (1,
\tfrac{11}{2} \right )\oplus\left ( \tfrac{1}{2},5 \right
)\oplus\left ( \tfrac{1}{2},4 \right )\oplus2\left ( 0,\tfrac{9}{2}
\right )\oplus\left (0, \tfrac{7}{2} \right )\oplus\left (
0,\tfrac{5}{2} \right ). \eea

\item\underline{\bf $2B+5F$:} This curve is of genus $3$. The spin
content is given by \bea\nn \left ( \tfrac{3}{2},7 \right
)\oplus\left (1, \tfrac{13}{2} \right )\oplus\left (1, \tfrac{11}{2}
\right )\oplus2\left ( \tfrac{1}{2},6 \right )\oplus\left (
\tfrac{1}{2},5 \right )\oplus2\left ( 0,\tfrac{11}{2} \right
)\oplus\left ( \tfrac{1}{2},4 \right )\oplus2\left ( 0,\tfrac{9}{2}
\right ) \oplus\left ( 0,\tfrac{7}{2} \right )\oplus\left (0,
\tfrac{5}{2} \right ). \eea

\item\underline{\bf $3B+3F$:} This curve is of genus $1$. The spin content is given by \bea \nn\left (
\tfrac{1}{2},\tfrac{9}{2}\right ) \oplus\left ( 0,3 \right). \eea
Note that this is also the spin content of the curve $3H$ in local
$\mathbb{P}^{2}$.

\item\underline{\bf $3B+4F$:} This curve is of genus $3$. The spin content is given by \bea\nn \left (
\tfrac{3}{2},\tfrac{13}{2} \right )\oplus\left (1,6 \right
)\oplus\left (1,5 \right )\oplus2\left(
\tfrac{1}{2},\tfrac{11}{2}\right )\oplus \left(0,6 \right
)\oplus2\left (\tfrac{1}{2},\tfrac{9}{2}\right )\oplus\left (0,5
\right )\oplus\left ( \tfrac{1}{2},\tfrac{7}{2}\right
)\oplus\left(0,4 \right )\oplus\left(0,3\right )\oplus\left(0,2
\right ) \eea

\item\underline{\bf $3B+5F$:} This curve is of genus $5$. The spin
content is given by
\begin{eqnarray}\nonumber
&&\left ( \tfrac{5}{2},\tfrac{17}{2}\right)\oplus\left (2,8
\right)\oplus\left (2,7 \right)\oplus3\left (
\tfrac{3}{2},\tfrac{15}{2}\right)\oplus\left (1,8
\right)\oplus2\left ( \tfrac{3}{2},\tfrac{13}{2}\right)\oplus3\left
(1,7 \right)\oplus\left (\tfrac{1}{2},\tfrac{15}{2}
\right)\oplus\left (\tfrac{3}{2},\tfrac{11}{2} \right)\\
\nonumber&&\oplus\,4\left (1,6 \right)\oplus5\left
(\tfrac{1}{2},\tfrac{13}{2} \right)\oplus2\left (0,7
\right)\oplus2\left (1,5 \right)\oplus5\left (
\tfrac{1}{2},\tfrac{11}{2}\right)\oplus3\left (0,6
\right)\oplus\left (1,4 \right)\oplus4\left (
\tfrac{1}{2},\tfrac{9}{2}\right)\oplus5\left (0,5 \right)\\
\nonumber &&\oplus\,2\left (
\tfrac{1}{2},\tfrac{7}{2}\right)\oplus3\left (0,4 \right)\oplus\left
( \tfrac{1}{2},\tfrac{5}{2}\right)\oplus3\left (0,3
\right)\oplus\left (0,2 \right)\oplus\left (0,1 \right).
\end{eqnarray}
\item\underline{\bf $4B+4F$:} This curve has genus $3$. The spin
content is given by \bea\nn
&&\left(\tfrac{3}{2},7\right)\oplus\left(1,\tfrac{11}{2}\right)
\oplus\left(\tfrac{1}{2},6\right) \oplus\left(\tfrac{1}{2},5\right)
\oplus\left(\tfrac{1}{2},4\right)\oplus\left(0,\tfrac{13}{2}\right)
\oplus\left(0,\tfrac{9}{2}\right) \oplus\left(0,\tfrac{5}{2}\right)
\eea

This is also the spin content of the curve $4H$ in local
$\mathbb{P}^{2}$.

\item\underline{\bf $5B+5F$:} This curve has genus $6$. The spin
content is given by \bea\nn
&&(3,10)+\left(\tfrac{5}{2},\tfrac{17}{2}\right)\oplus(2,9)\oplus(2,8)\oplus(2,7)\oplus\left(\tfrac{3}{2},\tfrac{19}{2}\right)
\oplus\left(\tfrac{3}{2},\tfrac{17}{2}\right)
\oplus\,2\left(\tfrac{3}{2},\tfrac{15}{2}\right)\\\nn
&&\oplus\left(\tfrac{3}{2},\tfrac{13}{2}\right)\oplus\left(\tfrac{3}{2},\tfrac{11}{2}\right)
\oplus(1,9)\oplus\,2(1,8)\oplus\,2(1,7)\oplus\,2(1,6)\oplus(1,5)\oplus(1,4)\\\nn
&&\oplus\left(\tfrac{1}{2},\tfrac{17}{2}\right)
\oplus\,2\left(\tfrac{1}{2},\tfrac{15}{2}\right)\oplus\,3\left(\tfrac{1}{2},\tfrac{13}{2}\right)\oplus\,2\left(\tfrac{1}{2},\tfrac{11}{2}\right)\oplus
\left(\tfrac{1}{2},\tfrac{7}{2}\right)\oplus\left(\tfrac{1}{2},\tfrac{5}{2}\right)\oplus(0,8)\\\nn
&&\oplus\,2(0,7)\oplus\,2(0,6)\oplus\,2(0,5)\oplus\,(0,4)
\oplus(0,3)\oplus(0,1)\,. \eea This is also the spin content of the
curve $5H$ in local $\mathbb{P}^{2}$
\end{itemize}

From this example it is clear that although the refined vertex can
only be used for a certain kind of geometries (those giving rise to
gauge theories) the spin content of BPS states for any toric
CY3-fold can be obtained by embedding this toric CY3-fold in another
toric CY3-fold which does have a gauge theory interpretation. For
example, the refined vertex can not be used to determine the refined
partition function for local $\mathbb{P}^{2}$ but since one point
blowup of local $\mathbb{P}^{2}$ is local $\mathbb{F}_{1}$ which
does have a gauge theory interpretation therefore spin content of
BPS states coming from local $\mathbb{P}^{2}$ can be obtained from
the refined partition function of local $\mathbb{F}_{1}$. We list
the spin content of first few BPS states for local $\mathbb{P}^{2}$
in the table below.

\begin{pspicture}(0,-3)(5,5)

\psline[unit=0.5cm,linewidth=1pt,linecolor=myorange](1.8,9)(30,9)
\psline[unit=0.5cm,linewidth=1pt,linecolor=myorange](1.8,7.4)(30,7.4)
\put(1.5,4){$C_{n}=nH$}
\put(8,4){$\sum_{j_{L},j_{R}}N^{(j_{L},j_{R})}_{C}(j_{L},j_{R})$}
\psline[unit=0.5cm,linewidth=1pt,linecolor=myorange](2,9)(2,-7)
\psline[unit=0.5cm,linewidth=1pt,linecolor=myorange](1.8,9)(1.8,-7)
\psline[unit=0.5cm,linewidth=1pt,linecolor=myorange](30,9)(30,-7)
\psline[unit=0.5cm,linewidth=1pt,linecolor=myorange](29.8,9)(29.8,-7)
\psline[unit=0.5cm,linewidth=1pt,linecolor=myorange](8.1,9)(8.1,-7)
\put(1.1,3.2){$H$}\put(9,3.2){$(0,1)$}
\psline[unit=0.5cm,linewidth=1pt,linecolor=myorange](2,5.8)(30,5.8)
\put(1.1,2.45){$2H$}\put(9,2.45){$(0,\tfrac{5}{2})$}

\psline[unit=0.5cm,linewidth=1pt,linecolor=myorange](2,4.3)(30,4.3)
\put(1.1,1.7){$3H$}\put(8.2,1.7){$\left (
\tfrac{1}{2},\tfrac{9}{2}\right ) \oplus\left ( 0,3 \right)$}
\psline[unit=0.5cm,linewidth=1pt,linecolor=myorange](2,2.8)(30,2.8)

\put(1.1,0.9){$4H$}\put(6,0.9){$\left(\tfrac{3}{2},7\right)\oplus\left(1,\tfrac{11}{2}\right)
\oplus\left(\tfrac{1}{2},6\right)
\oplus\left(\tfrac{1}{2},5\right)$}\put(6,0.3){$
\oplus\left(\tfrac{1}{2},4\right)\oplus\left(0,\tfrac{13}{2}\right)
\oplus\left(0,\tfrac{9}{2}\right)
\oplus\left(0,\tfrac{5}{2}\right)$}

\psline[unit=0.5cm,linewidth=1pt,linecolor=myorange](2,-0.2)(30,-0.2)

\put(1.1,-0.6){$5H$}\put(4.5,-0.6){$(3,10)+\left(\tfrac{5}{2},\tfrac{17}{2}\right)\oplus(2,9)\oplus(2,8)\oplus(2,7)\oplus\left(\tfrac{3}{2},\tfrac{19}{2}
\right)$}\put(4.5,-1.2){$
\oplus\left(\tfrac{3}{2},\tfrac{17}{2}\right)
\oplus\,2\left(\tfrac{3}{2},\tfrac{15}{2}\right)\oplus\left(\tfrac{3}{2},\tfrac{13}{2}\right)\oplus\left(\tfrac{3}{2},\tfrac{11}{2}\right)
\oplus(1,9)\oplus\,2(1,8)$}
\put(4.5,-1.8){$\oplus\,2(1,7)\oplus\,2(1,6)\oplus(1,5)\oplus(1,4)\oplus\left(\tfrac{1}{2},\tfrac{17}{2}\right)
\oplus\,2\left(\tfrac{1}{2},\tfrac{15}{2}\right)$}
\put(4.5,-2.4){$\oplus\,3\left(\tfrac{1}{2},\tfrac{13}{2}\right)\oplus\,2\left(\tfrac{1}{2},\tfrac{11}{2}\right)\oplus
\left(\tfrac{1}{2},\tfrac{7}{2}\right)\oplus\left(\tfrac{1}{2},\tfrac{5}{2}\right)\oplus(0,8)\oplus\,2(0,7)$}
\put(4.5,-3){$\oplus\,2(0,6)\oplus\,2(0,5)\oplus\,(0,4)
\oplus(0,3)\oplus(0,1)\,$}

\psline[unit=0.5cm,linewidth=1pt,linecolor=myorange](2,-7)(30,-7)
\end{pspicture}

\subsection{An $SU(3)$ geometry}
In this section, we will use the refined vertex to calculate the
partition function of a certain CY3-fold which gives rise to
compactified 5D supersymmetric $SU(3)$ gauge theory with
Chern-Simons coefficient $m$.

The refined partition function is given by \bea
Z=\sum_{\nu_{1},\nu_{2},\nu_{3}}Q_{b1}^{|\nu_{1}|}Q_{b_{2}}^{|\nu_{2}|}Q_{b3}^{|\nu_{3}|}Z_{\nu_{1},\nu_{2},\nu_{3}}(t,q)\,
(f^{m+2}_{\nu_{1}}(t,q)\,f^{-m}_{\nu_{2}}(q,t)\,f^{-m+2}_{\nu_{3}}(t,q))\,Z_{\nu_{3},\nu_{2},\nu_{1}}(q,t),
\eea where
 \bea \nn
Z_{\nu_{1},\nu_{2},\nu_{3}}&:=&\sum_{\lambda,\mu}(-Q_{1})^{|\lambda|}(-Q_{2})^{|\mu|}C_{\emptyset\,\lambda\,\nu_{1}}(t,q)f_{\lambda}(t,q)C_{\lambda^{t}\,\mu\,\nu_{2}}(t,q)
f_{\mu}(t,q)C_{\mu^{t}\,\emptyset\,\nu_{3}}(t,q)\\\nn &=&
(-Q_{1})^{|\lambda|}(-Q_{2})^{|\mu|}\Big[(\tfrac{q}{t})^{\frac{||\lambda||^2-|\lambda|}{2}}t^{\frac{\kappa(\lambda)}{2}}\,q^{\frac{||\nu_{1}||^2}{2}}\widetilde{Z}_{\nu_{1}}(t,q)s_{\lambda}(t^{-\nu_{1}^{t}}q^{-\rho})\Big]
f_{\lambda}(t,q)\\\nn &\times&
\Big[(\tfrac{q}{t})^{\frac{||\mu||^2-|\mu|}{2}}t^{\frac{\kappa(\mu)}{2}}\,q^{\frac{||\nu_{2}||^2}{2}}\widetilde{Z}_{\nu_{2}}(t,q)
\sum_{\eta}s_{\lambda/\eta}(t^{-\rho}q^{-\nu_{2}})s_{\mu/\eta}(t^{-\nu_{2}^{t}}q^{-\rho})\Big]\,f_{\mu}(t,q)\\\nn
&\times&
\Big[q^{\frac{||\nu_{3}||^2}{2}}\widetilde{Z}_{\nu_{3}}(t,q)s_{\mu}(t^{-\rho}q^{-\nu_{3}})\Big]\\\nn
&=&q^{\frac{||\nu_{1}||^2+||\nu_{2}||^2+||\nu_{3}||^2}{2}}\widetilde{Z}_{\nu_{1}}(t,q)\widetilde{Z}_{\nu_{2}}(t,q)\widetilde{Z}_{\nu_{3}}(t,q)\\\nn
&\times&
\sum_{\eta}(\tfrac{q}{t})^{\frac{|\eta|}{2}}\Big(\sum_{\lambda}s_{\lambda}(-Q_{1}q^{-\rho}t^{-\nu_{1}^{t}})\,s_{\lambda/\eta}(t^{-\rho}q^{-\nu_{2}})\Big)
\Big(\sum_{\mu}s_{\mu}(-Q_{2}t^{-\rho}q^{-\nu_{3}})\,s_{\mu/\eta}(q^{-\rho}q^{-\nu_{2}^{t}}\Big)\\\nn
&=& q^{\frac{||\nu_{1}||^2+||\nu_{2}||^2+||\nu_{3}||^2}{2}}\widetilde{Z}_{\nu_{1}}(t,q)\widetilde{Z}_{\nu_{2}}(t,q)\widetilde{Z}_{\nu_{3}}(t,q)\prod_{i,j=1}^{\infty}(1-Q_{1}\,t^{j-1/2-\nu_{1,i}^{t}}q^{i-1/2-\nu_{2,j}})^{-1}\\\nn
&\times&(1-Q_{2}\,t^{j-1/2-\nu_{2,i}^{t}}q^{i-1/2-\nu_{3,j}})^{-1}(1-Q_{1}Q_{2}\,t^{j-1/2-\nu_{1,i}^{t}}q^{i-1/2-\nu_{3,j}})^{-1}
\eea
In the above expression $-\mbox{log}\,Q_{1}$ and
$-\mbox{log}\,Q_{2}$ are the K\"ahler classes of the
$\mathbb{P}^{1}$'s in the fiber and $Q_{b1,b2,b3}$ are given by
\cite{KI2}\bea
Q_{b1}&=&Q_{b}Q_{1}^{m+1}Q_{2}^{(m-1)(1-\delta_{m,0})}\,\\\nn
Q_{b2}&=&Q_{b}Q_{2}^{(m-1)(1-\delta_{m,0})}\\\nn
Q_{b3}&=&Q_{b}Q_{2}^{\delta_{m,0}}\,. \eea

This is exactly the K-theoretic version of the Nekrasov's partition
function as can be verified by using the weights of the torus action
given in \cite{NY1}.

\subsection{An $SU(3)$, $N_{f}=4$ geometry}
\begin{figure}\begin{center}
\includegraphics[width=2.5in]{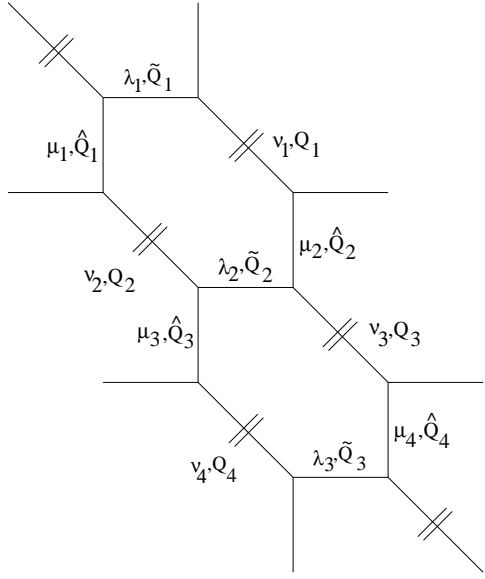}
\caption{The web diagram of the CY3-fold giving rise to $SU(3)$
gauge theory with $N_{f}=4$.} \label{U3}
\end{center}\end{figure}

In this section, we will compute the refined partition function for
the CY3-fold which gives rise to $SU(3)$ gauge theory with adjoint
matter via geometric engineering. This CY3-fold is a blowup of a
resolved $A_{2}$ singularity fibered over ${\mathbb P}^{1}$. The
toric diagram of this geometry and the choice of the preferred
direction for each vertex is shown in \figref{U3}.

The refined partition function for this geometry is given by \bea\nn
Z&=&\sum_{\{\mu_{i}\},\{\nu_{i}\},\{\lambda_{i}\}}(-\hat{Q}_{1})^{|\mu_{1}|}\mathellipsis
(-\hat{Q}_{4})^{|\mu_{4}|}(-Q_{1})^{|\nu_{1}|}\mathellipsis
(-Q_{4})^{|\nu_{4}|}(-\tilde{Q}_{1})^{|\lambda_{1}|}\mathellipsis
(-\tilde{Q}_{3})^{|\lambda_{3}|}\\ \nn \\ \nn &\times&
C_{\lambda_{1}\mu_{1}\emptyset}(t,q)C_{\lambda_{1}^{t}\emptyset\nu_{1}}(q,t)C_{\emptyset\mu_{1}^{t}\nu_{2}}(q,t)C_{\emptyset\mu_{2}\nu_{1}^{t}}(t,q)C_{\lambda_{2}\mu_{3}\nu_{2}^{t}}(t,q)C_{\lambda_{2}^{t}\mu_{2}^{t}\nu_{3}}(q,t)C_{\emptyset\mu_{3}^{t}\nu_{4}}(q,t)\\
\nn
&\times&C_{\emptyset\mu_{4}\nu_{3}^{t}}(t,q)C_{\lambda_{3}\emptyset\nu_{4}^{t}}(t,q)C_{\lambda_{3}^{t}\mu_{4}^{t}\emptyset}(q,t)
\\\nn\\ \nn
&=&\sum_{\{\mu_{i}\},\{\nu_{i}\},\{\lambda_{i}\},\{\eta_{i}\}}(-\hat{Q}_{1})^{|\mu_{1}|}\mathellipsis
(-\hat{Q}_{4})^{|\mu_{4}|}(-Q_{1})^{|\nu_{1}|}\mathellipsis
(-Q_{4})^{|\nu_{4}|}(-\tilde{Q}_{1})^{|\lambda_{1}|}\mathellipsis
(-\tilde{Q}_{3})^{|\lambda_{3}|} \\ \nn \\ \nn &\times&
q^{\frac{\|\nu_{1}^{t}\|^{2}+\mathellipsis+\|\nu_{4}^{t}\|^{2}}{2}}t^{\frac{\|\nu_{1}\|^{2}+\mathellipsis+\|\nu_{4}\|^{2}}{2}}\left
( \frac{q}{t}\right
)^{\frac{|\eta_{1}|+|\eta_{2}|-|\eta_{3}|-|\eta_{4}|}{2}}\widetilde{Z}_{\nu_{1}}(q,t)\widetilde{Z}_{\nu_{1}^{t}}(t,q)\mathellipsis\widetilde{Z}_{\nu_{4}}(q,t)\widetilde{Z}_{\nu_{4}^{t}}(t,q)\\
\nn &\times&
s_{\lambda_{1}^{t}/\eta_{1}}(t^{-\rho})s_{\mu_{1}/\eta_{1}}(q^{-\rho})s_{\lambda_{1}}(q^{-\rho}t^{-\nu_{1}})s_{\mu_{1}^{t}}(q^{-\nu_{2}^{t}}t^{-\rho})s_{\mu_{2}}(t^{-\nu_{1}}q^{-\rho})
s_{\lambda_{2}^{t}/\eta_{2}}(t^{-\rho}q^{-\nu_{2}^{t}})s_{\mu_{3}/\eta_{2}}(t^{-\nu_{2}}q^{-\rho})\\
\nn &\times&
s_{\lambda_{2}/\eta_{3}}(q^{-\rho}t^{-\nu_{3}})s_{\mu_{2}^{t}/\eta_{3}}(q^{-\nu_{3}^{t}}t^{-\rho})s_{\mu_{3}^{t}}(q^{-\nu_{4}^{t}}t^{-\rho})s_{\mu_{4}}(t^{-\nu_{3}}q^{-\rho})s_{\lambda_{3}^{t}}(t^{-\rho}q^{-\nu_{4}^{t}})s_{\lambda_{3}/\eta_{4}}(q^{-\rho})s_{\mu_{4}^{t}/\eta_{4}}(t^{-\rho})\\
\nn\\ \nn &=&\sum_{\{\nu_{i}\}}(-Q_{1})^{|\nu_{1}|}\mathellipsis
(-Q_{4})^{|\nu_{4}|}q^{\frac{\|\nu_{1}^{t}\|^{2}+\mathellipsis+\|\nu_{4}^{t}\|^{2}}{2}}t^{\frac{\|\nu_{1}\|^{2}+\mathellipsis+\|\nu_{4}\|^{2}}{2}}\widetilde{Z}_{\nu_{1}}(q,t)\widetilde{Z}_{\nu_{1}^{t}}(t,q)\mathellipsis\widetilde{Z}_{\nu_{4}}(q,t)\widetilde{Z}_{\nu_{4}^{t}}(t,q)\\
\nn &\times&
\prod_{i,j=1}^{\infty}\frac{(1-\tilde{Q}_{1}q^{-\rho_{i}}t^{-\nu_{1,i}-\rho_{j}})(1-\hat{Q}_{1}q^{-\nu_{2,i}^{t}-\rho_{j}}t^{-\rho_{i}})(1-\hat{Q}_{2}q^{-\nu_{3,j}^{t}-\rho_{i}}t^{-\nu_{1,i}-\rho_{j}})(1-\hat{Q}_{3}q^{-\nu_{4,i}^{t}-\rho_{j}}t^{-\nu_{2,j}-\rho_{i}})}{(1-\tilde{Q}_{1}\hat{Q}_{1}q^{-\nu_{2,j}^{t}-\rho_{i}+1/2}t^{-\nu_{1,i}-\rho_{j}-1/2})(1-\tilde{Q}_{3}\hat{Q}_{4}q^{-\nu_{4,j}^{t}-\rho_{i}-1/2}t^{-\nu_{3,i}-\rho_{j}+1/2})}\\
\nn &\times&
\frac{(1-\hat{Q}_{4}q^{-\rho_{i}}t^{-\nu_{3,i}-\rho_{j}})(1-\tilde{Q}_{3}q^{-\nu_{4,i}^{t}-\rho_{j}}t^{-\rho_{i}})(1-\tilde{Q}_{2}q^{-\nu_{2,i}^{t}-\rho_{j}}t^{-\nu_{3,j}-\rho_{i}})(1-\tilde{Q}_{2}\hat{Q}_{2}\hat{Q}_{3}q^{-\nu_{4,j}^{t}-\rho_{i}}t^{-\nu_{1,i}-\rho_{j}})}{(1-\tilde{Q}_{2}\hat{Q}_{3}q^{-\nu_{4,i}^{t}-\rho_{j}+1/2}t^{-\nu_{3,j}-\rho_{i}-1/2})(1-\tilde{Q}_{2}\hat{Q}_{2}q^{-\nu_{2,j}^{t}-\rho_{i}-1/2}t^{-\nu_{1,i}-\rho_{j}+1/2})}
\eea

Using the results of \cite{Nekrasov:2002qd}, it is easy to show that
the above partition function is the same as the compactified 5D
gauge theory partition function.

\subsection{Slicing independence of the partition function}
\begin{figure}
$\begin{array}{c@{\hspace{1in}}c} \multicolumn{1}{l}{\mbox{}} &
    \multicolumn{1}{l}{\mbox{}} \\ [-0.53cm]
 \includegraphics[width=5in]{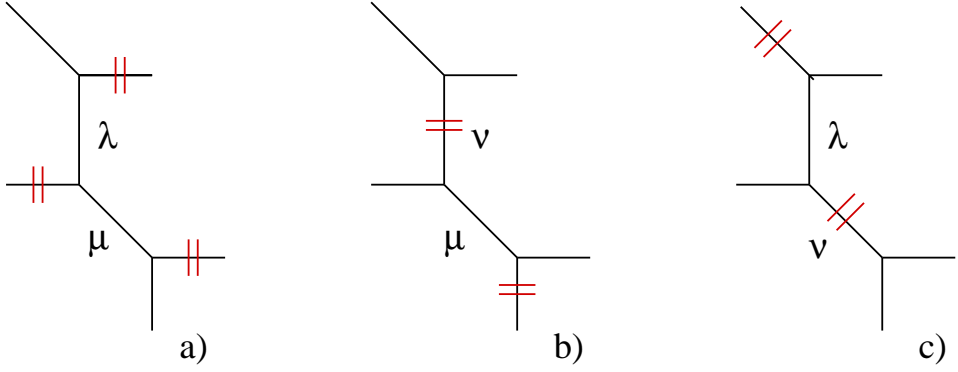}
\\ [0.4cm] \mbox{} & \mbox{}
\end{array}$
\caption{Three different slicings of the same toric diagram. Short
red lines indicate the ``instanton'' direction.} \label{indep}
\end{figure}

To show that the partition functions defined by the refined vertex
are independent of the chosen ``instanton'' or the preferred direction
consider the toric diagrams shown in \figref{indep}. For this diagram
we can choose the preferred direction in three different ways.
Partition function for \figref{indep}(a) is given by \bea
Z&=&\sum_{\lambda,
\mu}(-Q_{1})^{|\lambda|}(-Q_{2})^{|\mu|}C_{\emptyset\,\lambda\,\emptyset}(t,q)\,C_{\mu\,\lambda^{t}\,\emptyset}(q,t)\,
C_{\mu^{t}\,\emptyset\,\emptyset}(t,q)\\\nn &=&
\sum_{\lambda,\mu}(-Q_{1})^{|\lambda|}(-Q_{2})^{|\mu|}s_{\lambda}(q^{-\rho})\,\left(\sum_{\eta}\Big(\frac{t}{q}\Big)^{\frac{|\eta|}{2}}s_{\mu^{t}/\eta}(q^{-\rho})s_{\lambda^{t}/\eta}(t^{-\rho})\right)
s_{\mu}(t^{-\rho})\\\nn &=& \sum_{\eta}
\Big(\frac{t}{q}\Big)^{\frac{|\eta|}{2}}\left(\sum_{\lambda}s_{\lambda}(-Q_{1}q^{-\rho})s_{\lambda^{t}/\eta}(t^{-\rho})\right)\left
(\sum_{\mu}s_{\mu}(-Q_{2}t^{-\rho})s_{\mu^{t}/\eta}(q^{-\rho})\right)\\\nn
&=&\prod_{i,j=1}^{\infty}(1-Q_{1}\,q^{-\rho_{i}}t^{-\rho_{j}})(1-Q_{2}\,q^{-\rho_{i}}t^{-\rho_{j}})
\sum_{\eta}\Big(\frac{t}{q}\Big)^{\frac{|\eta|}{2}}\,s_{\eta^{t}}(-Q_{1}q^{-\rho})s_{\eta^{t}}(-Q_{2}t^{-\rho})\\\nn
&=&\prod_{i,j=1}^{\infty}\frac{(1-Q_{1}\,q^{-\rho_{i}}t^{-\rho_{j}})(1-Q_{2}\,q^{-\rho_{i}}t^{-\rho_{j}})}{1-Q_{1}Q_{2}\,
\sqrt{\frac{t}{q}}\,q^{-\rho_{i}}t^{-\rho_{j}}}.\eea

The partition function for \figref{indep}(b) is given by \bea
Z&=&\sum_{\nu,\lambda}(-Q_{1})^{|\nu|}(-Q_{2})^{|\lambda|}C_{\emptyset\,\emptyset\,\nu}(t,q)\,C_{\emptyset\,\lambda\,\nu^{t}}(q,t)
C_{\emptyset\,\lambda^{t}\,\emptyset}(t,q)\\\nn &=&
\sum_{\nu}(-Q_{1})^{|\nu|}\,q^{\frac{||\nu||^{2}}{2}}\,t^{\frac{||\nu^{t}||^{2}}{2}}\,
\widetilde{Z}_{\nu}(t,q)\,\widetilde{Z}_{\nu^{t}}(q,t)\,\Big(\sum_{\lambda}(-Q_{2})^{|\lambda|}s_{\lambda}(t^{-\rho}\,q^{-\nu})s_{\lambda^{t}}(q^{-\rho})\Big)\\\nn
&=&\sum_{\nu}(-Q_{1})^{|\nu|}\,q^{\frac{||\nu||^{2}}{2}}\,t^{\frac{||\nu^{t}||^{2}}{2}}\,
\widetilde{Z}_{\nu}(t,q)\,\widetilde{Z}_{\nu^{t}}(q,t)\,\prod_{i,j=1}^{\infty}(1-Q_{2}t^{-\rho_{i}}q^{-\rho_{j}-\nu_{i}})\\\nn
&=&\prod_{i,j=1}^{\infty}(1-Q_{2}\,q^{\rho_{i}}\,t^{-\rho_{j}})\,\sum_{\nu}(-Q_{1})^{|\nu|}\,q^{\frac{||\nu||^{2}}{2}}\,t^{\frac{||\nu^{t}||^{2}}{2}}\,
\widetilde{Z}_{\nu}(t,q)\,\widetilde{Z}_{\nu^{t}}(q,t)\,\prod_{(i,j)\in
\nu}(1-Q_{2}t^{-\rho_{i}}q^{-\rho_{j}-\nu_{i}}).\\\nn \eea

The partition function for \figref{indep}(c) is the same as that of
\figref{indep}(b) after changing $\nu\mapsto \nu^{t}$.

Thus for the partition function to be independent of the preferred
direction requires the following identity: \bea\nn
\sum_{\nu}(-Q_{1})^{|\nu|}\,q^{\frac{||\nu||^{2}}{2}}\,t^{\frac{||\nu^{t}||^{2}}{2}}\,
\widetilde{Z}_{\nu}(t,q)\,\widetilde{Z}_{\nu^{t}}(q,t)\,\prod_{(i,j)\in
\nu}(1-Q_{2}\,t^{-\rho_{i}}q^{-\rho_{j}-\nu_{i}})=\prod_{i,j=1}^{\infty}\frac{1-Q_{1}\,q^{-\rho_{i}}t^{-\rho_{j}}}{1-Q_{1}Q_{2}\,
\sqrt{\frac{t}{q}}\,q^{-\rho_{i}}t^{-\rho_{j}}} \eea which can be
written in terms of Macdonald function $P_{\nu}({\bf x};t,q)$ as
\bea\nn
\sum_{\nu}P_{\nu}(-Q_{1}\,t^{-\rho};q,t)\,P_{\nu^{t}}(q^{-\rho};t,q)\,\prod_{(i,j)\in
\nu}(1-Q_{2}\,t^{-\rho_{i}}q^{-\rho_{j}-\nu_{i}})=\prod_{i,j=1}^{\infty}\frac{1-Q_{1}\,q^{-\rho_{i}}t^{-\rho_{j}}}{1-Q_{1}Q_{2}\,
\sqrt{\frac{t}{q}}\,q^{-\rho_{i}}t^{-\rho_{j}}} \eea

For $Q_{2}=0$ this is a well known identity (Example 6, page 352 of
\cite{macdonald}). For $Q_{2}\neq 0$ we have verified the above
identity up to $Q_{1}^3$. This irrelevance of the chosen preferred
direction is the manifestation of duality between supersymmetric
$\mathcal{N}=2$ gauge theories with gauge groups $SU(M)^{N-1}$ and
$SU(N)^{M-1}$ as conjectured in \cite{Katz:1997eq}.

\section{Conclusion}
In this paper we constructed a refined topological vertex which was
used to determine the generalized partition function encoding
left-right spin content information. The derivation of the refined
topological vertex depended upon insights from the instanton
calculus. From the very beginning it was clear that the cyclic
symmetry of the topological vertex will have to be sacrificed in
order to obtain a refined vertex if the instanton calculus is to be
our guide. Whether a refined vertex exists which is cyclically
symmetric and can be used for all toric geometries, unlike the
refined vertex which is not suitable for geometries that do  not
give rise to gauge theories, remains to be seen.

\section*{Acknowledgments}
We would like to thank Sergei Gukov and Sheldon Katz for many
valuable discussions.  AI would also like to thank Charles Doran,
Andreas Karch and Matthew Strassler for many valuable discussions.
CK would also like to thank Andrew O'Bannon for many valuable
discussions. We all would also like to thank the Stony Brook physics
department and fourth Simons Workshop in Mathematics and Physics for
their hospitality while this project was in progress. The research
of C.V. is supported in part by NSF grants PHY-0244821 and
DMS-0244464

\section{Appendix A: Derivation of the Refined Topological Vertex}

In this section, we will review the correspondence between the 3D
partition and the topological vertex following \cite{ORV,OR} and use
the formalism of 3D partitions to define a ``vertex'' which depends on
infinitely many parameters. A specialization of this ``vertex'' will
be the refined vertex.

\subsection{Young diagrams and skew partitions}

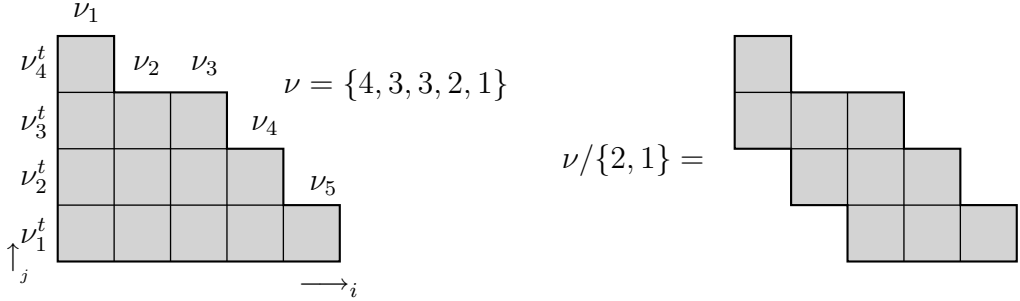
\begin{figure}\begin{center}
$\begin{array}{c@{\hspace{1in}}c} \multicolumn{1}{l}{\mbox{}} &
    \multicolumn{1}{l}{\mbox{}} \\ [-0.53cm]
{
\begin{pspicture}(4,0)(6,4)
\pspolygon[unit=0.75cm,linecolor=black, fillstyle=solid,
fillcolor=lightgray](0,1)(5,1)(5,2)(4,2)(4,3)(3,3)(3,4)(1,4)(1,5)(0,5)

\psline[unit=0.75cm, linecolor=black,linewidth=0.5pt](0,2)(4,2)

\psline[unit=0.75cm, linecolor=black,linewidth=0.5pt](0,3)(3,3)

\psline[unit=0.75cm, linecolor=black,linewidth=0.5pt](0,4)(1,4)

\psline[unit=0.75cm, linecolor=black,linewidth=0.5pt](1,1)(1,5)
\psline[unit=0.75cm, linecolor=black,linewidth=0.5pt](2,1)(2,4)
\psline[unit=0.75cm, linecolor=black,linewidth=0.5pt](3,1)(3,3)
\psline[unit=0.75cm, linecolor=black,linewidth=0.5pt](4,1)(4,2)

\put(3.2,0.35){$\longrightarrow_{\Large{i}}$}
\put(-0.7,0.7){$\uparrow$} \put(-0.5,0.5){\tiny{$j$}}

\put(0.2,4){$\nu_{1}$}\put(1,3.3){$\nu_{2}$}
\put(1.77,3.3){$\nu_{3}$}\put(2.57,2.5){$\nu_{4}$}\put(3.35,1.7){$\nu_{5}$}

\put(-0.5,3.3){$\nu^{t}_{4}$} \put(-0.5,2.5){$\nu^{t}_{3}$}
\put(-0.5,1.77){$\nu^{t}_{2}$}\put(-0.5,1){$\nu^{t}_{1}$}

\put(3,3){$\nu=\{4,3,3,2,1\}$}

\pspolygon[unit=0.75cm,linecolor=black, fillstyle=solid,
fillcolor=lightgray](14,1)(17,1)(17,2)(16,2)(16,3)(15,3)(15,4)(13,4)(13,5)(12,5)(12,3)(13,3)(13,2)(14,2)(14,1)

\psline[unit=0.75cm, linecolor=black,linewidth=0.5pt](14,2)(17,2)

\psline[unit=0.75cm, linecolor=black,linewidth=0.5pt](13,3)(15,3)

\psline[unit=0.75cm, linecolor=black,linewidth=0.5pt](12,4)(13,4)

\psline[unit=0.75cm, linecolor=black,linewidth=0.5pt](13,4)(13,2)
\psline[unit=0.75cm, linecolor=black,linewidth=0.5pt](14,2)(14,4)
\psline[unit=0.75cm, linecolor=black,linewidth=0.5pt](15,3)(15,1)
\psline[unit=0.75cm, linecolor=black,linewidth=0.5pt](16,1)(16,2)

\put(6.7,2){$\nu/\{2,1\}=$}

 \label{young1} \end{pspicture}
}
\\ [-0.5cm] \mbox{} & \mbox{}
\end{array}$\\
\caption{(a) Young diagram of the partition $\{4,3,3,2,1\}$. (b) The
skew Young diagram of $\{4,3,3,2,1\}/\{2,1\}$.}
\label{f2}\end{center}
\end{figure}

Let $\nu =\{\nu_{1}\geq \nu_{2}\geq \nu_{3}\geq\cdots|\,\nu_{i}\geq
0\}$ be a Young diagram, \textit{i.e.}, a 2D partition. We denote by
$|\nu|$ the size of the partition, $|\nu|=\sum_{i}\nu_{i}$, and by
$\ell(\nu)$ the number of non-zero $\nu_{i}$. A pictorial
representation can be obtained by placing $\nu_{i}$ boxes at the
$i^{th}$ position, as shown in \figref{f2} for $\nu=\{4,3,3,2,1\}$. The
height of the columns either stays the same or decreases as we move
to the right. The transpose of $\nu$ is denoted by $\nu^{t}$, \bea
\nu^{t}=\{\nu_{1}^{t},\nu_{2}^{t},\cdots\}\,,\,\,\,\,
\nu_{j}^{t}=\#\{i\,|\,\nu_{i}\geq j\}\,.\eea We denote by $(i,j)\in
\nu$ the box whose upper right corner has coordinates $(i,j)$. If
$(i,j)\in \nu$, then it is clear that $(j,i)\in \nu^{t}$. Given two
partitions $\lambda$ and $\nu$ we say $\lambda\subseteq \nu$, if
$(i,j)\in \lambda$ implies $(i,j)\in \nu$.

Given two partitions $\lambda$ and $\nu$ such that $\lambda\subseteq
\nu$ a skew partition denoted by $\nu/\lambda$ consists of all
boxes of $\nu$ which are not in $\lambda$, \bea
\nu/\lambda=\{(i,j)\in \nu\,|\,(i,j)\notin\lambda\}.\eea A skew
partition $\nu/\lambda$ for $\nu=\{4,3,2,2,1\}$ and
$\lambda=\{2,1\}$ is shown in \figref{f2}(b).

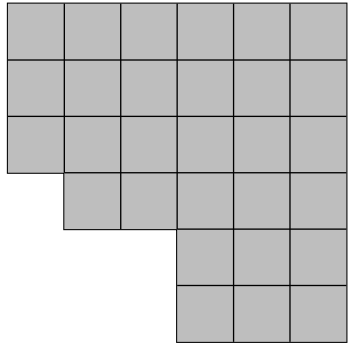
\begin{figure}\begin{center}
$\begin{array}{c@{\hspace{1in}}c} \multicolumn{1}{l}{\mbox{}} &
    \multicolumn{1}{l}{\mbox{}} \\ [-0.53cm]
{

\begin{pspicture}(1,0)(4,4)

\pspolygon[unit=0.75cm,linecolor=black, fillstyle=solid,
fillcolor=lightgray](3,0)(6,0)(6,6)(0,6)(0,3)(1,3)(1,2)(3,2)(3,0)

\psframe[unit=0.75cm,linecolor=gray, fillstyle=solid,
fillcolor=gray](3,0)(6,6)

\psframe[unit=0.75cm, linecolor=gray, fillstyle=solid,
fillcolor=gray](2,2)(3,6)

\psframe[unit=0.75cm, linecolor=gray, fillstyle=solid,
fillcolor=gray](1,2)(2,6)

\psframe[unit=0.75cm, linecolor=gray, fillstyle=solid,
fillcolor=gray](0,3)(1,6)

\psline[unit=0.75cm, linecolor=black,linewidth=0.5pt](3,1)(6,1)
\psline[unit=0.75cm, linecolor=black,linewidth=0.5pt](3,2)(6,2)

\psline[unit=0.75cm, linecolor=black,linewidth=0.5pt](1,3)(6,3)

\psline[unit=0.75cm, linecolor=black,linewidth=0.5pt](0,4)(6,4)
\psline[unit=0.75cm, linecolor=black,linewidth=0.5pt](0,5)(6,5)
\psline[unit=0.75cm, linecolor=black,linewidth=0.5pt](3,2)(3,6)

\psline[unit=0.75cm, linecolor=black,linewidth=0.5pt](1,3)(1,6)
\psline[unit=0.75cm, linecolor=black,linewidth=0.5pt](2,2)(2,6)

\psline[unit=0.75cm, linecolor=black,linewidth=0.5pt](4,0)(4,6)
\psline[unit=0.75cm, linecolor=black,linewidth=0.5pt](5,0)(5,6)

 \label{skew2} \end{pspicture}
}
\\ [-0.5cm] \mbox{} & \mbox{}
\end{array}$
\caption{For a partition $\sigma(N)$ and an arbitrary 2$D$
partition $\nu$, the skew partition $\sigma(N)/\nu$ is always a 2D
partition, provided $N\geq max(\nu_{1},\nu^{t}_{1})$. }
\label{f4}\end{center}
\end{figure}

In general, $\nu/\lambda$ is not a 2D partition, \textit{i.e.}, not a Young
diagram. But if $\nu$ is such that it has $N$ boxes in each row
and $N$ rows then for $N\geq \mbox{max}(\lambda_{1},\lambda_{1}^{t})$ the
skew diagram $\nu/\lambda$ is a 2D partition. We will denote by
$\sigma(N)$ the 2D partition for which $\ell(\sigma(N))=N$ and
$\sigma_{1}(N)=\sigma_{2}(N)=\cdots \sigma_{N}(N)=N$. In this paper,
we will only consider skew partitions of the form $\sigma(N)/\nu$
and will denote this 2D partition by $\nu^{c}$, \bea
\nu^{c}_{i}=N-\nu_{N-i+1}\,,\,\,i=1,2,\cdots, N \eea For
$\nu=\{3,2,2\}$ and $\sigma(6)=\{6,6,6,6,6,6\}$, \figref{f4} shows
the skew partition $\sigma(6)/\nu$.

\subsection{Plane partitions and skew plane partitions} 
\begin{figure}
$\begin{array}{c@{\hspace{1in}}c} \multicolumn{1}{l}{\mbox{}} &
    \multicolumn{1}{l}{\mbox{}} \\ [-0.53cm]
 \includegraphics[width=2.5in]{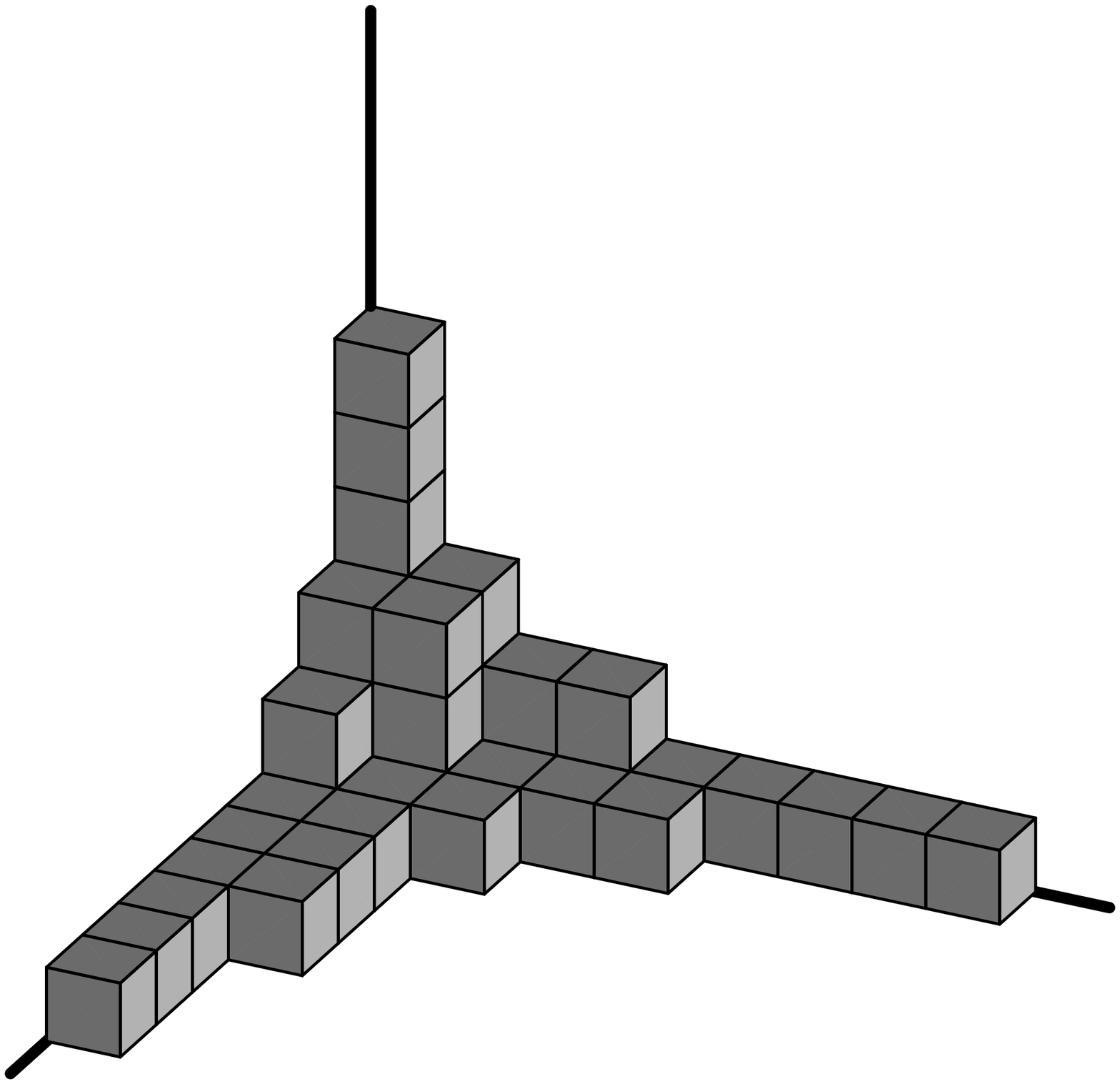} &
\includegraphics[width=2in]{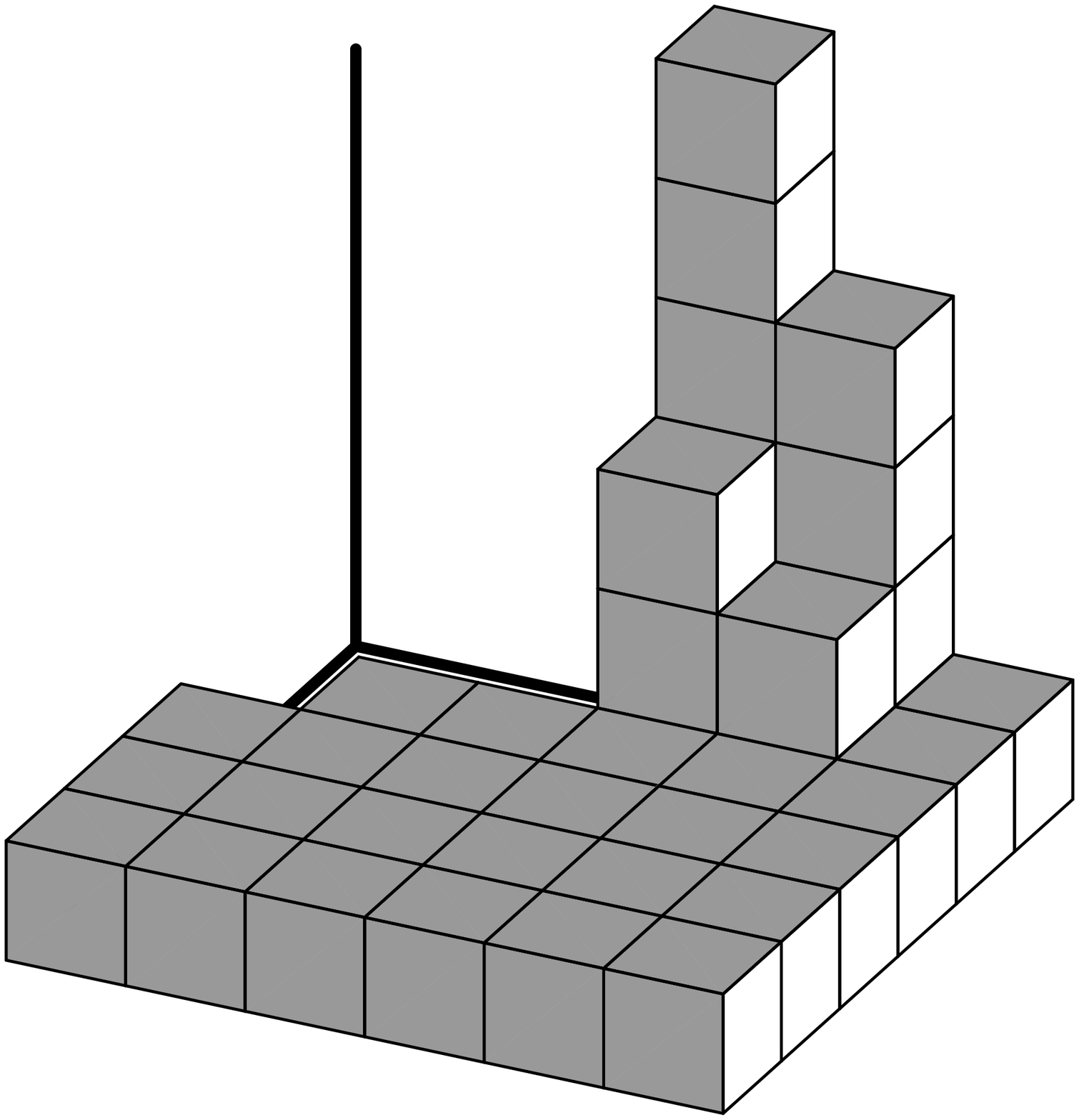}\\
\\ [0.4cm] \mbox{(a)} & \mbox{(b)}
\end{array}$
\caption{(a) A 3D partition, (b) a skew 3D partition.}
\label{f6}
\end{figure}

A
\textit{plane partition} is an array of non-negative integers
$\{\pi_{i,j}\,|\,i,j\geq 1\}$ such that \bea \pi_{i,j}\geq
\pi_{i+r,j+s}\,,\,\,\,\,r,s\geq 0 \eea Placing $\pi_{i,j}$ cubes at
the $(i,j)$ position gives a pictorial representation of the plane
partition. In this sense, plane partitions can be regarded as a 3
dimensional generalization of the Young diagrams, and they are
also known as 3D partitions. The total number of cubes is given by
$|\pi|=\sum_{i,j}\pi_{i,j}$. \figref{f6}(a) shows an example of a
3D partition.

A skew 3D partition of shape $\nu/\lambda$ is an array of
non-negative integers $\{\pi_{i,j}\,\,|\,\,(i,j)\in \nu/\lambda\}$
such that \bea \pi_{i,j}\geq \pi_{i+r,j+s}\,,\,\,\,r,s\geq 0.\eea An
example of a skew plane partition is shown in \figref{f6}(b)

A skew plane partition of shape $\nu^{c}$ will be denoted by
$\pi(\nu)$. It is clear that $\pi(\nu)$ is just a semi-standard
Young tableau (SSYT) of shape $\nu^{c}$ except that we have to subtract the
minimal semi-standard Young tableau of the same shape. Since the sum
of entries of a minimal semi-standard Young diagram of shape
$\lambda$ is given by $m(\lambda)=\sum_{i}i\lambda_{i}$, the
generating function for the number of skew plane partitions of shape
$\nu^{c}$ is given by \bea Z_{\nu}(q):=\sum_{\pi(\nu)}q^{|\pi|}&=&
q^{-m(\nu^{c})}\sum_{SSYT, {\it T}}x_{1}^{\# \mbox{of 1's}}x_{2}^{\#
\mbox{of 2's}}\cdots\\\nn
&=&q^{-m(\nu^{c})}s_{\nu^{c}}(x_{1},x_{2},\cdots)\,,\,\,\,x_{i}=q^{i}\\\nn
&=&q^{-m(\nu^{c})}s_{\nu^{c}}(q,q^{2},q^{3},\cdots)\\\nn
&=&\prod_{(i,j)\in \,\nu^{c}}(1-q^{\widehat{h}(i,j)})^{-1} \eea
Where $\widehat{h}(i,j)=j-\nu_{i}+i-\nu_{j}^{t}-1$ is the hook
length. For $\nu=\emptyset$ we get the number of 3D partitions in
a box of size $N\times N\times \infty $. In the limit $N\mapsto
\infty$ this becomes the MacMahon function,
$\prod_{k=1}^{\infty}(1-q^{k})^{-k}$ \footnote{The generating
function of 3D partitions is given by the product of q-deformed
hook length, $[h(s)]_{q}:=(1-q^{h(s)})^{-1}$, over the infinite 2D
partition, $\sigma(\infty)$. It is easy to see that the generating
function of 2D partitions $\prod_{k=1}^{\infty}(1-q^{k})^{-1}$ is
given by the product of q-deformed hook lengths over the infinite
1D partition. However, the product of q-deformed hook lengths over
the infinite 3D partition is not the generating function of 4D
partitions.}. From now on we will take the limit $N\mapsto \infty$.
In this limit the function\bea \widetilde{Z}_{\nu}(q):=
\frac{Z_{\nu}(q)}{Z_{\emptyset}(q)}\eea can be written as a product
over $\nu$ of q-deformed hook lengths\footnote{Proof of this is
given in Appendix C for the two parameter generalization, this
identity follows by setting $q=t$.}, \bea
\widetilde{Z}_{\nu}(q)&=&\prod_{(i,j)\in
\nu}(1-q^{h(i,j)})^{-1}\,,\,\,\,\,h(i,j)=\nu_{i}-j+\nu_{j}^{t}-i+1\,\\\nn
&=&
q^{-m(\nu)}s_{\nu}(q,q^2,q^3,\cdots)=q^{-\frac{||\nu||^2}{2}}s_{\nu^t}(q^{1/2},q^{3/2},q^{5/2}\cdots)\,.\eea

\begin{figure}\begin{center}
$\begin{array}{c@{\hspace{1in}}c} \multicolumn{1}{l}{\mbox{}} &
    \multicolumn{1}{l}{\mbox{}} \\ [-0.53cm]
{\begin{pspicture}(7,0)(5,5)

\psframe[unit=0.5cm, linecolor=white, fillstyle=solid,
fillcolor=lightgray](3,3)(4,6)

\psframe[unit=0.5cm, linecolor=white, fillstyle=solid,
fillcolor=lightgray](4,3)(6,5)

\psframe[unit=0.5cm, linecolor=white, fillstyle=solid,
fillcolor=lightgray](6,3)(7,4)

\psframe[unit=0.5cm, linecolor=white, fillstyle=solid,
fillcolor=lightgray](2,3)(3,8)

\psframe[unit=0.5cm, linecolor=white,
fillstyle=crosshatch,hatchwidth=0.3pt,hatchsep=2.5pt,
fillcolor=lightgray](5,6)(6,7)

\psline[unit=0.5cm, linecolor=red,linewidth=2pt](5.5,6.5)(5.5,5)
\psline[unit=0.5cm, linecolor=red,linewidth=2pt](5.5,6.5)(3,6.5)
 \psgrid[unit=0.5cm, subgriddiv=1,
gridcolor=myorange, %
gridlabelcolor=white]%
(2,3)(9,10)

\psframe[unit=0.5cm, linecolor=white, fillstyle=solid,
fillcolor=lightgray](17,3)(19,6)

\psframe[unit=0.5cm, linecolor=white, fillstyle=solid,
fillcolor=lightgray](17,7)(18,8)

\psframe[unit=0.5cm, linecolor=white, fillstyle=solid,
fillcolor=lightgray](19,3)(21,5)

\psframe[unit=0.5cm, linecolor=white, fillstyle=solid,
fillcolor=lightgray](21,3)(22,4)

\psframe[unit=0.5cm, linecolor=white, fillstyle=solid,
fillcolor=lightgray](17,5)(18,7)

\psframe[unit=0.5cm, linecolor=white,
fillstyle=crosshatch,hatchwidth=0.3pt,hatchsep=2.5pt,
fillcolor=lightgray](18,4)(19,5)

\psline[unit=0.5cm, linecolor=red,linewidth=2pt](18.5,4.5)(18.5,6)
\psline[unit=0.5cm, linecolor=red,linewidth=2pt](18.5,4.5)(21,4.5)

\psgrid[unit=0.5cm, subgriddiv=1,
gridcolor=myorange, %
gridlabelcolor=white]%
(17,3)(24,10) \put(0,0.8){\mbox{(a)\, The hook length of a box
$(i,j)\in \nu^{c}$,}} \put(1,0.3){\mbox{
$\widehat{h}(i,j)=j-\nu_{i}+i-\nu_{j}^{t}-1$}}

\put(8.5,0.8){\mbox{(b)\, The hook length of a box $(i,j)\in \nu$,}}
\put(9.5,0.3){\mbox{ $h(i,j)=\nu_{i}-j+\nu_{j}^{t}-i+1$}}
\end{pspicture}}
\\ [-0.5cm] \mbox{} & \mbox{}
\end{array}$
\caption{The hook length for $\nu$ and $\nu^{c}$ is defined in the
usual way. Note that the orientation of the hook in $\nu$ and
$\nu^{c}$ agree if we rotate $\nu^{c}$ by $180^{\circ}$ }
\label{f7}\end{center}
\end{figure}

Apart from the framing factor $q^{\frac{||\nu||^{2}}{2}}$, the
function $\widetilde{Z}_{\nu}(q)$ is just the one-partition
topological vertex, \bea C_{\emptyset\,\emptyset\,\nu}(q)=
q^{\frac{||\nu||^{2}}{2}}\widetilde{Z}_{\nu}(q)\,.\eea As discussed
in \cite{ORV}, the topological vertex with all three non-empty
partitions is related to the combinatorics of skew 3D partitions
in which the ``hole'' in the partition is along all three axes. More
specifically, we imagine that the region behind the asymptotic 2D
partition in all three directions is excised.

The boxes are placed in the positive octant $O^{+}$ of ${\mathbb
R}^{3}$ whose coordinates we are going to denote by
$(x,y,z)$. Let us associate the $(x,y)$ plane
with the $(i,j)$ plane. Given a 3D partition $\pi$ as a stack of
cubes in the positive octant $O^{+}$ of $R^{3}$ we can reconstruct
the array of non-negative numbers $\pi_{i,j}$ as the height of the
stack of cubes, \textit{i.e.}, as a height function defined on the
$(x,y)$ plane. However, we can obtain a different array of
non-negative numbers $\pi^{t}_{j,k}$ ($\pi^{tt}_{i,k}$) by
considering the height of the stack relative to the $(y,z)$
($(x,z)$) plane. This transformation is the analog of the
transpose for the 2D partitions.

We can define a generalized skew plane partition
$\pi(\lambda,\mu,\nu)$ as an ordinary 3D partition from which
cubes at $(i,j,k)$ are removed if $(i,j)\in\nu$, or $(j,k)\in \mu$
or $(k,i)\in \lambda$, \bea
\pi(\lambda,\mu,\nu)=\pi\smallsetminus\{(i,j,k)|(i,j)\in\nu\}\cup\{(i,j,k)|(j,k)\in\mu\}\cup\{(i,j,k)|(k,i)\in
\lambda\}\,.\eea

Then we can define the generating function for the number of
generalized skew plane partitions of shape $(\lambda,\mu, \nu)$,
\bea\label{pf}
Z_{\lambda\,\mu\,\nu}(q)=\sum_{\pi(\lambda,\mu,\nu)}q^{|\pi(\lambda,\mu,\nu)|}\,.
\eea It is clear that
$Z_{\lambda\,\mu\,\nu}(q)=Z_{\mu\,\lambda\,\nu^{t}}(q)$. The cyclic
symmetry of the function is related to the transpose of the 3D
partition, \bea
\pi(\lambda,\mu,\nu)&=&\pi^{t}(\mu,\nu,\lambda)\,\,\Rightarrow
\,\,Z_{\lambda\,\mu\,\nu}(q)=Z_{\mu\,\nu\,\lambda}(q)\\\nn
\pi(\lambda,\mu,\nu)&=&\pi^{tt}(\nu,\lambda,\mu)\,\,\Rightarrow
\,\,Z_{\lambda\,\mu\,\nu}(q)=Z_{\nu\,\lambda\,\mu}(q) \eea

Apart from the framing factors
$Z_{\lambda\,\mu\,\nu}(q)/Z_{\emptyset\,\emptyset\,\emptyset}(q)$ is
equal to the topological vertex \cite{ORV}. To calculate
$Z_{\lambda\,\mu\,\nu}(q)$ we use the transfer matrix approach
following \cite{ORV,OR}.

\subsection{Transfer matrix approach and Schur functions} To a 3D
partition $\pi$ we can associate a sequence of 2D  partitions,
$\{\eta(a)\,|\,a\in \mathbb{Z}\}$,  by taking diagonal slices of
$\pi$ as shown in \figref{f7}(a), \bea \eta(a)=\{\pi_{i+a,i}\,|\,i\geq
max(1,-a+1)\}\,.\eea These diagonal slices are perpendicular to the
$(x,y)$ plane and their projections on the base are given by
a set of equations parameterized by $a\in \mathbb{Z}$:
$x-y=a$.

\begin{figure}\begin{center}
$\begin{array}{c@{\hspace{1in}}c} \multicolumn{1}{l}{\mbox{}} &
    \multicolumn{1}{l}{\mbox{}} \\ [-0.53cm]
 \includegraphics[width=2in]{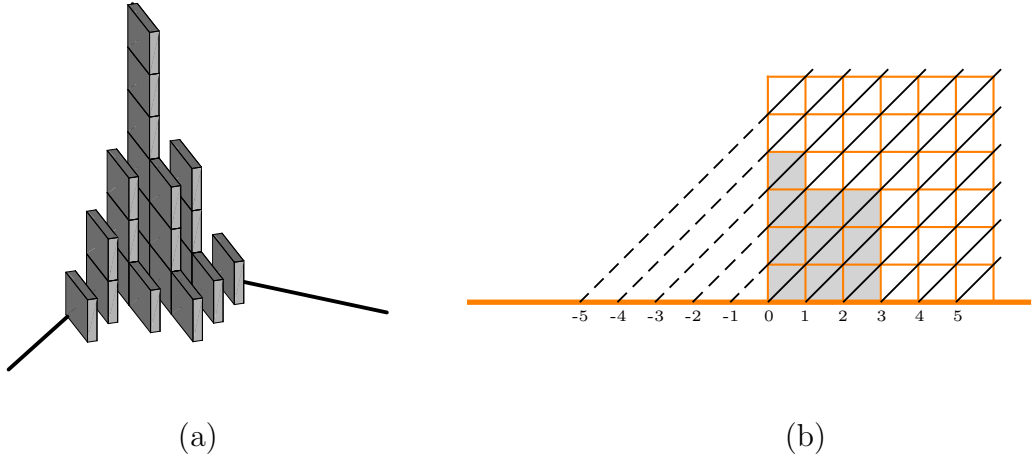} &
{\begin{pspicture}(0,0)(6,6)
 \pspolygon[unit=0.5cm, linecolor=white,
fillstyle=solid, fillcolor=lightgray](5,2)(8,2)(8,5)(6,5)(6,6)(5,6)

\psline[unit=0.5cm, linecolor=myorange,linewidth=2pt](-3,2)(12,2)
\psgrid[unit=0.5cm, subgriddiv=1,
gridcolor=myorange, %
gridlabelcolor=white]%
(5,2)(11,8)

\psline[unit=0.5cm, linecolor=black,linewidth=0.7pt](5,2)(11.2,8.2)
\psline[unit=0.5cm, linecolor=black,linewidth=0.7pt](6,2)(11.2,7.2)
\psline[unit=0.5cm, linecolor=black,linewidth=0.7pt](7,2)(11.2,6.2)
\psline[unit=0.5cm,
linecolor=black,linewidth=0.7pt](8,2)(11.2,5.2)\psline[unit=0.5cm,
linecolor=black,linewidth=0.7pt](9,2)(11.2,4.2)\psline[unit=0.5cm,
linecolor=black,linewidth=0.7pt](10,2)(11.2,3.2)

\psline[unit=0.5cm, linecolor=black,linewidth=0.7pt](5,3)(10.2,8.2)
\psline[unit=0.5cm,
linecolor=black,linewidth=0.7pt](5,4)(9.2,8.2)\psline[unit=0.5cm,
linecolor=black,linewidth=0.7pt](5,5)(8.2,8.2)\psline[unit=0.5cm,
linecolor=black,linewidth=0.7pt](5,6)(7.2,8.2)\psline[unit=0.5cm,
linecolor=black,linewidth=0.7pt](5,7)(6.2,8.2)

\psline[unit=0.5cm,
linestyle=dashed,linecolor=black,linewidth=0.7pt](5,3)(4,2)
\psline[unit=0.5cm,
linestyle=dashed,linecolor=black,linewidth=0.7pt](5,4)(3,2)\psline[unit=0.5cm,
linestyle=dashed,linecolor=black,linewidth=0.7pt](5,5)(2,2)\psline[unit=0.5cm,
linestyle=dashed,linecolor=black,linewidth=0.7pt](5,6)(1,2)\psline[unit=0.5cm,
linestyle=dashed,linecolor=black,linewidth=0.7pt](5,7)(0,2)

\put(2.45,0.75){{\tiny 0}} \put(2.93,0.75){{\tiny 1}}
\put(3.43,0.75){{\tiny 2}} \put(3.95,0.75){{\tiny 3}}
\put(4.45,0.75){{\tiny 4}}\put(4.96,0.75){{\tiny 5}}

\put(1.9,0.75){{\tiny -1}} \put(1.4,0.75){{\tiny -2}}
\put(0.9,0.75){{\tiny -3}} \put(0.4,0.75){{\tiny -4}}
\put(-0.1,0.75){{\tiny -5}}

\end{pspicture}}
\\ [0.4cm] \mbox{(a)} & \mbox{(b)}
\end{array}$
\caption{(a)\,The diagonal slicing of the plane partition: we end up
with a series of 2D  partitions that obey the interlacing condition
Eq. (\ref{interlace}),  (b) The slices are parametrized by integers}
\label{f7}\end{center}
\end{figure}

Each slice obtained from the plane partition will be a 2D partition.
Since these 2D partitions come  from a plane partition, they satisfy
the interlacing condition. Two 2D partitions $\mu$ and $\nu$
interlace, $\mu\succ\nu$, if: \bea
\mu_{1}\geq\nu_{1}\geq\mu_{2}\geq\nu_{2}\geq\mathellipsis \eea The
diagonal slices $\{\eta(a)\,|\,a\in \mathbb{Z}\}$ of a 3D partition
$\pi$ are such that \bea \label{interlace}\eta(a+1)&\succ&
\eta(a)\,,\,\,\,\,\,a<0\,,\\\nn \eta(a)&\succ& \eta(a+1)\,,\,\,a\geq
0\,. \eea There exists a very useful set of coordinates to describe
the 2D partitions called the \textit{Frobenius coordinates}: \bea
a_{i}=\mu_{i}-i+\frac{1}{2}\,,\,\,\,b_{i}=\mu_{i}^{t}-i+\frac{1}{2}\,,\,\,i=1,2,\cdots,d(\mu)
\eea where $d(\mu)$ is the number of squares along the diagonal of
$\mu$. In terms of Frobenius coordinates, one can relate certain
fermionic states to the 2D partition \bea
|\mu\rangle=\prod_{i=1}^{d}\psi_{a_{i}}^{*}\psi_{b_{i}}|0\rangle
\eea where $\psi_{a}$, $\psi_{a}^{*}$, $a\in {\mathbb Z}+1/2$ are
the generators of the Clifford algebra satisfying the following
anti-commutation relations: \bea \{\psi_{a},\psi_{b}\}=0,\,\,
\{\psi_{a}^{*},\psi_{b}^{*}\}=0,\,\,
\{\psi_{a},\psi_{b}^{*}\}=\delta_{ab}. \eea One can define operators
analogous to creation and annihilation operators which can be
written in terms of the modes $J_{n}$ of the fermionic current
$\psi^{*}\psi$, \bea \Gamma_{\pm}(z)=\exp \left ( \sum_{n>0}
\frac{z^{n}J_{\pm n}}{n} \right ). \eea The modes $J_{n}$ of the
fermionic bilinear are such that \bea J_{n}&=&\sum_{k\in
\mathbb{Z}+\frac{1}{2}}\psi_{k+n}\psi^{*}_{k}\,,\,\,\,n=\pm 1,\pm
2,\cdots,\eea and satisfying the commutation relations\bea  [J_{n},
J_{m}]=-n\delta_{n+m,0}\,,\,\,[J_{n},\psi_{k}]=\psi_{k+n}\,,\,\,
[J_{n},\psi^{*}_{k}]=-\psi^{*}_{k-n}\,. \eea The operators
$\Gamma_{\pm}(x)$ satisfy the following commutation relation, \bea
\Gamma_{+}(x)\Gamma_{-}(y)=(1-xy)\Gamma_{-}(y)\Gamma_{+}(x)\eea

The relevance to the creation and annihilation operators becomes
more evident if their action on a state corresponding to a 2D
partition is considered:\bea
\label{ac}\prod_{i}\Gamma_{+}(x_{i})|\lambda \rangle
=\sum_{\mu}s_{\mu/\lambda}(x_{1},x_{2},\cdots)|\mu\rangle\\\nn
\prod_{i}\Gamma_{-}(x_{i})|\lambda \rangle
=\sum_{\mu}s_{\lambda/\mu}(x_{1},x_{2},\cdots)|\mu\rangle \eea Since
\bea s_{\lambda/\mu}(1)=\left\{
                    \begin{array}{ll}
                      1, & \hbox{if}\,\lambda \succ \mu \\
                      0, & \hbox{otherwise.}
                    \end{array}
                  \right\}\,,
\eea it follows from Eq. (\ref{ac}) that
\begin{eqnarray}\nonumber
\Gamma_{+}(1)|\lambda\rangle &=& \sum_{\mu\succ\lambda}|\mu\rangle, \\
\Gamma_{-}(1)|\lambda\rangle &=& \sum_{\lambda\succ\mu}|\mu\rangle .
\end{eqnarray}
The generating function of the number of plane partitions
$Z_{3D}(q):=Z_{\emptyset\,\emptyset\,\emptyset}(q)$  can be now
expressed using $\Gamma_{\pm}$ as \bea
Z_{\emptyset\,\emptyset\,\emptyset}(q)=\langle \emptyset|\left
(\prod_{t=0}^{\infty}q^{L_{0}}\Gamma_{-}(1)\right )q^{L_{0}}\left (
\prod_{t=-\infty}^{-1}\Gamma_{+}(1)q^{L_{0}}\right
)|\emptyset\rangle \eea where $L_{0}$ is the Hamiltonian such that
the operator $q^{L_{0}}$ moves a diagonal slice by one unit. The
action of the operator $q^{L_{0}}$ on a state corresponding to a
2D partition is defined as \bea
q^{L_{0}}|\mu\rangle=q^{|\mu|}|\mu\rangle \eea The intuitive way of
understanding the form of the partition function in terms of the
creation and annihilation operators is straightforward: we start
with the slice at $a=-\infty$ with the empty set and act on this
slice with $\Gamma_{+}(1)$ to create all possible partitions as a
sum. On the next slice (as we go from $a=-\infty$ to 0), we apply the
creation operator on this sum, we again create all possible
partitions such that they interlace the partitions in the previous
slice. We keep acting with $\Gamma_{+}(1)$, until we hit the main
slice $a=0$. The main slice is where we start applying the
annihilation operator $\Gamma_{-}(1)$ which destroys the previously
created partitions, essentially by ``creating'' 2D partitions on
the slice $a$ that are interlaced by the partitions on the previous
slice $a-1$, for positive $a$'s. This procedure, with the operators
$q^{L_{0}}$'s, gives the sum of $q^{|\pi|}$ over all possible 3D
partitions satisfying the interlacing condition that we stated
before. Note that $\Gamma_{-}$ acting on the vacuum gives zero, so
we can move the $\Gamma_{-}$'s to the right to act on the vacuum,
and use the commutation relations between $\Gamma_{\pm}$'s each time
we pass them through each other. In \cite{ORV}, it is shown that
$Z_{3D}(q)$ is actually the McMahon function; \bea
Z_{3D}(q)=\prod_{n>0}\frac{1}{(1-q^{n})^{n}}. \eea

\subsection{Partition function with an infinite number of parameters}

\begin{figure}\begin{center}
$\begin{array}{c@{\hspace{1in}}c} \multicolumn{1}{l}{\mbox{}} &
    \multicolumn{1}{l}{\mbox{}} \\ [-0.53cm]
{\begin{pspicture}(0,1)(6,3)
 \pspolygon[unit=0.5cm, linecolor=white,
fillstyle=solid, fillcolor=lightgray](5,2)(8,2)(8,5)(6,5)(6,6)(5,6)

\psline[unit=0.5cm, linecolor=myorange,linewidth=2pt](-3,2)(12,2)
\psgrid[unit=0.5cm, subgriddiv=1,
gridcolor=myorange, %
gridlabelcolor=white]%
(5,2)(11,8)

\psline[unit=0.5cm, linecolor=black,linewidth=0.7pt](5,2)(11.2,8.2)
\psline[unit=0.5cm,
linecolor=black,linewidth=0.7pt](8,2)(11.2,5.2)

\psline[unit=0.5cm,
linecolor=black,linewidth=0.7pt](5,4)(9.2,8.2)\psline[unit=0.5cm,
linecolor=black,linewidth=0.7pt](5,5)(8.2,8.2)\psline[unit=0.5cm,
linecolor=black,linewidth=0.7pt](5,6)(7.2,8.2)

\psline[unit=0.5cm,linestyle=dashed,linecolor=black,linewidth=0.7pt](5,4)(3,2)
\psline[unit=0.5cm,linestyle=dashed,linecolor=black,linewidth=0.7pt](5,5)(2,2)
\psline[unit=0.5cm,linestyle=dashed,linecolor=black,linewidth=0.7pt](5,6)(1,2)

\put(2.45,0.75){{\tiny 0}} \put(2.93,0.75){{\tiny 1}}
\put(3.43,0.75){{\tiny 2}} \put(3.95,0.75){{\tiny 3}}
\put(4.45,0.75){{\tiny 4}}\put(4.96,0.75){{\tiny 5}}

\put(1.9,0.75){{\tiny -1}} \put(1.4,0.75){{\tiny -2}}
\put(0.9,0.75){{\tiny -3}} \put(0.4,0.75){{\tiny -4}}
\put(-0.1,0.75){{\tiny -5}}

\put(6,3.6){Three inner corners: $v_0,v_1,v_2$} \put(6,3.1){Two
outer corners: $u_{0},u_{1}$}
\put(6,2.1){$\{v_{1},v_{2},v_{3}\}=\{-4,-2,3\}$}
\put(6,1.6){$\{u_{1},u_{2}\}=\{-3,0\}$}
\put(6,1){$\sum_{i}u_{i}=\sum_{i}v_{i}$}
\end{pspicture}}
\\ [0.4cm] \mbox{} & \mbox{}
\end{array}$
\caption{Inner and Outer corners of the partition  $\nu=(4,3,3)$. } \label{ff8}\end{center}
\end{figure}
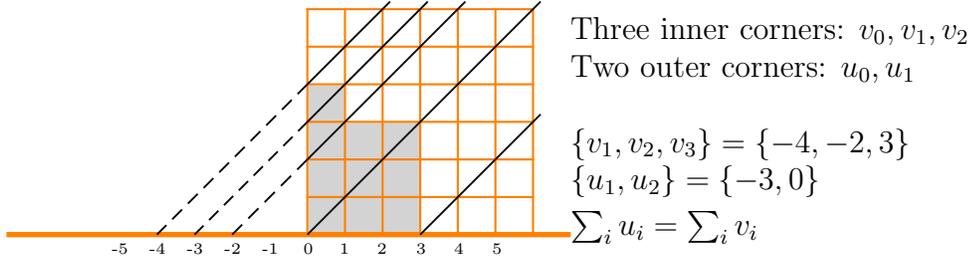

In the previous section, we briefly outlined the systematic way to
compute the partition function $Z_{3D}(q)$. We assumed that the
partition at each slice is counted with the same parameter $q$. In
this section, following \cite{OR} we want to describe the
generalization of this to an infinite number of parameters and show
that this generalization gives the same partition function when the
different parameters on each slice are set to be equal to each
other. Let us begin with the case when $\lambda=\mu=\emptyset$ and
then we will allow $\lambda$ and $\mu$ to be non-trivial. We keep
our convention from the previous section that an integer $a$ is used
to describe each slice and we associate the parameter $q_{a}$ to
each slice. For a 2D partition $\nu$, we can divide the
corners of the pictorial representation of the corresponding
partition into two groups: inner and outer corners. We parametrize
the inner and outer corners by their coordinates, $v_{i}$ and
$u_{i}$ respectively, of their projection onto the real line as
shown in \figref{ff8}. It is easy to see that \cite{OR} \bea
\sum_{i=0}^{M}v_{i}=\sum_{i=0}^{M-1}u_{i}\,,\,\,\,M=\mbox{$\#$ of
outer corners}\,. \eea

It is convenient to introduce
another set of parameters $\{x^{\pm}_{m}\,|\,m\in
\mathbb{Z}+\tfrac{1}{2}\}$ and identify them with $q_{a}$'s in the
following shape dependent way \cite{OR},\bea
\frac{x^{+}_{m+1}}{x^{+}_{m}}&=&q_{m+\frac{1}{2}}\,,\,\,\,\,m>v_{M}\,\,\mbox{or}\,\,\,u_{i}-1>m>v_{i}\,,\\\nn
x^{+}_{u_{i}-\frac{1}{2}}x^{-}_{u_{i}+\frac{1}{2}}&=&
q_{u_{i}}^{-1}\,,\\\nn
x^{-}_{v_{i}-\frac{1}{2}}x^{+}_{v_{i}+\frac{1}{2}}&=&q_{v_{i}}\,,\\\nn
\frac{x^{-}_{m}}{x^{-}_{m+1}}&=&q_{m+\frac{1}{2}}\,,\,\,m<v_{1}\,\,\mbox{or}\,\,\,v_{i+1}-1>m>u_{i}\,.
\eea The generating function of the 3D partitions is then given by \cite{OR}
\bea Z_{3D}({\bf q})=\sum_{\pi}\prod_{a\in
\mathbb{Z}}q_{a}^{|\eta_a|}&=&\langle
0|\prod_{k=-\infty}^{0}\Gamma_{-}(x^{+}_{-k+\frac{1}{2}})\prod_{k=1}^{\infty}
\Gamma_{+}(x^{-}_{-k+\frac{1}{2}})|0\rangle\\\nn
&=&\prod_{k_{1}=1}^{\infty}\prod_{k_{2}=1}^{\infty}\Big(1-x^{+}_{k_{1}-\frac{1}{2}}x^{-}_{-k_{2}+\frac{1}{2}}\Big)^{-1}
\eea \bea x^{+}_{k-\frac{1}{2}}&=&\prod_{i=0}^{k-1}q_{i}\,,\,\,k\geq
1\,,\\\nn x^{-}_{-\frac{1}{2}}&=&1\,,\,\,\,x^{-}_{-k+\frac{1}{2}}=
\prod_{i=1}^{k-1} q_{-i}\,,\,\,\,k\geq 2\,. \eea \bea Z_{3D}({\bf
q})=\prod_{k_{1}=1}^{\infty}\prod_{k_{2}=1}^{\infty}\Big(1-\prod_{i=0}^{k_{1}-1}q_{i}\prod_{j=1}^{k_{2}-1}q_{-j}\Big)^{-1}
\eea For $q_{i}=q\,,\,i\in \mathbb{Z}$ we get \bea
Z_{3D}(q)=\prod_{k=1}^{\infty}(1-q^{k})^{-k}\,.\eea

Having shown that setting all parameters equal to each other agrees
with what we originally obtained, we can continue to develop the
generalization to non-trivial $\lambda$ and $\mu$. The partition
function with $\lambda=\mu=\emptyset$ is given by ($\nu$ being the
2D partition in the preferred direction) \bea   Z_{\nu}({\bf
q})&=&P_{\emptyset\,\emptyset\,\nu}(\mathbf{q})=\prod_{(i,j)\in
\nu^{c}}(1-q_{(i,j)})^{-1}\,,\\\nn q_{(i,j)}&=&\prod_{(a,b)\in
H(i,j)}q_{b-a}\,, \eea where $H(i,j)$ is the set of boxes which form
the hook of $(i,j)$.

The partition function $Z_{3D}({\bf q})$ can be written as a product
over boxes of $\sigma(\infty)$ such that each box contributes a
factor of $(1-x)^{-1}$ where $x$ is the product of parameters
$q_{i}$ intersected by the hook length\footnote{$Z_{3D}({\bf
q})=\prod_{i,j=1}^{\infty}(1-(\prod_{a=1}^{i}q_{a-j})(\prod_{b=1}^{j-1}q_{i-b}))^{-1}$.}.
A similar interpretation in terms of hook length exists for the
generating function of skew plane partitions.

We define $P_{\lambda\mu\nu}({\bf q})$ as \cite{OR} \bea
P_{\lambda\mu\nu}({\bf q})=\sum_{\pi\diagdown \pi_{0}}\prod_{k\in
\mathbf{Z}}q_{k}^{|\lambda(k)|} \eea Where the sum is over all 3D
partitions $\pi$ such that $\pi$ asymptotically approaches
$\lambda,\mu$ and $\nu$ along the three axes and $\pi_{0}$ is a 3D
partition with the least number of boxes. Each such partition $\pi$
can be sliced along the diagonal such that we get a 2D partition
$\pi(a)$ along the diagonal passing through $(0,a)$. In defining
$P_{\lambda\mu\nu}(\mathbf{q})$ we weight each slice with a different parameter
$q_{a}$.

For non-trivial $\lambda$ and $\mu$ the partition function is given
by \cite{OR}\bea P_{\lambda\mu\nu}({\bf q})=Z_{\nu}({\bf
q})\,\sum_{\eta}s_{\lambda^{t}/\eta}({\bf x}^{+})\,s_{\mu/\eta}({\bf
x}^{-})\,,\eea where ${\bf x}^{\pm}=\{x^{\pm}_{m}\,|\,m\in
\mathbb{Z}+\tfrac{1}{2}\}$. This is the most general partition
function in which each diagonal slice is counted with a different
parameter.

At the end of this section, we would like to start talking about how
to assign $q$ and $t$ to the slices depending on the shape of $\nu$
to get the generalized partition function from the generalized plane
partitions. We will leave the physical motivation for the particular
choice and the details to the next section.

\begin{figure}\begin{center}
$\begin{array}{c@{\hspace{1in}}c} \multicolumn{1}{l}{\mbox{}} &
    \multicolumn{1}{l}{\mbox{}} \\ [-0.53cm]
{
\begin{pspicture}(7,0)(6,6)
\psline[unit=0.5cm, linecolor=gray,linewidth=1pt](11,3)(30,3)
\psline[unit=0.5cm, linecolor=gray,linewidth=1pt](11,4)(30,4)

\psframe[unit=0.5cm,linecolor=black,fillstyle=solid,fillcolor=gray,linewidth=0.7pt](13,3)(14,4)
\psframe[unit=0.5cm,linecolor=black,fillstyle=solid,fillcolor=gray,linewidth=0.7pt](14,3)(15,4)
\psframe[unit=0.5cm,linecolor=black,fillstyle=solid,fillcolor=gray,linewidth=0.7pt](15,3)(16,4)
\psframe[unit=0.5cm,linecolor=black,fillstyle=solid,fillcolor=white,linewidth=0.7pt](16,3)(17,4)
\psframe[unit=0.5cm,linecolor=black,fillstyle=solid,fillcolor=gray,linewidth=0.7pt](17,3)(18,4)
\psframe[unit=0.5cm,linecolor=black,fillstyle=solid,fillcolor=gray,linewidth=0.7pt](18,3)(19,4)
\psframe[unit=0.5cm,linecolor=black,fillstyle=solid,fillcolor=white,linewidth=0.7pt](19,3)(20,4)
\psframe[unit=0.5cm,linecolor=black,fillstyle=solid,fillcolor=white,linewidth=0.7pt](20,3)(21,4)
\psframe[unit=0.5cm,linecolor=black,fillstyle=solid,fillcolor=gray,linewidth=0.7pt](21,3)(22,4)
\psframe[unit=0.5cm,linecolor=black,fillstyle=solid,fillcolor=white,linewidth=0.7pt](22,3)(23,4)
\psframe[unit=0.5cm,linecolor=black,fillstyle=solid,fillcolor=white,linewidth=0.7pt](23,3)(24,4)
\psframe[unit=0.5cm,linecolor=black,fillstyle=solid,fillcolor=gray,linewidth=0.7pt](24,3)(25,4)
\psframe[unit=0.5cm,linecolor=black,fillstyle=solid,fillcolor=white,linewidth=0.7pt](25,3)(26,4)
\psframe[unit=0.5cm,linecolor=black,fillstyle=solid,fillcolor=white,linewidth=0.7pt](26,3)(27,4)

\put(5.25,1.65){$\cdots \cdots$} \put(13.8,1.65){$\cdots\cdots$}

\put(10,4){$\downarrow\,\,=$} \put(9.815,3.3){$\rightarrow\,\,=$}

\psframe[unit=0.5cm,linecolor=black,fillstyle=solid,fillcolor=gray,linewidth=0.7pt](22,7.8)(23,8.8)
\psframe[unit=0.5cm,linecolor=black,fillstyle=solid,fillcolor=white,linewidth=0.7pt](22,6.5)(23,7.5)

\pspolygon[unit=0.5cm,linecolor=white,fillstyle=solid,fillcolor=lightgray]
(2,3)(7,3)(7,4)(5,4)(5,5)(3,5)(3,7)(2,7)

\psline[unit=0.5cm,arrows=->,linecolor=black](2,10)(2,9)
\psline[unit=0.5cm,arrows=->,linecolor=black](2,9)(2,8)
\psline[unit=0.5cm,arrows=->,linecolor=black](2,8)(2,7)

\psline[unit=0.5cm,arrows=->,linecolor=black](2,7)(3,7)
\psline[unit=0.5cm,arrows=->,linecolor=black](3,7)(3,6)
\psline[unit=0.5cm,arrows=->,linecolor=black](3,6)(3,5)

\psline[unit=0.5cm,arrows=->,linecolor=black](3,5)(4,5)
\psline[unit=0.5cm,arrows=->,linecolor=black](4,5)(5,5)

\psline[unit=0.5cm,arrows=->,linecolor=black](5,5)(5,4)
\psline[unit=0.5cm,arrows=->,linecolor=black](5,4)(6,4)
\psline[unit=0.5cm,arrows=->,linecolor=black](6,4)(7,4)
\psline[unit=0.5cm,arrows=->,linecolor=black](7,4)(7,3)

\psline[unit=0.5cm,arrows=->,linecolor=black](7,3)(8,3)
\psline[unit=0.5cm,arrows=->,linecolor=black](8,3)(9,3)
\psgrid[unit=0.5cm, subgriddiv=1,
gridcolor=myorange, %
gridlabelcolor=white]%
(2,3)(9,10)

\put(2.5,0.5){(a)}
\put(10,0.5){(b)}
\end{pspicture}}
\end{array}$
\caption{a) We can trace the profile of a particular 2D partition
starting at $j=\infty$ and going to $i=\infty$ as depicted in the
figure. b) To each vertical pass we associate a black box, and a
white one to each horizontal pass. If we put these boxes in a row,
then we get a unique ``finger print'' to a partition. The
coordinates of the centers are given by the sets $D^{\pm}$. }
\label{f9}\end{center}
\end{figure}

\figref{f9} illustrates the idea behind our choice: while following
the arrows on the boundary of the 2D partition, every time we have
an arrow pointing down, we assign a black box (\figref{f9}(b), and
every time we have an arrow pointing to the right, we assign a white
box. The coordinates of the center of these boxes are given by \bea
\mbox{Black
Boxes}:\,\,\{\mu_{i}-i+\frac{1}{2}\,|\,i=1,2,\cdots\}\\\nn
\mbox{White
Boxes}:\,\,\{j-\mu^{t}_{j}-\frac{1}{2}\,|\,j=1,2,\cdots\} \eea

These coordinates are closely related to the Frobenius coordinates.
Note that if we count the number of black boxes to the left of the
$i^{th}$ white box, we get $\nu_{i}$. Similarly, if we count the number
of white boxes to the right of the $j^{th}$ black box, we get
$\nu_{j}^{t}$.

We can divide the half-integers into two sets using the function
$\epsilon(n)$ defined as \bea \epsilon(n)&=&+\,\, \textrm{if}\,\,
v_{i}<n<u_{i} \\\nn \epsilon(n)&=&-\,\, \textrm{if}\,\,
u_{i}<n<v_{i+1} \eea for $0\leq i\leq M-1$\footnote{$M$ is the
number of outer corners.}: $D^{+}=\{n|\epsilon(n)=+\}$ and similarly
$D^{-}=\{n|\epsilon(n)=-\}$. The sets $D^{+}$ and $D^{-}$ are
actually the same as the sets consisting of the coordinates of the
center of the black and white boxes, respectively.

\subsection{Equivariant parameters, boundary of the Young
diagram and instanton calculus}

The vertex we have obtained so far counts each slice with a
different parameter and therefore depends on infinitely many
parameters. The usual vertex can be obtained by setting all the
parameters equal to $q$ \cite{ORV, OR}. It is clear that we can
obtain the vertex which depends on two parameters by some choice of
identification between $q_{a}$ and $t,q$. It is not clear a priori
what the map $\{q_{a}\,|\,a\in \mathbb{Z}\} \mapsto \{t,q\}$ should
be.

However, the relation between the instanton partition functions and
A-model topological string partition function, via geometric
engineering, provides some insight into the possible map between the
parameters. Recall that the partition function of the 5D
compactified $U(1)$ theory can be written as \cite{ON}\bea
Z(\epsilon_{1},\epsilon_{2},\beta)=\sum_{\nu}\mbox{exp}\Big(-\frac{1}{4}\int_{x\neq
y}f_{\nu}''(\epsilon_{1},\epsilon_{2}|x)f_{\nu}''(\epsilon_{1},\epsilon_{2}|y)\gamma_{\epsilon_{1},\epsilon_{2}}(x-y|\beta,\Lambda)\Big)\,,\eea
where\bea\nn \gamma_{\epsilon_{1},\epsilon_{2}}(x|\beta,\Lambda)=
\frac{1}{2\epsilon_{1}\epsilon_{2}}\Big[-\frac{\beta}{6}\Big(x+\frac{1}{2}(\epsilon_{1}+\epsilon_{2})\Big)^3+x^2\mbox{log}(\beta\Lambda)\Big]+\sum_{n=1}^{\infty}\frac{1}{n}\frac{e^{-\beta
n x}}{(1-e^{\beta n \epsilon_{1}})(1-e^{\beta n \epsilon_{2}})}\eea
and $f_{\nu}(\epsilon_{1},\epsilon_{2}|x)$ is the profile of the
partition $\nu$ ($\epsilon_{2}>0>\epsilon_{1}$),\bea\nn
f_{\nu}(x|\epsilon_{1},\epsilon_{2})=|x|+\sum_{i=1}^{\infty}\Big(|x+\epsilon_{1}-\epsilon_{2}\nu_{i}-\epsilon_{1}\,i|-
|x-\epsilon_{1}-\epsilon_{1}\,i|-|x-\epsilon_{2}\nu_{i}-\epsilon_{1}i|+|x-\epsilon_{1}i|\Big)\,.\eea
The profile of the partition controls the contribution of the
partition to the partition function. The parameters $-\epsilon_{1}$
and $\epsilon_{2}$ are the height and the width of the boxes in the
partition as shown in \figref{f10}. Since these 2D partitions on the
edges are the boundaries of the 3D partitions therefore the height
and the width of the 3D box is exactly $-\epsilon_{1},\epsilon_{2}$
as shown in \figref{f11}.

Hence in constructing the 3D partition from the diagonal slices as
we move the slice towards the left we move it an amount
$-\epsilon_{1}$ and as we move it upward we move it an amount
$\epsilon_{2}$. In the transfer matrix formalism this implies that
different diagonal slices are counted with different parameters
$e^{-\epsilon_{1}}$ and $e^{\epsilon_{2}}$. Since the shape of the
partition $\nu$ in the $z$ direction determines the left-ward and
upward motion of the slice therefore slices are counted with
$e^{-\epsilon_{1}}$ and $e^{\epsilon_{2}}$ according to the shape of
the partition $\nu$.

\begin{figure}\begin{center}
$\begin{array}{c@{\hspace{1in}}c} \multicolumn{1}{l}{\mbox{}} &
    \multicolumn{1}{l}{\mbox{}} \\ [-0.53cm]
{\begin{pspicture}(6,0)(6,6)
\psline[unit=0.3cm,linecolor=myorange,linewidth=2pt](5,1)(25,1)

\pspolygon[unit=0.3cm,linecolor=black,fillstyle=solid,fillcolor=lightgray]
(15,1)(11,5)(12,6)(13,5)(15,7)(16,6)(17,7)(18,6)(19,7)(20,6)

\psline[unit=0.3cm,linecolor=black,linewidth=2pt](20,6)(23,9)
\psline[unit=0.3cm,linecolor=black,linewidth=2pt](11,5)(7,9)

\psline[unit=0.3cm,linecolor=black,linewidth=0.6pt](16,2)(12,6)
\psline[unit=0.3cm,linecolor=black,linewidth=0.6pt](17,3)(14,6)
\psline[unit=0.3cm,linecolor=black,linewidth=0.6pt](18,4)(16,6)
\psline[unit=0.3cm,linecolor=black,linewidth=0.6pt](19,5)(18,6)

\psline[unit=0.3cm,linecolor=black,linewidth=0.6pt](14,2)(19,7)
\psline[unit=0.3cm,linecolor=black,linewidth=0.6pt](13,3)(16,6)
\psline[unit=0.3cm,linecolor=black,linewidth=0.6pt](12,4)(13,5)

\psline[unit=0.3cm,linecolor=black,linewidth=2pt](11,5)(12,6)(13,5)(15,7)(16,6)(17,7)(18,6)(19,7)(20,6)

\end{pspicture}} &
{\begin{pspicture}(1,0)(6,6)
\psline[unit=0.3cm,linecolor=myorange,linewidth=2pt](5,1)(25,1)

\pspolygon[unit=0.3cm,linecolor=black,fillstyle=solid,fillcolor=lightgray]
(15,1)(7,9)(7.5,9.5)(9.5,7.5)(10.5,8.5)(12.5,6.5)(13,7)(15,5)(15.5,5.5)(17.5,3.5)

\psline[unit=0.3cm,linecolor=black,linewidth=2pt](17.5,3.5)(24,10)
\psline[unit=0.3cm,linecolor=black,linewidth=2pt](7,9)(6,10)

\psline[unit=0.3cm,linecolor=black,linewidth=0.6pt](13,3)(15,5)
\psline[unit=0.3cm,linecolor=black,linewidth=0.6pt](11,5)(13,7)
\psline[unit=0.3cm,linecolor=black,linewidth=0.6pt](9,7)(9.5,7.5)

\psline[unit=0.3cm,linecolor=black,linewidth=0.6pt](15.5,1.5)(9.5,7.5)

\psline[unit=0.3cm,linecolor=black,linewidth=0.6pt](16,2)(10,8)
\psline[unit=0.3cm,linecolor=black,linewidth=0.6pt](16.5,2.5)(12.5,6.5)

\psline[unit=0.3cm,linecolor=black,linewidth=0.6pt](17,3)(15,5)

\psline[unit=0.3cm,linecolor=black,linewidth=2pt](7,9)(7.5,9.5)(9.5,7.5)(10.5,8.5)(12.5,6.5)(13,7)(15,5)(15.5,5.5)(17.5,3.5)

\put(-3.3,-0.25){(a)}
\put(4.3,-0.25){(b)}

\end{pspicture}}
\end{array}$
\caption{The profile of the partition is drawn bold; a) shows the
2D partition for the self-dual case $\epsilon_{2}=-\epsilon_{1}$,
whereas b) shows the same partition for the non self-dual case
$\epsilon_{1}\ne\epsilon_{2}$ $(\epsilon_{2}=-2\epsilon_{1})$ }
\label{f10}\end{center}
\end{figure}
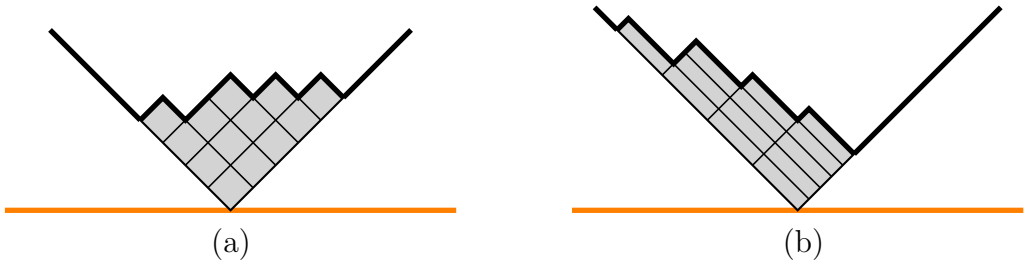

\begin{figure}\begin{center}
$\begin{array}{c@{\hspace{1in}}c} \multicolumn{1}{l}{\mbox{}} &
    \multicolumn{1}{l}{\mbox{}} \\ [-0.53cm]
{\begin{pspicture}(9,-2,)(10,5)

\pspolygon[unit=0.3cm,linecolor=black,fillstyle=solid,fillcolor=lightgray]
(15,5)(12,2)(12,3)(13,4)(13,5)(14,6)(14,8)(15,9)

\psline[unit=0.3cm,linecolor=black,linewidth=0.6pt](15,5)(22,5)
\psline[unit=0.3cm,linecolor=black,linewidth=0.6pt](23,5)(30,5)
\psline[unit=0.3cm,linecolor=black,linewidth=0.6pt](15,5)(15,12)
\psline[unit=0.3cm,linecolor=black,linewidth=0.6pt](30,5)(30,-0.5)
\psline[unit=0.3cm,linecolor=black,linewidth=0.6pt](15,5)(10,0)

\psline[unit=0.3cm,linecolor=black,linewidth=1pt](12,2)(12,3)
\psline[unit=0.3cm,linecolor=black,linewidth=1pt](12,3)(13,4)
\psline[unit=0.3cm,linecolor=black,linewidth=0.6pt](13,4)(15,6)
\psline[unit=0.3cm,linecolor=black,linewidth=0.6pt](13,3)(13,4)
\psline[unit=0.3cm,linecolor=black,linewidth=1pt](13,4)(13,5)
\psline[unit=0.3cm,linecolor=black,linewidth=0.6pt](13,5)(15,7)
\psline[unit=0.3cm,linecolor=black,linewidth=0.6pt](14,4)(14,6)
\psline[unit=0.3cm,linecolor=black,linewidth=1pt](14,6)(14,8)
\psline[unit=0.3cm,linecolor=black,linewidth=1pt](14,8)(15,9)
\psline[unit=0.3cm,linecolor=black,linewidth=0.6pt](14,7)(15,8)

\psline[unit=0.3cm,linecolor=black,linewidth=0.6pt](30,5)(35,10)

\psline[unit=0.3cm,linecolor=black,linewidth=0.6pt](19,2)(19,3)
\psline[unit=0.3cm,linecolor=black,linewidth=0.6pt](19,3)(20,4)
\psline[unit=0.3cm,linecolor=black,linewidth=0.6pt](20,4)(22,6)
\psline[unit=0.3cm,linecolor=black,linewidth=0.6pt](20,3)(20,4)
\psline[unit=0.3cm,linecolor=black,linewidth=0.6pt](20,4)(20,5)
\psline[unit=0.3cm,linecolor=black,linewidth=0.6pt](20,5)(22,7)
\psline[unit=0.3cm,linecolor=black,linewidth=0.6pt](21,4)(21,6)
\psline[unit=0.3cm,linecolor=black,linewidth=0.6pt](21,6)(21,8)
\psline[unit=0.3cm,linecolor=black,linewidth=0.6pt](21,8)(22,9)
\psline[unit=0.3cm,linecolor=black,linewidth=0.6pt](21,7)(22,8)
\psline[unit=0.3cm,linecolor=black,linewidth=0.6pt](22,5)(22,9)

\psline[unit=0.3cm,linecolor=black,linewidth=0.6pt](22,5)(19,2)

\psline[unit=0.3cm,linecolor=black,linewidth=0.3pt,linestyle=dashed](15,9)(22,9)
\psline[unit=0.3cm,linecolor=black,linewidth=0.3pt,linestyle=dashed](14,8)(21,8)
\psline[unit=0.3cm,linecolor=black,linewidth=0.3pt,linestyle=dashed](14,6)(21,6)
\psline[unit=0.3cm,linecolor=black,linewidth=0.3pt,linestyle=dashed](13,5)(20,5)
\psline[unit=0.3cm,linecolor=black,linewidth=0.3pt,linestyle=dashed](13,4)(20,4)
\psline[unit=0.3cm,linecolor=black,linewidth=0.3pt,linestyle=dashed](12,3)(19,3)
\psline[unit=0.3cm,linecolor=black,linewidth=0.3pt,linestyle=dashed](12,2)(19,2)

\put(6.5,1.35){\,\,$\wr\,\wr$}

\put(5,-1){$\epsilon_{2}=-\epsilon_{1}$}
\end{pspicture}} &

{\begin{pspicture}(5,-1.9)(6,5)
\pspolygon[unit=0.3cm,linecolor=black,fillstyle=solid,fillcolor=lightgray]
(15,5)(10.5,0.5)(10.5,1.5)(12,3)(12,4)(13.5,5.5)(13.5,7.5)(15,9)

\psline[unit=0.3cm,linecolor=black,linewidth=0.6pt](15,5)(22,5)
\psline[unit=0.3cm,linecolor=black,linewidth=0.6pt](23,5)(30,5)
\psline[unit=0.3cm,linecolor=black,linewidth=0.6pt](15,5)(15,12)
\psline[unit=0.3cm,linecolor=black,linewidth=0.6pt](30,5)(30,-0.5)
\psline[unit=0.3cm,linecolor=black,linewidth=0.6pt](15,5)(10,0)
\psline[unit=0.3cm,linecolor=black,linewidth=0.6pt](30,5)(35,10)

\psline[unit=0.3cm,linecolor=black,linewidth=0.6pt](12,3)(15,6)
\psline[unit=0.3cm,linecolor=black,linewidth=0.6pt](13.5,5.5)(15,7)
\psline[unit=0.3cm,linecolor=black,linewidth=0.6pt](13.5,6.5)(15,8)
\psline[unit=0.3cm,linecolor=black,linewidth=0.6pt](12,2)(12,3)
\psline[unit=0.3cm,linecolor=black,linewidth=0.6pt](13.5,3.5)(13.5,6)

\pspolygon[unit=0.3cm,linecolor=black,fillstyle=none,fillcolor=lightgray]
(22,5)(17.5,0.5)(17.5,1.5)(19,3)(19,4)(20.5,5.5)(20.5,7.5)(22,9)

\psline[unit=0.3cm,linecolor=black,linewidth=0.6pt](19,3)(22,6)
\psline[unit=0.3cm,linecolor=black,linewidth=0.6pt](20.5,5.5)(22,7)
\psline[unit=0.3cm,linecolor=black,linewidth=0.6pt](20.5,6.5)(22,8)
\psline[unit=0.3cm,linecolor=black,linewidth=0.6pt](19,2)(19,3)
\psline[unit=0.3cm,linecolor=black,linewidth=0.6pt](20.5,3.5)(20.5,6)

\psline[unit=0.3cm,linecolor=black,linewidth=0.3pt,linestyle=dashed](15,9)(22,9)
\psline[unit=0.3cm,linecolor=black,linewidth=0.3pt,linestyle=dashed](13.5,7.5)(20.5,7.5)
\psline[unit=0.3cm,linecolor=black,linewidth=0.3pt,linestyle=dashed](13.5,6.5)(20.5,6.5)
\psline[unit=0.3cm,linecolor=black,linewidth=0.3pt,linestyle=dashed](13.5,5.5)(20.5,5.5)
\psline[unit=0.3cm,linecolor=black,linewidth=0.3pt,linestyle=dashed](12,4)(19,4)
\psline[unit=0.3cm,linecolor=black,linewidth=0.3pt,linestyle=dashed](12,3)(19,3)
\psline[unit=0.3cm,linecolor=black,linewidth=0.3pt,linestyle=dashed](10.5,1.5)(17.5,1.5)
\psline[unit=0.3cm,linecolor=black,linewidth=0.3pt,linestyle=dashed](10.5,0.5)(17.5,0.5)

\put(-1,-2){(a)}
\put(6.5,-2){(b)}

\put(6.5,1.35){\,\,$\wr\,\wr$}

\put(5,-1){$\epsilon_{2}=-2\epsilon_{1}$}
\end{pspicture}}

\end{array}$
\caption{a) The figure shows the partition along the preffered
direction for the self-dual case for the toric diagram ${\cal
O}(-1)\oplus {\cal O}(-1)\mapsto \mathbb{P}^{1}$. b) the same as in
a) but for $\epsilon_{2}=-2\epsilon_{1}$. } \label{f11}\end{center}
\end{figure}
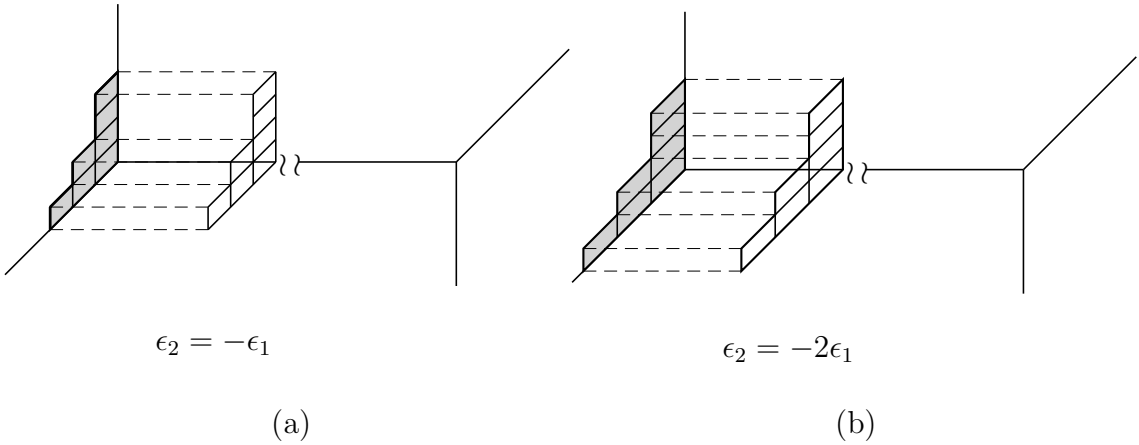

\subsection{$q,t$ slices and the boundary of the Young diagram}
The generating function of 3D partitions is not difficult to
calculate since we need only to specialize the parameters $q_{a}$.
From the discussion in the previous discussion it follows that for
$\nu=\emptyset$,\bea q_{a}=\left\{
                                 \begin{array}{ll}
                                   t, & \,\,a\geq 0 \\
                                   q, & \,\,a<0\,.
                                 \end{array}
                               \right.\eea

The partition function $Z_{3D}({\bf q})$ becomes \bea
Z_{3D}(t,q)=\prod_{i,j=1}^{\infty}(1-t^{i}q^{j-1})\,. \eea

But, in general, the shape of $\nu$ will determine whether a slice
is counted with parameter $t$ or parameter $q$.

\begin{figure}\begin{center}
$\begin{array}{c@{\hspace{1in}}c} \multicolumn{1}{l}{\mbox{}} &
    \multicolumn{1}{l}{\mbox{}} \\ [-0.53cm]
{
\begin{pspicture}(7,0)(4,5)
\psframe[unit=0.75cm, linestyle=none, fillstyle=solid,
fillcolor=lightgray](0,0)(1,1) \psframe[unit=0.75cm, linestyle=none,
fillstyle=solid, fillcolor=lightgray](1,0)(2,1)
\psframe[unit=0.75cm, linestyle=none, fillstyle=solid,
fillcolor=lightgray](0,1)(1,2) \psframe[unit=0.75cm, linestyle=none,
fillstyle=solid, fillcolor=lightgray](1,1)(2,2)
\psframe[unit=0.75cm, linestyle=none, fillstyle=solid,
fillcolor=lightgray](2,0)(3,1)\psframe[unit=0.75cm, linestyle=none,
fillstyle=solid, fillcolor=lightgray](0,2)(1,3)
\psframe[unit=0.75cm, linestyle=none, fillstyle=solid,
fillcolor=lightgray](0,3)(1,4) \psframe[unit=0.75cm, linestyle=none,
fillstyle=solid, fillcolor=lightgray](1,2)(2,3)

 \psgrid[unit=0.75cm, subgriddiv=1,
gridcolor=myorange, %
gridlabelcolor=white]%
(0,0)(6,6) \psline[unit=0.75cm,
linecolor=red,linestyle=dashed](5,0)(6.2,1.2) \psline[unit=0.75cm,
linecolor=red, linestyle=dashed](4,0)(6.2,2.2) \psline[unit=0.75cm,
linecolor=red,linestyle=dashed](3,0)(6.2,3.2) \psline[unit=0.75cm,
linecolor=blue](3,1)(6.2,4.2) \psline[unit=0.75cm,
linecolor=red,linestyle=dashed](2,1)(6.2,5.2) \psline[unit=0.75cm,
linecolor=blue](2,2)(6.2,6.2) \psline[unit=0.75cm,
linecolor=blue](2,3)(5.2,6.2) \psline[unit=0.75cm,
linecolor=red,linestyle=dashed](1,3)(4.2,6.2) \psline[unit=0.75cm,
linecolor=blue](1,4)(3.2,6.2) \psline[unit=0.75cm,
linecolor=red,linestyle=dashed](0,4)(2.2,6.2) \psline[unit=0.75cm,
linecolor=blue](0,5)(1.2,6.2) \psline[unit=0.75cm,
linecolor=red,linestyle=dashed](6,0)(6.2,0.2) \psline[unit=0.75cm,
linecolor=blue](0,6)(.2,6.2) \psline[unit=0.75cm,
linecolor=myorange](0,0)(0,7) \psline[unit=0.75cm,
linecolor=myorange](0,0)(7,0)

\put(5,3){$\pi\rightarrow
\,\prod_{a}q_{a}^{|\pi(a)|}=q^{\sum_{i=1}^{\infty}|\pi(\nu^{t}_{i}-i)|}\,t^{\sum_{j=1}^{\infty}|\pi(-\nu_{j}+j-1)|}$}
\put(5,2){$q=\mbox{blue}\,(\mbox{solid
line}),\,t=\mbox{red}\,(\mbox{dashed line})$}

\put(5,1){$\nu=(4,3,1)$}
\end{pspicture}
}
\\ [-0.5cm] \mbox{} & \mbox{}
\end{array}$
\caption{Slices of the 3D partitions are counted with parameters
$t$ and $q$ depending on the shape of $\nu$.}
\label{f12}\end{center}
\end{figure}
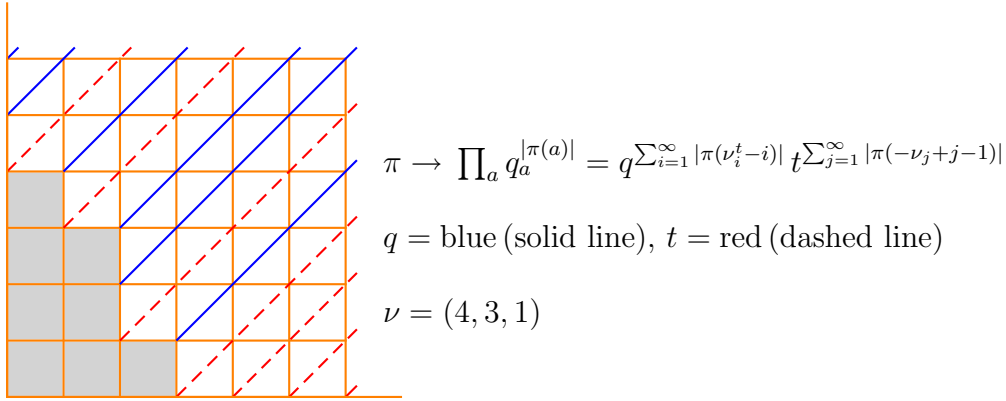
\newpage

For a non-trivial $\nu$, the map between $\{x^{\pm}_{m}\,|\,m\in
\mathbb{Z}+\tfrac{1}{2}\}$ and $\{t,q\}$ is given by \bea
\{x^{+}_{m}|m\in
D^{+}\}&=&\{t^{i}q^{-\nu_{i}}|i=1,2,3,\cdots\}\,,\\\nn
\{x^{-}_{m}|m\in
D^{-}\}&=&\{q^{j-1}t^{-\nu^{t}_{j}}|j=1,2,3,\cdots\}\,, \eea where
$D^{+}$ is the set of black boxes and $D^{-}$ is the set of white
boxes in the Maya diagram (\figref{f9}(b)) of $\nu$. If we consider
the $i^{th}$ white box from the left side, the number of black boxes to
the right of this box is given by $\nu_{i}$. This implies that there
is one to one correspondence, \bea \{(m_{1},m_{2})|m_{1}\in
D^{-}\,,m_{2}\in D^{+}\,,m_{1}\geq m_{2}\}\mapsto \{(i,j)\in
\nu\},\eea and therefore \bea \{(m_{1},m_{2})|m_{1}\in
D^{-}\,,m_{2}\in D^{+}\,,m_{1}< m_{2}\}\simeq \{(i,j)\notin
\nu\},\eea which implies that \bea Z_{\nu}=
\prod_{m_{1}<m_{2},m_{1,2}\in
D^{\pm}}(1-x_{m_{2}}^{+}x_{m_{1}}^{-})^{-1}=\prod_{(i,j)\notin\nu}(1-q^{j-\nu_{i}-1}t^{i-\nu^{t}_{j}})^{-1}\eea

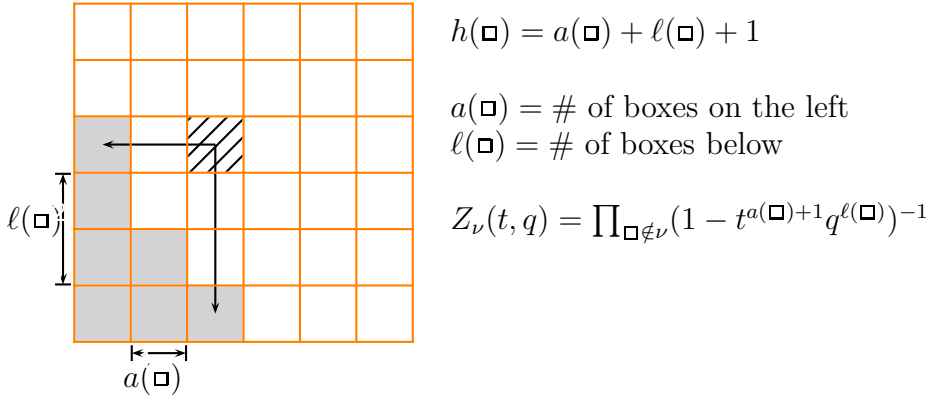
\begin{figure}\begin{center}
$\begin{array}{c@{\hspace{1in}}c} \multicolumn{1}{l}{\mbox{}} &
    \multicolumn{1}{l}{\mbox{}} \\ [-0.53cm]
{
\begin{pspicture}(5,0)(4,4)
\psframe[unit=0.75cm, linestyle=none, fillstyle=solid,
fillcolor=lightgray](0,0)(1,1) \psframe[unit=0.75cm, linestyle=none,
fillstyle=solid, fillcolor=lightgray](1,0)(2,1)
\psframe[unit=0.75cm, linestyle=none, fillstyle=solid,
fillcolor=lightgray](0,1)(1,2) \psframe[unit=0.75cm, linestyle=none,
fillstyle=solid, fillcolor=lightgray](1,1)(2,2)
\psframe[unit=0.75cm, linestyle=none, fillstyle=solid,
fillcolor=lightgray](2,0)(3,1)\psframe[unit=0.75cm, linestyle=none,
fillstyle=solid, fillcolor=lightgray](0,2)(1,3)
\psframe[unit=0.75cm, linestyle=none, fillstyle=solid,
fillcolor=lightgray](0,3)(1,4)

\psline[unit=0.75cm, arrows=->](2.5,3.5)(0.5,3.5)

\psline[unit=0.75cm, arrows=->](2.5,3.5)(2.5,0.5)
\psframe[unit=0.75cm, linestyle=none, fillstyle=hlines,
fillcolor=lightgray](2,3)(3,4)
\put(5,4){$h(\tableau{1})=a(\tableau{1})+\ell(\tableau{1})+1$}

\put(5,3){$a(\tableau{1})= \#$ of boxes on the left}
\put(5,2.5){$\ell(\tableau{1})=\#$ of boxes below}

\put(5,1.5){$Z_{\nu}(t,q)=\prod_{\tableau{1}\notin
\nu}(1-t^{a(\tableau{1})+1}q^{\ell(\tableau{1})})^{-1}$}

\psline[unit=0.75cm, arrows=|<->|](-0.2,1)(-0.2,3)
\put(-0.9,1.5){$\ell(\tableau{1})$}

\psline[unit=0.75cm, arrows=|<->|](1,-0.2)(2,-0.2)
\put(0.65,-0.6){$a(\tableau{1})$} \psgrid[unit=0.75cm, subgriddiv=1,
gridcolor=myorange, %
gridlabelcolor=white]%
(0,0)(6,6)
\end{pspicture}
}
\\ [0.0cm] \mbox{} & \mbox{}
\end{array}$
\caption{$Z_{\nu}(t,q)$ is the Hook series of the complement of
$\nu$.} \label{f13}\end{center}
\end{figure}

 For $\nu=\emptyset$, \bea
Z_{\emptyset}(t,q)&:=&M(t,q)=\prod_{i,j=1}^{\infty}(1-t^{i}q^{j-1})^{-1}\,.\eea
$M(t,q)$ is a two parameter generalization of the MacMahon function.

If we define $q=e^{i\epsilon_{1}}, t=e^{-i\epsilon_{2}}$ then
$\mbox{log}\,M(t,q)$ is symmetric in $\epsilon_{1},\epsilon_{2}$
(upto an infinite constant) \bea\nn
\mbox{log}\,M(t,q)&=&\frac{\zeta(3)}{\epsilon_{1}\epsilon_{2}}-i\frac{\zeta(2)}{2}\Big(\frac{\epsilon_{1}+\epsilon_{2}}{\epsilon_{1}\epsilon_{2}}\Big)
+\frac{\zeta(1)}{12}\Big(\frac{(\epsilon_{1}+\epsilon_{2})^{2}+\epsilon_{1}\epsilon_{2}}{\epsilon_{1}\epsilon_{2}}\Big)
+i\frac{\zeta(0)}{24}(\epsilon_{1}+\epsilon_{2})\\\nn
&&+\sum_{g_{1}+g_{2}\geq
2}(-1)^{g_{1}+g_{2}}\frac{B_{2g_{1}}B_{2g_{2}}B_{2g_{1}+2g_{2}-2}}{(2g_{1})!(2g_{2})!(2g_{1}+2g_{2}-2)}\,\epsilon_{1}^{2g_{1}-1}\epsilon_{2}^{2g_{2}-1}\,.
\eea

It is easy to show that (Appendix C) if we define \bea
 \widetilde{Z}_{\nu}(t,q)&:=&\frac{Z_{\nu}(t,q)} {Z_{\emptyset}(t,q)}\,,\eea
then $\widetilde{Z}_{\nu}(t,q)$ can be written as a product over
boxes of $\nu$,
 \bea
 \widetilde{Z}_{\nu}(t,q)=\prod_{s\in
\nu}(1-t^{a(s)+1}q^{\ell(s)})^{-1}=\prod_{s\in
\nu^{t}}(1-t^{\ell(s)+1}q^{a(s)})^{-1}\,. \eea The function
$\widetilde{Z}_{\nu}(t,q)$ is a specialization of the Macdonald
symmetric function $P({\bf x};q,t)$ \cite{macdonald}, \bea
\widetilde{Z}_{\nu}(t,q)=\,t^{-\frac{||\nu||^{2}}{2}}\,P_{\nu^t}(t^{-\rho};q,t)\,.
\eea Thus we see that this particular specialization of the
Macdonald function can be interpreted as counting skew plane
partitions such that the shape of $\nu$ determines whether to count
a box with $t$ or $q$.
\begin{figure}\begin{center}
$\begin{array}{c@{\hspace{1in}}c} \multicolumn{1}{l}{\mbox{}} &
    \multicolumn{1}{l}{\mbox{}} \\ [-0.53cm]
{
\begin{pspicture}(7,0)(4,4)
\psframe[unit=0.75cm, linestyle=none, fillstyle=solid,
fillcolor=lightgray](0,0)(1,1) \psframe[unit=0.75cm, linestyle=none,
fillstyle=solid, fillcolor=lightgray](1,0)(2,1)
\psframe[unit=0.75cm, linestyle=none, fillstyle=hlines*,
fillcolor=lightgray](0,1)(1,2) \psframe[unit=0.75cm, linestyle=none,
fillstyle=solid, fillcolor=lightgray](1,1)(2,2)
\psframe[unit=0.75cm, linestyle=none, fillstyle=solid,
fillcolor=lightgray](2,0)(3,1)\psframe[unit=0.75cm, linestyle=none,
fillstyle=solid, fillcolor=lightgray](0,2)(1,3)
\psframe[unit=0.75cm, linestyle=none, fillstyle=solid,
fillcolor=lightgray](0,3)(1,4) \psline[unit=0.75cm,
arrows=->](0.5,1.5)(3.5,1.5) \psline[unit=0.75cm,
arrows=->](0.5,1.5)(0.5,5)
\put(5,4){$h(\tableau{1})=a(\tableau{1})+\ell(\tableau{1})+1$}

\put(5,3){$a(\tableau{1})= \#$ of boxes on the right}
\put(5,2.5){$\ell(\tableau{1})=\#$ of boxes on top}

\put(5,1.5){$\prod_{\tableau{1}\in
\nu}(1-q^{h(\tableau{1})})^{-1}\,\,\rightarrow
\,\,=\prod_{\tableau{1}\in
\nu}(1-t^{a(\tableau{1})+1}q^{\ell(\tableau{1})})^{-1}$}
\put(5,0.5){$s_{(4\,2\,1)}(q)=\frac{1}{(1-q)^{3}(1-q^2)(1-q^3)(1-q^4)(1-q^6)}$}
\put(5,-0.2){$\widetilde{Z}_{(4\,2\,1)}(t,q)=\frac{1}{(1-t)^{3}(1-t\,q)(1-t^2\,q)(1-t^2\,q^2)(1-t^3\,q^3)}$}

\psline[unit=0.75cm, arrows=|<->|](-0.2,2)(-0.2,4)
\put(-1.3,2.1){$\ell(i,j)$} \psline[unit=0.75cm,
arrows=|<->|](1,-0.2)(2,-0.2) \put(0.65,-0.6){$a(i,j)$}
\psgrid[unit=0.75cm, subgriddiv=1,
gridcolor=myorange, %
gridlabelcolor=white]%
(0,0)(6,6)
\end{pspicture}}
\\ [0.0cm] \mbox{} & \mbox{}
\end{array}$
\caption{A Young diagram $\nu=(4\,2\,1)$.}
\label{f14}\end{center}
\end{figure}
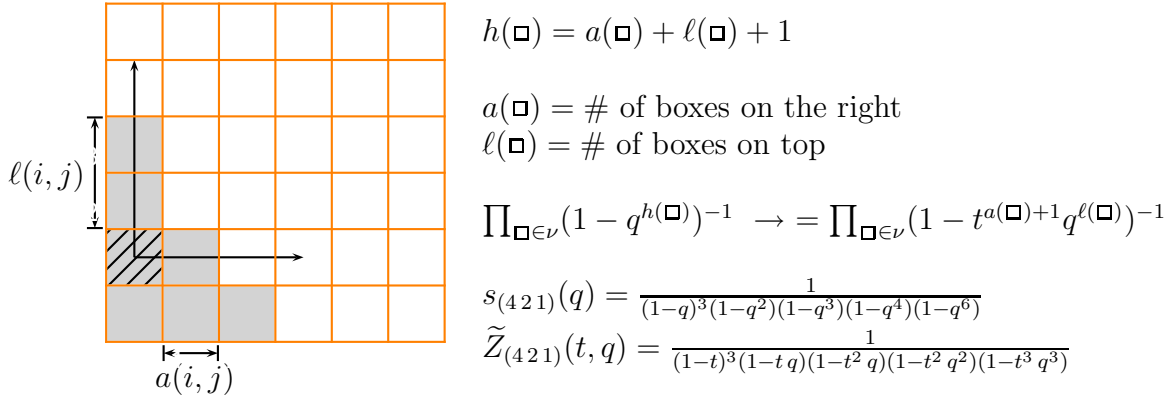

When all three partitions $(\lambda,\mu,\nu)$ are non-trivial, the
partition function (in diagonal slicing) is given by \bea
P_{diag}(\lambda\mu\nu)&=&\langle \lambda^{t}|\prod_{m\in
Z+\frac{1}{2}}\Gamma_{-\epsilon(m)}(x^{\epsilon(m)}_{m})|\mu
\rangle\,,\\\nn &=&\langle \lambda^{t}|\prod_{m\in
D^{+}}\Gamma_{-}(x^{+}_{m})\prod_{m\in
D^{-}}\Gamma_{+}(x^{-}_{m})|\mu\rangle\,\,\prod_{m_{1}<m_{2},m_{1}\in
D^{-},m_{2}\in D^{+}}(1-x^{+}_{m_{2}}x^{-}_{m_{1}})^{-1}\,,\\\nn
&=&t^{-|\lambda|}\,Z_{\nu}(t,q)\sum_{\eta}s_{\lambda^{t}/\eta}({\bf
x}^{+})s_{\mu/\eta}({\bf x}^{-})\\\nn
&=&\,t^{-\frac{|\lambda|}{2}}\,q^{-\frac{|\mu|}{2}}
Z_{\nu}(t,q)\sum_{\eta}\Big(\frac{q}{t}\Big)^{|\eta|/2}s_{\lambda^{t}/\eta}(t^{-\rho}q^{-\nu})s_{\mu/\eta}(t^{-\nu^{t}}q^{-\rho}),\eea
where $\rho=\{-\frac{1}{2},-\frac{3}{2},-\frac{5}{2},\cdots\}$.

To convert the above partition function in the diagonal slicing to
the partition function in the perpendicular slicing we multiply by
$q^{-n(\lambda^{t})}\,t^{-n(\mu)}$ \cite{ORV}. The perpendicular
partition function is then given by \bea
P_{\lambda\mu\nu}(t,q)&=&q^{-n(\lambda^{t})-\frac{|\mu|}{2}}\,t^{-n(\mu)-\frac{|\lambda|}{2}}\,
Z_{\nu}(t,q)\sum_{\eta}\Big(\frac{q}{t}\Big)^{|\eta|/2}s_{\lambda^{t}/\eta}(t^{-\rho}q^{-\nu})s_{\mu/\eta}(t^{-\nu^{t}}q^{-\rho})\\\nn
&=&q^{-\frac{||\lambda||^{2}}{2}}\,t^{-\frac{||\mu^{t}||^{2}}{2}}\,
Z_{\nu}(t,q)\sum_{\eta}\Big(\frac{q}{t}\Big)^{\frac{|\eta|+|\lambda|-|\mu|}{2}}s_{\lambda^{t}/\eta}(t^{-\rho}q^{-\nu})s_{\mu/\eta}(t^{-\nu^{t}}q^{-\rho}).\\\nn
\eea The refined topological vertex is given by \bea
C_{\lambda\mu\nu}(t,q)&=&q^{f(\nu)}\,t^{g(\nu)}\,q^{\frac{||\lambda||^2}{2}+\frac{||\mu||^{2}}{2}}
\frac{P_{\lambda\mu\nu}(t,q)}{M(t,q)}\\\nn &=&
q^{f(\nu)}\,t^{g(\nu)}\,\Big(\frac{q}{t}\Big)^{\frac{||\mu||^2}{2}}\,t^{\frac{\kappa(\mu)}{2}}\,
\frac{Z_{\nu}(t,q)}{M(t,q)}\,\sum_{\eta}\Big(\frac{q}{t}\Big)^{\frac{|\eta|+|\lambda|-|\mu|}{2}}
s_{\lambda^{t}/\eta}(t^{-\rho}q^{-\nu})s_{\mu/\eta}(t^{-\nu^{t}}q^{-\rho})\\\nn
&=&
q^{f(\nu)}\,t^{g(\nu)}\,\Big(\frac{q}{t}\Big)^{\frac{||\mu||^2}{2}}\,t^{\frac{\kappa(\mu)}{2}}\,
\widetilde{Z}_{\nu}(t,q)\,\sum_{\eta}\Big(\frac{q}{t}\Big)^{\frac{|\eta|+|\lambda|-|\mu|}{2}}
s_{\lambda^{t}/\eta}(t^{-\rho}q^{-\nu})s_{\mu/\eta}(t^{-\nu^{t}}q^{-\rho}).\eea

The functions $f(\nu)$ and $g(\nu)$ are such that \bea
f(\nu)+g(\nu)=\tfrac{||\nu||^2}{2}\,. \eea This one relation is not
enough to fix the two functions $f(\nu)$ and $g(\nu)$ therefore we
will make a choice here and take $g(\nu)=0$. The one partition
topological vertex is equal to a specialization of the Schur
function, $s_{\nu^{t}}(q^{-\rho})$, and with the above choice of
$g(\nu)$ the one partition refined topological vertex is equal to
the generalization of the Schur function,\bea
C_{\emptyset\,\emptyset\,\nu}(t,q)=q^{\frac{||\nu||^{2}}{2}}\,\widetilde{Z}_{\nu}(t,q)\,\left\{
                                                                                      \begin{array}{ll}
                                                                                       = q^{\frac{\kappa(\nu)}{2}}Q_{\nu}(q^{-\rho};t,q), \\
                                                                                       = \Big(\frac{q}{t}\Big)^{\frac{||\nu||^2}{2}}\,P_{\nu^{t}}(t^{-\rho};q,t)\,.
                                                                                      \end{array}
                                                                                    \right.
\eea where $P_{\nu}({\bf x};q,t)$ is the Macdonald function and
$Q_{\nu}({\bf x};q,t)$ is dual of the Macdonald function. As we will
show in the next section this choice gives correct A-model partition
functions. The refined vertex becomes, \bea\nn\shadowbox{$
C_{\lambda\,\mu\,\nu}(t,q)=\Big(\frac{q}{t}\Big)^{\frac{||\mu||^2+||\nu||^2}{2}}\,t^{\frac{\kappa(\mu)}{2}}\,
P_{\nu^{t}}(t^{-\rho};q,t)\,
\sum_{\eta}\Big(\frac{q}{t}\Big)^{\frac{|\eta|+|\lambda|-|\mu|}{2}}s_{\lambda^{t}/\eta}(t^{-\rho}q^{-\nu})
s_{\mu/\eta}(t^{-\nu^{t}}q^{-\rho})
 $}\eea
For $t=q$ the above reduces to the usual topological vertex since
$P_{\nu}(q^{-\rho};q,q)=s_{\nu}(q^{-\rho})$.

\subsection{Framing factors}

Recall that the framing factor arises whenever the ${\mathbb P}^{1}$
along which the two vertices are glued has a global geometry other
than ${\cal O}(-1)\oplus {\cal O}(-1)\mapsto {\mathbb P}^{1}$. For
$q=t$ the framing factor is given by $q^{-\frac{\kappa(\mu)}{2}}$.

Since the two directions orthogonal to the preferred direction
correspond to the parameters $t$ and $q$ therefore rotations along
these directions will be counted with these two parameters. Rotating
the $\nu$ diagram along the first row gives $n(\nu)$ extra boxes
which we count with the parameter $t$. Then, rotating the diagram
along the first column gives $n(\nu^{t})$ boxes which we count with
the parameter $q$. Thus, the framing factor along the preferred
direction is given by
 \bea
f_{\nu}(t,q):=(-1)^{|\nu|}\,t^{n(\nu)}\,q^{-n(\nu^{t})}=(-1)^{|\nu|}\,\Big(\frac{t}{q}\Big)^{n(\nu)}\,q^{-\frac{\kappa(\nu)}{2}}\,,
\eea where we have introduced a factor of $(-1)^{|\nu|}$ for later
convenience.

The framing factor when $t\neq q$ can also be calculated from the
geometry of instanton moduli spaces following \cite{tachikawa}. The
simplest case is to take the $U(1)$ theory with the charge $k$
instanton moduli space given by $\mbox{Sym}^{k}(\mathbb{C}^{2})$ and
consider a supersymmetric quantum mechanics on this moduli space
with coupling to an external gauge field. A shown in
\cite{tachikawa} the effect of this extra coupling is to introduce
an extra term, which is the framing factor, in the partition
function given by \bea
 e^{\sum_{(i,j)\in
\nu}(\epsilon_{1}(i-1)+\epsilon_{2}(j-1))}\,&=&\,t^{\sum_{(i,j)\in
\nu}(i-1)}\,q^{-\sum_{(i,j)\in \nu}(j-1)}\,\\\nn
&=&\,t^{n(\nu)}q^{-n(\nu^{t})}\,. \eea This is exactly the framing
factor one gets from the combinatorics of 3D partitions.

\section{Appendix B: Gromov-Witten Theory and Refined Partition Function:
The case of ${\cal O}(-1)\oplus {\cal O}(-1)\mapsto \mathbb{P}^{1}$}

It is interesting to consider the case of ${\cal O}(-1)\oplus {\cal
O}(-1)\mapsto \mathbb{P}^{1}$ from the point of view of
Gromov-Witten theory \cite{FP}. In this case the refined partition
function can be obtained from the Gromov-Witten theory and has a
interesting interpretation from the localization point of view which
might be useful for other toric CY3-folds.

The multi-cover contribution is given by \cite{FP}\bea
C(g,d)&=&\int_{[\overline{\cal {M}}_{g,0}(\mathbb{P}^{1},d)]^{vir}}c_{top}(R^{1}\pi_{*}\mu^{*}N)\\\nn
&=&d^{2g-3}\,\frac{|B_{2g}(2g-1)|}{(2g)!}\,,\,\,\,g\geq 0\,. \eea
Where $B_{n}$ are Bernoulli numbers defined as $\sum_{m\geq
0}\frac{B_{m}}{m!}t^{m}=\frac{t}{e^{t}-1}$.

The partition function can be calculated using the multicover
contribution and is given by \bea Z=\mbox{Exp}\Big(\sum_{g\geq
0}\sum_{d\geq
1}\lambda_{s}^{2g-2}Q^{d}C(g,d)\Big)=\prod_{n=1}^{\infty}(1-q^{n}\,Q)^{-n}\,,\,\,\,q=e^{-\lambda_{s}}\,.\eea

 However, using
localization $C(g,d)$ can also be written as \cite{FP,BP}\bea
C(g,d)&=&d^{2g-3}\sum_{g_{1}+g_{2}=g, g_{1,2}\geq
0}C_{g_{1},g_{2}}\,,\,\,\,\,\,\,C_{g_{1},g_{2}}=b_{g_{1}}b_{g_{2}}\\\nn
b_{g}&=&\left\{
        \begin{array}{ll}
          1, & \hbox{for}\,\,\,g=0 \\
          \int_{\overline{\cal{M}}_{g,1}}\psi_{1}^{2g-2}\lambda_{g}=\frac{2^{2g-1}-1}{2^{2g-1}}\,\frac{|B_{2g}|}{(2g)!}, &
\hbox{for}\,\,\,g\geq 1\,.
        \end{array}
      \right.
\eea Essentially the contribution $b_{g_{1}}b_{g_{2}}$ is from
degenerate worldsheets of genus $g_{1}+g_{2}$ such that the two
components of genus $g_{1}$ and $g_{2}$ map to the two fixed points
of $X$. By weighing the contribution the two fixed points
differently we get \bea
\widehat{Z}&=&\mbox{Exp}\Big(\sum_{g_{1},g_{2}\geq
0}\lambda_{1}^{2g_{1}-1}\lambda_{2}^{2g_{2}-1}Q^{d}\,d^{2g-3}\,b_{g_{1}}b_{g_{2}}\Big)\\\nn
&=&
\prod_{n,m=1}^{\infty}\Big(1-q^{n-\frac{1}{2}}t^{m-\frac{1}{2}}\,Q\Big)\,,\,\,\,q=e^{-\lambda_{1}}\,,t=e^{-\lambda_{2}}\,.\eea
Which is exactly the refined topological string partition function.

A similar calculation for the case of target space $\mathbb{C}^{3}$
gives the constant map contribution to the topological string
partition function, \textit{i.e.}, the MacMahon function.
 \bea
C_{g}:=\int_{\overline{\cal{M}}_{g}}\lambda_{g-1}^{3}&=&\frac{|B_{2g-2}|}{2g-2}\sum_{g_{1}+g_{2}=g}b_{g_{1}}b_{g_{2}}\\\nn
&=&|\zeta(3-2g)|\sum_{g_{1}+g_{2}=g}b_{g_{1}}b_{g_{2}}=\sum_{g_{1}+g_{2}=g}\widehat{C}_{g_{1},g_{2}}\,,\\\nn
\widehat{C}_{g_{1},g_{2}}&=&|\zeta(3-2g_{1}-2g_{2})|\,b_{g_{1}}\,b_{g_{2}}\,
\eea where $B_{n}$ is the $n^{th}$ Bernoulli number and we have used the
identity $B_{n}=(-1)^{n+1}n\,\zeta(1-n)$. \bea\nn
M(q)=\mbox{Exp}\Big(\sum_{g\geq
0}\lambda^{2g-2}C_{g}\Big)=\prod_{n\geq
1}(1-q^{n})^{-n}\,\,\,\,\Rrightarrow\,\,\,\,
\widetilde{M}(t,q)&=&\mbox{Exp}\Big(\sum_{g_{1},g_{2}\geq
0}\lambda_{1}^{2g_{1}-1}\lambda_{2}^{2g_{2}-1}\widehat{C}_{g_{1},g_{2}}\Big)\\\nn
&=&\prod_{n,m\geq 1}(1-q^{n-\frac{1}{2}}t^{m-\frac{1}{2}})^{-1}\,.
\eea The function $M(q)$ is the MacMahon function and is the
generating function of the number of 3D partitions, \bea
M(q)&=&\sum_{n=0}^{\infty}p(n)\,q^{n}\,\\\nn p(n)&=&\mbox{\# of
plane partitions of $n$}\,.\eea The function $\widetilde{M}(t,q)$ also
has a combinatorial interpretation in terms of 3D partitions, \bea
\widetilde{M}(t,q)=\sum_{\pi}p(|\pi_{+}|,|\pi_{-}|)t^{|\pi_{+}|}q^{|\pi_{-}|}\,,\eea
where $\pi_{+}$ and $\pi_{-}$ are two parts of the 3D partition
$\pi$ obtained by cutting $\pi$ by the plane $x=y$. $|\pi_{+}|$ and
$|\pi_{-}|$ are the volumes of the two parts such that
$|\pi_{+}|+|\pi_{-}|=|\pi|$ and $p(n,m)$ is the number of 3D
partitions with $(|\pi_{+}|,|\pi_{-}|)=(n,m)$.

\section{Appendix C: An Important Identity}
In this appendix we prove the identity \bea
\widetilde{Z}_{\nu}(t,q)=\frac{Z_{\nu}(t,q)}{M(t,q)}. \eea

\textbf{Proof:} Consider the following identity \cite{NY1}\bea
\sum_{i,j=1}^{\infty}\left (
t^{i-\nu_{1,j}}q^{j-\nu_{2,i}^{t}-1}-t^{i}q^{j-1}\right
)=\sum_{s\in\nu_{1}}t^{-\ell_{\nu_{1}}(s)}q^{-a_{\nu_{2}}(s)-1}+\sum_{s\in\nu_{2}}t^{\ell_{\nu_{2}}(s)+1}q^{a_{\nu_{1}}(s)}.
\eea Let us set $\nu_{1}=\nu_{2}=\nu^{t}$ \bea\nn
\sum_{i,j=1}^{\infty}\left ( t^{i-\nu_{j}^{t}}q^{j-\nu_{i}-1}-t^{i}q^{j-1}\right ) &=& \sum_{s\in\nu^{t}}\left ( t^{-\ell_{\nu^{t}}(s)}q^{-a_{\nu^{t}}(s)-1}+t^{\ell_{\nu^{t}}(s)+1}q^{a_{\nu^{t}}(s)}\right ) \\
&=&\sum_{s\in\nu}\left (
t^{-a_{\nu}(s)}q^{-\ell_{\nu}(s)-1}+t^{a_{\nu}(s)+1}q^{\ell_{\nu}(s)}\right
). \eea The substitutions $q\rightarrow q^{m}$ and $t\rightarrow
t^{m}$ will allow us to find a formal expansion of $\log$:
\begin{eqnarray}
\sum_{m=1}^{\infty}\frac{1}{m}\sum_{i,j=1}^{\infty}
t^{m(i-\nu_{j}^{t})}q^{m(j-\nu_{i}-1)}&-&\sum_{m=1}^{\infty}\frac{1}{m}\sum_{(i,j)\in\nu}t^{m(i-\nu_{j}^{t})}q^{m(j-\nu_{i}-1)}\\
\nonumber&=& \sum_{m=1}^{\infty}\frac{1}{m}\sum_{i,j=1}^{\infty}t^{m
i}q^{m(j-1)}+\sum_{m=1}^{\infty}\frac{1}{m}\sum_{s\in\nu}t^{m(a_{\nu}(s)+1)}q^{m\ell_{\nu}(s)}.
\end{eqnarray}

If the order of the $m-$summation is changed with the one following
it, one actually gets the identity we are trying to prove:
\begin{eqnarray}
\sum_{i,j=1}^{\infty}\log\left (
1-t^{i-\nu_{j}^{t}}q^{j-\nu_{i}-1}\right)&-&\sum_{(i,j)\in\nu}\log\left
(1-t^{i-\nu_{j}^{t}}q^{j-\nu_{i}-1} \right )\\
\nonumber&=&\sum_{i,j=1}^{\infty}\log\left (1-t^{i}q^{j-1}
\right)+\sum_{s\in\nu}\log\left(1-t^{a_{\nu}(s)+1}q^{\ell_{\nu}(s)}
\right).
\end{eqnarray}
This can be put in a more suggestive form by exponentiating both
sides and taking the inverse
\begin{eqnarray}
\frac{\prod_{i,j=1}^{\infty}\left (
1-t^{i-\nu_{j}^{t}}q^{j-\nu_{i}-1}\right)^{-1}}{\prod_{(i,j)\in\nu}\left
(
1-t^{i-\nu_{j}^{t}}q^{j-\nu_{i}-1}\right)^{-1}}&=&\prod_{(i,j)\notin\nu}\left
( 1-t^{i-\nu_{j}^{t}}q^{j-\nu_{i}-1}\right)^{-1} \\ \nonumber
&=&\prod_{i,j=1}^{\infty}\left (1-t^{i}q^{j-1}
\right)^{-1}\prod_{s\in\nu}\left(1-t^{a_{\nu}(s)+1}q^{\ell_{\nu}(s)}
\right)^{-1}.
\end{eqnarray}

\section{Appendix D: Schur Functions}
\par{This appendix should serve as a review of the definition and some properties we have used of the Schur functions. Before defining the Schur function, let us introduce the antisymmetric polynomial $a_{\alpha}$ of a finite number of variables $\{ x_{i}\}_{i=1}^{n}$}:
\bea a_{\alpha}(x_{1},\dots,x_{n})=\sum_{\omega\in S_{n}}
\epsilon(\omega)\ \omega(x^{\alpha}) \eea where $\epsilon(\omega)$
serves as the antisymmetrizer for an element $\omega$ of the symmetric
group $S_{n}$ and $x^{\alpha}$ is a shorthand notation for the
monomial $x_{1}^{\alpha_{1}}\dots x_{n}^{\alpha_{n}}$. We have a
non-vanishing polynomial $a_{\alpha}(x_{i})$ only if all
$\alpha_{i}$'s are different. That allows us to put the exponents of
the variables, without loss of generality, into a particular
ordering: $\alpha_{1}>\alpha_{2}>\dots>\alpha_{n}\geq0$. The freedom of
choosing such an ordering among $\alpha_{i}$'s  enables us to connect
the polynomials $a_{\alpha}(x_{i})$ to partitions, so we can write
$\alpha=\lambda+\delta$ for a partition $\lambda$ with length
$\ell(\lambda)\leq n$ and $\delta=(n-1,n-2,\dots,1,0)$. The
polynomial $a_{\alpha}(x_{i})$ can be now rewritten in terms of the
partition $\lambda$ as \bea
a_{\lambda+\delta}(x_{1},\dots,x_{n})=\sum_{\omega}\epsilon(\omega)\
\omega(x^{\lambda+\delta}). \eea This particular form of the
polynomial $a_{\alpha}( x_{i} )$ makes it more evident to express
this sum as a determinant \bea
a_{\lambda+\delta}(x_{1},\dots,x_{n})=det\left (
x_{i}^{\lambda_{j}+n-j}\right )_{1\leq i,j\leq n}. \eea This form of
$a_{\alpha}(x_{i})$ makes it evident that it is divisible in the
ring of polynomials in the variables $\{ x_{i}\}_{i=1}^{n}$ with
integer coefficients, ${\mathbb Z}[x_{1},\dots,x_{n}]$, by any
difference of the form $x_{i}-x_{j}$ with $1\leq i<j \leq n$. Then
it is divisible by their product as well, hence, by the Vandermonde
determinant \bea \prod_{1\leq i<j \leq n} (x_{i}-x_{j})=det\left (
x_{i}^{n-j}\right ). \eea Let us denote the above product by
$a_{\delta}$. Now we are ready to define the Schur function
$s_{\lambda}(x_{i})$ as a quotient \bea
s_{\lambda}(x_{1},\dots,x_{n})\equiv a_{\lambda+\delta}/a_{\delta}.
\eea Note that $s_{\lambda}(x_{i})$ is symmetric and its definition
makes sense as long as $\lambda\in {\mathbb Z}^{n}$ is an integer
vector such that $\lambda+\delta$ does not have any negative parts.
The Schur functions $s_{\lambda}(x_{i})$ form an orthonormal basis
for the symmetric polynomials which is a subring
$\Lambda_{n}={\mathbb Z}[x_{1},\dots,x_{n}]^{S_{n}}$. The
orthonormality requires the definition of the scalar product of
symmetric functions. Let us give first the definition and describe
later the individual ingredients we use. The scalar product on
$\Lambda$ is a ${\mathbb Z}$-valued bilinear form $\langle u,v
\rangle$ such that the bases $h_{\lambda}$ and $m_{\mu}$ are dual to
each other which is precisely that they satisfy the following
relationship: \bea \langle h_{\lambda},m_{\mu}\rangle
=\delta_{\lambda\mu} \eea with the Kronecker delta
$\delta_{\lambda\mu}$. Given a partition $\lambda$, $m_{\lambda}$ is
defined as the sum over all permutations of the parts of
$\lambda=(\lambda_{1},\dots,\lambda_{n})$ \bea
m_{\lambda}(x_{1},\dots,x_{n})=\sum_{\alpha}x^{\alpha}. \eea
$h_{\lambda}$ is defined in terms of the complete symmetric
functions $h_{r}$ as
$h_{\lambda}=h_{\lambda_{1}}h_{\lambda_{2}}\dots$, with \bea
h_{r}=\sum_{|\lambda|=r}m_{\lambda} \eea where $r$ is the degree of
$h_{r}$. Finally, $\Lambda$ is the free ${\mathbb Z}$ module which
is generated by the bases $m_{\lambda}$ for all $\lambda$. Any
symmetry function can be written as a linear combination of the
Schur functions with the coefficients calculable having the scalar
product defined.
\par{The skew Schur function $s_{\lambda/\mu}$ is defined by }
\bea \langle s_{\lambda/\mu},s_{\nu}\rangle = \langle
s_{\lambda},s_{\mu}s_{\nu}\rangle \eea where $\lambda$ interlaces
$\mu$. We can use the fact that the Schur functions form an
orthonormal basis and write the skew Schur function in another form
\bea s_{\lambda/\mu}=\sum_{\nu}c_{\mu\nu}^{\lambda}s_{\nu} \eea
where the $c_{\mu\nu}^{\lambda}$'s are defined by \bea
s_{\mu}s_{\nu}=\sum_{\lambda}c_{\mu\nu}^{\lambda}s_{\lambda} \eea
and are integers. An equivalent definition of the skew Schur
function can be given in terms of the semi-standard tableau, which
is obtained by assigning a positive integer to each box in a skew
partition such that the numbers weakly increase along the rows, and
strictly increase along the columns.

Having introduced the Schur and skew Schur functions, let us also
mention the identities we have made use of. If we sum two Schur
functions with two sets of independent variables
$x=(x_{1},x_{2},\dots)$ and $y=(y_{1},y_{2},\dots)$ over all
partitions, we get \bea
\sum_{\lambda}s_{\lambda}(x)s_{\lambda}(y)=\prod_{i,j}(1-x_{i}y_{j})^{-1}.
\eea Had we changed the partition from $\lambda$ to $\lambda^{t}$
in one of the Schur functions, we would end up with \bea
\sum_{\lambda}s_{\lambda^{t}}(x)s_{\lambda}(y)=\prod_{i,j}(1+x_{i}y_{j}).
\eea The Schur function of the variables $(1,q,q^{2},\dots)$ can be
expressed in terms of a product of terms which are dependent on the
hook length of the partition up to an overall factor: \bea
s_{\lambda}(1,q,q^{2},\cdots)=q^{n(\lambda)}\prod_{s\in
\lambda}(1-q^{h_{\lambda}(s)})^{-1} \eea where $n(\lambda)$ is
defined as \bea n(\lambda)\equiv\sum_{i}(i-1)\lambda_{i}. \eea It is
not hard to show that $n(\lambda)$ can be calculated alternatively
using the arm length as well as the leg lengths: \bea
n(\nu)=\sum_{i}(i-1)\nu_{i}=\frac{1}{2}\sum_{i}\nu^{t}_{i}(\nu^{t}_{i}-1)=\sum_{s\in
\nu} a'(s)=\sum_{s\in \nu}a_{\nu}(s)\,,\\
n(\nu^{t})=\sum_{i}(i-1)\nu^{t}_{i}=\frac{1}{2}\sum_{i}\nu_{i}(\nu_{i}-1)=\sum_{s\in
\nu} \ell'(s)=\sum_{s\in \nu}\ell_{\nu}(s)\,,\\\nn
 \eea
with the same $\ell_{\nu}(s)$ and $a_{\nu}(s)$ defined previously in
the text, and we introduce $\ell'(s)=j-1$ and
$a'(s)=i-1$\footnote{For the sake of completeness, let us also
mention some useful relations among $h(\lambda)$, $n(\lambda)$ and
$\kappa(\lambda)=2\sum_{(i,j)\in\lambda}(j-i)$: \bea \nn \sum_{s\in
\lambda}h_{\lambda}(s)&=&n(\lambda)+n(\lambda^{t})+|\lambda|=2\,n(\lambda)+\frac{1}{2}\kappa(\lambda)+|\lambda|\\\nn
\kappa(\lambda)&=&2(n(\lambda^{t})-n(\lambda))\,.\eea }.
\par{Two skew Schur functions $s_{\lambda/\nu}(x)$ and $s_{\nu/\mu}(y)$ can be summed over all possible partitions satisfying $\mu\prec\nu\prec\lambda$ to give another skew Schur function}
\bea
s_{\lambda/\mu}(x,y)=\sum_{\nu}s_{\lambda/\nu}(x)s_{\nu/\mu}(y).
\eea The above sum can be generalized to multiple sums in the
following way \bea
s_{\lambda/\mu}(x^{(1)},\dots,x^{(n)})=\sum_{(\nu)}\prod_{i=1}^{n}s_{\nu^{(i)}/\nu^{(i-1)}}(x^{(i)})
\eea where the summation is again over all partitions
$(\nu)=(\nu^{(0)},\dots,\nu^{(n)})$ satisfying the same interlacing
condition generalized to more partitions,
$\mu=\nu^{(0)}\prec\nu^{(1)}\prec\dots\prec\nu^{(n-1)}\prec\nu^{(n)}=\lambda$.

\bibliography{physics}

\end{document}